\documentclass[twocolumn, dvipdfmx]{aastex631}

\usepackage{amsmath}
\usepackage{color}
\usepackage{comment}
\usepackage{url}

\usepackage{graphicx}

\received{March 30, 2026}
\accepted{October 20, 2026}
\submitjournal{AAS (ApJ)}

\graphicspath{{./}{figures/}}

\begin{document}



\title{FAUST XXXII. Chemically layered structures and ionisation in the disk and envelope of NGC 1333 IRAS~4C}

\correspondingauthor{Shota Notsu}
\email{shota.notsu@eps.s.u-tokyo.ac.jp}

\author[0000-0003-2493-912X]{Shota Notsu}
\affiliation{Department of Earth and Planetary Science, Graduate School of Science, The University of Tokyo, 7-3-1 Hongo, Bunkyo-ku, Tokyo 113-0033, Japan}
\affiliation{Star and Planet Formation Laboratory, Pioneering Research Institute, RIKEN, 2-1 Hirosawa, Wako, Saitama 351-0198, Japan}

\author[0000-0002-3297-4497]{Nami Sakai}
\affiliation{Star and Planet Formation Laboratory, Pioneering Research Institute, RIKEN, 2-1 Hirosawa, Wako, Saitama 351-0198, Japan}

\author[0000-0001-7511-0034]{Yichen Zhang}
\affiliation{Department of Astronomy, School of Physics and Astronomy, Shanghai Jiao Tong University, 800 Dongchuan Road, Shanghai 200240, People's Republic of China}
\affiliation{State Key Laboratory of Dark Matter Physics, School of Physics and Astronomy, Shanghai Jiao Tong University, 800 Dongchuan Rd., Shanghai 200240, People's Republic of China}

\author[0000-0002-6729-3640]{Ana L\'{o}pez-Sepulcre}
\affiliation{Institut de Plan\'{e}tologie et d'Astrophysique de Grenoble (IPAG), Universit\'{e} Grenoble Alpes, CNRS, F-38000 Grenoble, France}
%
%
\affiliation{Institut de Radioastronomie Millim\'{e}trique (IRAM), F-38406 Saint-Martin d'H\'{e}res, France}

\author[0000-0002-7570-5596]{Claire J. Chandler}
\affiliation{National Radio Astronomy Observatory (NRAO), PO Box O, Socorro, NM 87801, USA}

\author[0000-0002-9927-2705]{Ziwei E. Zhang}
\affiliation{Star and Planet Formation Laboratory, Pioneering Research Institute, RIKEN, 2-1 Hirosawa, Wako, Saitama 351-0198, Japan}

\author[0000-0001-9664-6292]{Cecilia Ceccarelli}
\affiliation{Institut de Plan\'{e}tologie et d'Astrophysique de Grenoble (IPAG), Universit\'{e} Grenoble Alpes, CNRS, F-38000 Grenoble, France}

\author[0000-0003-1514-3074]{Claudio Codella}
\affiliation{Istituto Nazionale di Astrofisica (INAF), Osservatorio Astrofisico di Arcetri, Largo E. Fermi 5, I-50125 Florence, Italy}
\affiliation{Institut de Plan\'{e}tologie et d'Astrophysique de Grenoble (IPAG), Universit\'{e} Grenoble Alpes, CNRS, F-38000 Grenoble, France}

\author[0000-0003-3283-6884]{Yuri Aikawa}
\affiliation{Department of Astronomy, Graduate School of Science, The University of Tokyo, 7-3-1 Hongo, Bunkyo-ku, Tokyo 113-0033, Japan}

\author[0000-0002-3972-1978]{Jaime E. Pineda}
\affiliation{Center for Astrochemical Studies, Max-Planck-Institut f\"{u}r extraterrestrische Physik (MPE), Gie$\beta$enbachstr. 1, D-85741 Garching, Germany}
\author[0000-0002-8502-6431]{Brian Svoboda}
\affiliation{National Radio Astronomy Observatory (NRAO), PO Box O, Socorro, NM 87801, USA}
\author[0000-0002-7945-064X]{Felipe O. Alves}
\affiliation{Institut de Radioastronomie Millim\'{e}trique (IRAM), F-38406 Saint-Martin d'H\'{e}res, France}

\author[0000-0001-5121-5683]{Nadia Balucani}
\affiliation{Department of Chemistry, Biology, and Biotechnology, The University of Perugia, Via Elce di Sotto 8, 06123 Perugia, Italy}

\author[0000-0001-9249-7082]{Eleonora Bianchi}
\affiliation{Istituto Nazionale di Astrofisica (INAF), Osservatorio Astrofisico di Arcetri, Largo E. Fermi 5, I-50125 Florence, Italy}
\affiliation{Institut de Plan\'{e}tologie et d'Astrophysique de Grenoble (IPAG), Universit\'{e} Grenoble Alpes, CNRS, F-38000 Grenoble, France}

\author[0000-0003-0167-0746]{Mathilde Bouvier}
\affiliation{Transdisciplinary Research area (TRA) `Matter' - Hertz chair for Astrochemistry, Universit\"{a}t Bonn, 53121 Bonn, Germany}
\affiliation{Leiden Observatory, Leiden University, PO Box 9513, 2300 RA, Leiden, The Netherlands}
\author[0000-0002-2189-6278]{Gemma Busquet}
\affiliation{Departament de F\'{i}sica Qu\`{a}ntica i Astrof\'{i}sica, Institut de Ci\`{e}ncies del Cosmos, Universitat de Barcelona (IEEC-UB), Mart\'{i} Franq\`{e}s
1, 08028 Barcelona, Catalunya, Spain}
\author[0000-0003-1481-7911]{Paola Caselli}
\affiliation{Center for Astrochemical Studies, Max-Planck-Institut f\"{u}r extraterrestrische Physik (MPE), Gie$\beta$enbachstr. 1, D-85741 Garching, Germany}
\author[0000-0002-4463-6663]{Emmanuel Caux}
\affiliation{IRAP, Universit\'{e} de Toulouse, CNRS, CNES, UPS, Toulouse, France}
\author[0000-0001-6752-5109]{Steven Charnley}
\affiliation{Astrochemistry Laboratory, Code 691, NASA Goddard Space Flight Center, 8800 Greenbelt Road, Greenbelt, MD 20771, USA}
\author[0000-0002-7497-2713]{Spandan Choudhury}
\affiliation{Center for Astrochemical Studies, Max-Planck-Institut f\"{u}r extraterrestrische Physik (MPE), Gie$\beta$enbachstr. 1, D-85741 Garching, Germany}
\author[0000-0003-3713-8073]{Nicol\'{a}s Cuello}
\affiliation{Institut de Plan\'{e}tologie et d'Astrophysique de Grenoble (IPAG), Universit\'{e} Grenoble Alpes, CNRS, F-38000 Grenoble, France}
\author[0000-0001-5659-0140]{Marta De Simone}
\affiliation{European Southern Observatory, Karl-Schwarzschild Str. 2, 85748 Garching bei M\"{u}nchen, Germany}
\affiliation{Istituto Nazionale di Astrofisica (INAF), Osservatorio Astrofisico di Arcetri, Largo E. Fermi 5, I-50125 Florence, Italy}
\author[0000-0001-6981-0421]{Francois Dulieu}
\affiliation{CY Cergy Paris Universit\'{e}, Sorbonne Universit\'{e}, Observatoire de Paris, PSL University, CNRS, LERMA, F-95000, Cergy, France}
\author[0000-0002-7622-5762]{Aurora Dur\'{a}n}
\affiliation{Instituto de Radioastronom\'{i}a y Astrof\'{i}sica , Universidad Nacional Aut\'{o}noma de M\'{e}xico, A.P. 3-72 (Xangari), 8701, Morelia, Mexico}
\author[0009-0006-1929-3896]{Lucy Evans}
\affiliation{School of Physics and Astronomy, University of Leeds, Leeds, LS2 9JT, UK}
\affiliation{IRAP, Universit\'{e} de Toulouse, CNRS, CNES, UPS, Toulouse, France}
\affiliation{Istituto Nazionale di Astrofisica (INAF), Osservatorio Astrofisico di Arcetri, Largo E. Fermi 5, I-50125 Florence, Italy}
\author[0000-0002-5789-6931]{C\'{e}cile Favre}
\affiliation{Institut de Plan\'{e}tologie et d'Astrophysique de Grenoble (IPAG), Universit\'{e} Grenoble Alpes, CNRS, F-38000 Grenoble, France}
\author[0000-0002-4707-8409]{Siyi Feng}
\affiliation{Department of Astronomy, Xiamen University, Xiamen, Fujian 361005, People's Republic of China}
\author[0000-0001-6156-0034]{Davide Fedele}
\affiliation{Istituto Nazionale di Astrofisica (INAF), Osservatorio Astrofisico di Arcetri, Largo E. Fermi 5, I-50125 Florence, Italy}
\affiliation{Istituto Nazionale di Astrofisica (INAF), Osservatorio Astrofisico di Torino, Via Osservatorio 20, 10025, Pino Torinese, Italy}
\author[0000-0003-0348-3418]{Francesco Fontani}
\affiliation{Istituto Nazionale di Astrofisica (INAF), Osservatorio Astrofisico di Arcetri, Largo E. Fermi 5, I-50125, Florence, Italy}
\affiliation{Center for Astrochemical Studies, Max-Planck-Institut f\"{u}r extraterrestrische Physik (MPE), Gie$\beta$enbachstr. 1, D-85741 Garching, Germany}
\affiliation{Laboratoire d'Etudes du Rayonnement et de la Mati\'{e}re en Astrophysique et Atmosph\'{e}res (LERMA), Observatoire de Paris, Meudon, France with: LUX, Observatoire de Paris, Universit\'{e} PSL, Sorbonne Universit\'{e}, CNRS, 75014 Paris, France}
%
%
\author[0000-0002-4991-4044]{Tetsuya Hama}
\affiliation{Komaba Institute for Science, The University of Tokyo, 3-8-1 Komaba, Meguro, Tokyo 153-8902, Japan}
\affiliation{Department of Basic Science, The University of Tokyo, 3-8-1 Komaba, Meguro, Tokyo 153-8902, Japan}
\author[0000-0002-7538-581X]{Tomoyuki Hanawa}
\affiliation{Center for Frontier Science, Chiba University, 1-33 Yayoi-cho, Inage-ku, Chiba 263-8522, Japan}
\author[0000-0002-4649-2536]{Eric Herbst}
\affiliation{Department of Chemistry, University of Virginia, McCormick Road, PO Box 400319, Charlottesville, VA 22904, USA}
\author[0000-0002-4317-767X]{Shingo Hirano}
\affiliation{Department of Applied Physics, Faculty of Engineering, Kanagawa University, 3-27-1 Rokukakubashi, Kanagawa-ku, Yokohama, Kanagawa 221-8686, Japan}
\author[0000-0003-1659-095X]{Tomoya Hirota}
\affiliation{Mizusawa VLBI Observatory, National Astronomical Observatory of Japan (NAOJ), 2-12 Hoshiga-oka, Mizusawa, Oshu-shi, Iwate 023-0861, Japan}
\affiliation{The Graduate University for Advanced Studies, SOKENDAI, 2-21-1 Osawa, Mitaka, Tokyo 181-8588, Japan}
\affiliation{Department of Astronomy, Graduate School of Science, The University of Tokyo, 7-3-1 Hongo, Bunkyo-ku, Tokyo 113-0033, Japan}
\author[0000-0001-8061-2207]{Andrea Isella}
\affiliation{Department of Physics and Astronomy, Rice University, 6100 Main Street, MS-108, Houston, TX 77005, USA}
\author[0000-0003-4493-8714]{Izaskun J\'{i}menez-Serra}
\affiliation{Centro de Astrobiolog\'{\i}a (CSIC-INTA), Ctra. de Torrej\'on a Ajalvir, km 4, 28850, Torrej\'on de Ardoz, Spain}
\author[0000-0002-6773-459X]{Doug Johnstone}
\affiliation{NRC Herzberg Astronomy and Astrophysics, 5071 West Saanich Road, Victoria, BC, V9E 2E7, Canada}
\affiliation{Department of Physics and Astronomy, University of Victoria, Victoria, BC, V8P 5C2, Canada}
\author[0000-0003-1691-4686]{Claudine Kahane}
\affiliation{Institut de Plan\'{e}tologie et d'Astrophysique de Grenoble (IPAG), Universit\'{e} Grenoble Alpes, CNRS, F-38000 Grenoble, France}
\author[0000-0003-1837-3772]{Romane Le Gal}
\affiliation{Institut de Plan\'{e}tologie et d'Astrophysique de Grenoble (IPAG), Universit\'{e} Grenoble Alpes, CNRS, F-38000 Grenoble, France}
%
%
\affiliation{Institut de Radioastronomie Millim\'{e}trique (IRAM), F-38406 Saint-Martin d'H\'{e}res, France}
\author[0000-0002-9397-3826]{Bertrand Lefloch}
\affiliation{Institut de Plan\'{e}tologie et d'Astrophysique de Grenoble (IPAG), Universit\'{e} Grenoble Alpes, CNRS, F-38000 Grenoble, France}
\affiliation{Laboratoire d'Astrophysique de Bordeaux, B18N all\'{e}e Geoffroy St-Hilaire, 33615 Pessac, France}
\author[0000-0002-5635-3345]{Laurent Loinard}
\affiliation{Instituto de Radioastronom\'{i}a y Astrof\'{i}sica , Universidad Nacional Aut\'{o}noma de M\'{e}xico, 58341 Morelia, Mexico}
\affiliation{Black Hole Initiative at Harvard University, 20 Garden Street, Cambridge, MA 02138, USA}
\author[0000-0003-2300-2626]{Hauyu Baobab Liu}
\affiliation{Department of Physics, National Sun Yat-Sen University, No. 70, Lien-Hai Road, Kaohsiung City 80424, Taiwan, R.O.C.}
\affiliation{Center of Astronomy and Gravitation, National Taiwan Normal University, Taipei 116, Taiwan, R.O.C.}
\author[0000-0001-5191-2075]{Antonio Mart\'{\i}nez-Henares}
\affiliation{Star and Planet Formation Laboratory, Pioneering Research Institute, RIKEN, 2-1 Hirosawa, Wako, Saitama 351-0198, Japan}
\affiliation{Centro de Astrobiolog\'{\i}a (CSIC-INTA), Ctra. de Torrej\'on a Ajalvir, km 4, 28850, Torrej\'on de Ardoz, Spain}
\author[0000-0002-7675-3565]{Luke T. Maud}
\affiliation{European Southern Observatory, Karl-Schwarzschild Str. 2, 85748 Garching bei M\"{u}nchen, Germany}
\author[0000-0002-7026-8163]{Maria Jose Maureira}
\affiliation{Center for Astrochemical Studies, Max-Planck-Institut f\"{u}r extraterrestrische Physik (MPE), Gie$\beta$enbachstr. 1, D-85741 Garching, Germany}
\author[0000-0002-1637-7393]{Francois Menard}
\affiliation{Institut de Plan\'{e}tologie et d'Astrophysique de Grenoble (IPAG), Universit\'{e} Grenoble Alpes, CNRS, F-38000 Grenoble, France}
\author{Seyma Mercimek}
\affiliation{Istituto Nazionale di Astrofisica (INAF), Osservatorio Astrofisico di Arcetri, Largo E. Fermi 5, I-50125 Florence, Italy}
\affiliation{Universit$\grave{a}$ degli Studi di Florence, Dipartimento di Fisica e Astronomia, via G. Sansone 1, 50019 Sesto Fiorentino, Italy}
\author[0000-0002-7997-2528]{Anna Miotello}
\affiliation{European Southern Observatory, Karl-Schwarzschild Str. 2, 85748 Garching bei M\"{u}nchen, Germany}
\author[0000-0002-3296-8134]{George Moellenbrock}
\affiliation{National Radio Astronomy Observatory, PO Box O, Socorro, NM 87801, USA}
\author[0000-0002-7002-939X]{Shoji Mori}
\affiliation{Institute for Advanced Study, Tsinghua University, Beijing 100084, People's Republic of China}
\affiliation{Astronomical Institute, Tohoku University, 6-3 Aramaki, Aoba-ku, Sendai 980-8578, Japan}
\author[0009-0001-6486-6909]{Nadia M. Murillo}
\affiliation{Department of Physics, Tamkang University, S320, No.151, Yingzhuan Rd., Tamsui Dist., New Taipei City 251301, Taiwan, R.O.C.}
\affiliation{Instituto de Astronom\'{i}a, Universidad Nacional Aut\'{o}noma de M\'{e}xico, AP106, Ensenada CP 22830, B.C., M\'{e}xico}
\affiliation{Star and Planet Formation Laboratory, Pioneering Research Institute, RIKEN, 2-1 Hirosawa, Wako, Saitama 351-0198, Japan}
\author[0000-0002-1803-0203]{Riouhei Nakatani}
\affiliation{Dipartimento di Fisica, Universit\'{a} degli Studi di Milano, Via Celoria, 16, I-20133 Milano, Italy}
\affiliation{NASA Jet Propulsion Laboratory, California Institute of Technology, 4800 Oak Grove Drive, Pasadena, CA 91109, USA}
\affiliation{Star and Planet Formation Laboratory, Pioneering Research Institute, RIKEN, 2-1 Hirosawa, Wako, Saitama 351-0198, Japan}
\author[0000-0002-7058-7682]{Hideko Nomura}
\affiliation{National Astronomical Observatory of Japan (NAOJ), 2-21-1 Osawa, Mitaka, Tokyo 181-8588, Japan}
\affiliation{The Graduate University for Advanced Studies, SOKENDAI, 2-21-1 Osawa, Mitaka, Tokyo 181-8588, Japan}
\author[0000-0002-6852-3604]{Yasuhiro Oba}
\affiliation{Institute of Low Temperature Science, Hokkaido University, N19W8, Kita-ku, Sapporo, Hokkaido 060-0819, Japan}
\author[0000-0002-5317-6304]{Ross O'Donoghue}
\affiliation{Department of Physics and Astronomy, University College London, Gower Street, London, WC1E 6BT, UK}
\author[0000-0002-9661-7958]{Satoshi Ohashi}
\affiliation{Department of Astronomy, Graduate School of Science, The University of Tokyo, 7-3-1 Hongo, Bunkyo-ku, Tokyo 113-0033, Japan}
\affiliation{National Astronomical Observatory of Japan (NAOJ), 2-21-1 Osawa, Mitaka, Tokyo 181-8588, Japan}
\affiliation{Star and Planet Formation Laboratory, Pioneering Research Institute, RIKEN, 2-1 Hirosawa, Wako, Saitama 351-0198, Japan}
\author[0000-0003-3655-5270]{Yuki Okoda}
\affiliation{NRC Herzberg Astronomy and Astrophysics, 5071 West Saanich Road, Victoria, BC, V9E 2E7, Canada}
\affiliation{Star and Planet Formation Laboratory, Pioneering Research Institute, RIKEN, 2-1 Hirosawa, Wako, Saitama 351-0198, Japan}
\author{Juan Ospina-Zamudio}
\affiliation{Institut de Plan\'{e}tologie et d'Astrophysique de Grenoble (IPAG), Universit\'{e} Grenoble Alpes, CNRS, F-38000 Grenoble, France}
\author[0000-0002-0197-8751]{Yoko Oya}
\affiliation{Yukawa Institute for Theoretical Physics, Kyoto University, Oiwake-cho, Kitashirakawa, Sakyo-ku, Kyoto-shi, Kyoto-fu 606-8502, Japan}
\author[0000-0003-2733-5372]{Linda Podio}
\affiliation{Istituto Nazionale di Astrofisica (INAF), Osservatorio Astrofisico di Arcetri, Largo E. Fermi 5, I-50125 Florence, Italy}
\author[0000-0002-9637-4554]{Albert Rimola}
\affiliation{Departament de Qu\'{i}mica, Universitat Aut$\grave{o}$noma de Barcelona, 08193 Bellaterra, Spain}
\author[0000-0002-6428-9806]{Giovanni Sabatini}
\affiliation{Istituto Nazionale di Astrofisica (INAF), Osservatorio Astrofisico di Arcetri, Largo E. Fermi 5, I-50125 Florence, Italy}
\author[0000-0003-4521-7492]{Takeshi Sakai}
\affiliation{Graduate School of Informatics and Engineering, The University of Electro-Communications, Chofu, Tokyo 182-8585, Japan}
\author[0000-0003-3172-6763]{Dominique M. Segura-Cox}
\affiliation{Department of Astronomy, University of Illinois, Urbana, IL, USA}
\affiliation{Center for Astrochemical Studies, Max-Planck-Institut f\"{u}r extraterrestrische Physik (MPE), Gie$\beta$enbachstr. 1, D-85741 Garching, Germany}
\author{Yancy Shirley}
\affiliation{Steward Observatory, 933 N Cherry Ave., Tucson, AZ 85721 USA}
\author[0000-0003-1859-3070]{Leonardo Testi}
\affiliation{European Southern Observatory, Karl-Schwarzschild Str. 2, 85748 Garching bei M\"{u}nchen, Germany}
\affiliation{Istituto Nazionale di Astrofisica (INAF), Osservatorio Astrofisico di Arcetri, Largo E. Fermi 5, I-50125 Florence, Italy}
\author[0000-0001-8211-6469]{Charlotte Vastel}
\affiliation{IRAP, Universit\'{e} de Toulouse, CNRS, CNES, UPS, Toulouse, France}
\author[0000-0001-8504-8844]{Serena Viti}
\affiliation{Leiden Observatory, Leiden University, PO Box 9513, 2300 RA, Leiden, The Netherlands}
\author[0000-0001-8408-2872]{Naoki Watanabe}
\affiliation{Institute of Low Temperature Science, Hokkaido University, N19W8, Kita-ku, Sapporo, Hokkaido 060-0819, Japan}
\author[0000-0002-9668-3592]{Yoshimasa Watanabe}
\affiliation{Materials Science and Engineering, College of Engineering, Shibaura Institute of Technology, 3-7-5 Toyosu, Koto-ku, Tokyo 135-8548, Japan}
\affiliation{Star and Planet Formation Laboratory, Pioneering Research Institute, RIKEN, 2-1 Hirosawa, Wako, Saitama 351-0198, Japan}
\author{Arezu Witzel}
\affiliation{Institut de Plan\'{e}tologie et d'Astrophysique de Grenoble (IPAG), Universit\'{e} Grenoble Alpes, CNRS, F-38000 Grenoble, France}
\author[0000-0003-2760-2119]{Ci Xue}
\affiliation{Department of Chemistry, University of Virginia, McCormick Road, PO Box 400319, Charlottesville, VA 22904, USA}
\author{Bo Zhao}
\affiliation{Center for Astrochemical Studies, Max-Planck-Institut f\"{u}r extraterrestrische Physik (MPE), Gie$\beta$enbachstr. 1, D-85741 Garching, Germany}

\author[0000-0002-9865-0970]{Satoshi Yamamoto}
\affiliation{The Graduate University for Advanced Studies (SOKENDAI), Shonan Village, Hayama, Kanagawa 240-0193, Japan}
\affiliation{Department of Physics, Graduate School of Science, The University of Tokyo, 7-3-1, Hongo, Bunkyo-ku, Tokyo 113-0033, Japan}
\affiliation{Research Center for the Early Universe, Graduate School of Science, The University of Tokyo, 7-3-1, Hongo, Bunkyo-ku, Tokyo 113-0033, Japan}

\begin{abstract}
Recent molecular line observations with high spatial resolutions have revealed
vertical chemical stratification in Class I/II disks (i.e. chemical composition changing with height in the vertical direction of the disk), as well as a radial variation of ionisation rates.
However, such chemical layered structures in Class 0 disk and envelope systems have been investigated only for limited objects, such as L1527.
In the context of the ALMA Large Program FAUST, we report the results of molecular line observations toward the disk and envelope around the low-mass Class 0 protostar NGC 1333 IRAS~4C in the Perseus Molecular Cloud.
Previous observations have shown that 
the disk rotation axis of this object is largely tilted to the line of sight,
making it suitable for studying the molecular abundance distribution as a function of disk height.
We found that the molecules known as gas density tracers such as C$^{18}$O and H$_{2}$CO are distributed around the disk midplane.
In addition, molecules known as tracers of photochemical reactions such as CCH and c-C$_{3}$H$_{2}$ are distributed in the low-density upper layers of the disk, unlike the distribution in L1527.
Moreover, the column density ratio of $\mathrm{HCO}^{\mathrm{+}}$/CO derived from the H$^{13}$CO$^{+}$ and C$^{18}$O line emission suggests that the ionisation degree and the ionisation rate vary with the disk radius and they are larger in the outer region of the disk and envelope.
\\ \\

\end{abstract}




\section{Introduction}\label{sec:1}
%
The physical and chemical conditions within protoplanetary disks determine the properties of forming planets, including mass and chemical composition (e.g., \citealt{Endres2016, Eistrup2018, Cridland2020, Oberg2021, Oberg2021MAPS, Oberg2023, Notsu2020, Notsu2022, Molliere2022}).
In order to understand the origin of the chemical composition in both the solar system and exoplanets, it is necessary to investigate disk chemical structure at high spatial resolution around forming stars, from Class 0/I to Class II sources.
\\ \\
Along the vertical direction in the disks, gases are in hydrostatic equilibrium and the density decreases towards the disk surface.
Based on thermo-chemical models, protoplanetary disks can be simplified into three chemical layers: the hot disk surface, the inner warm molecular layer, and the central cold freeze-out layer (e.g., \citealt{Aikawa2002, Aikawa2024, Woitke2009, Dutrey2014, Walsh2010, Walsh2012, Notsu2016, Notsu2017, Oberg2023}, see also Section \ref{sec:4-1}).
The most direct way to investigate the distribution and origin of molecules is to observe these vertical structure of disks so as to resolve the distinct chemical layers. 
\\ \\
Recently, spatially resolved observations of some Class I-II edge-on disks (i.e. disks whose rotational axis is perpendicular to the line of sight) have been conducted with molecular lines of simple molecules such as CO, CS, SO, H$_{2}$CO, and c-C$_{3}$H$_{2}$ (e.g., \citealt{Dutrey2017, Dutrey2025, Louvet2018, Podio2020, Teague2020, vantHoff2020, vantHoff2023, Lin2023, Foucher2025}).
\citet{Podio2020} reported a resolved disk vertical structure with multiple molecular tracers, in the edge-on disk of the Class I protostar IRAS 04302+2247, using Atacama Large Millimeter/Submillimeter Array (ALMA) observations from the ALMA-DOT program (ALMA chemical survey of Disk-Outflow sources in Taurus, see e.g., \citealt{Garufi2020}).
They showed that molecular emission is vertically stratified, with $^{12}$CO observed at greater disk heights compared to both CS and H$_{2}$CO, which are nearly co-spatial.
In the outer midplane, the line emission decreases due to molecular freeze-out onto dust grains (freeze-out layer). 
\citet{vantHoff2020} also showed the vertically stratified structures in this disk from their ALMA observations that C$^{17}$O line emission originates in the disk midplane and H$_{2}$CO line emission traces more surface layers. They reported the presence of CO freeze-out in the outermost region of the disk midplane ($\gtrsim100$ au).
\citet{Lin2023} also confirmed these chemical layered structures in the same disk using ALMA eDisk observations, and noted that the dust has yet to settle significantly in this disk. They reported a dust scale height of $\sim$6 au at a disk radius of 100 au, which is comparable to the gas scale height at the same radius.
\citet{Villenave2023} also reported a similar dust scale height based on the analyses of ALMA and VLA\footnote{Very Large Array} observations.
\citet{Villenave2024} showed from their JWST\footnote{James Webb Space Telescope} imaging observations from 2 to 21$\mu$m that the dark lane in dust continuum emission does not vary significantly with wavelength, which are interpreted as an indication for intermediate-sized ($\sim10$ $\mu$m) grains in the upper layers of the disk.
\\ \\
\citet{vantHoff2023} conducted high-spatial resolution ($\sim24$ au) observations of molecular line emission toward the Class 0 protostar L1527 IRS (also known as IRAS 04368+2557; hereafter L1527) as part of the ALMA Large Program ``Early Planet Formation in Embedded Disks (eDisk)'' (e.g., \citealt{OhashiN2023}). Previous ALMA observations revealed that CCH and c-C$_{3}$H$_{2}$ emission originates primarily from the inner envelope and disappears inside the disk \citep{Sakai2014a, Sakai2014b, Sakai2017}.
Consistent with this picture, \citet{vantHoff2023} showed that C$^{18}$O, H$_{2}$CO, and SO trace the disk and inner envelope (see also \citealt{Zhang2024} for SO), with H$_{2}$CO emission arising predominantly in the disk surface layers. 
They also confirmed that c-C$_{3}$H$_{2}$ is absent in the disk and that its emission originates in the envelope (see also \citealt{vantHoff2026}).
It remains an open question when this kind of chemical layered structure forms during the evolution from Class 0 to Class I/II sources.
\\ \\
In addition to the vertical layered structure, the radial chemical gradients driven by the radial variations in disk physical structures are also crucial for understanding the disk chemical evolution processes and the chemical origin of the composition in both the solar system and exoplanets.
The disk thermal structure profiles determine the snowline and sootline\footnote{In protoplanetary disks, within the sootline, refractory carbon-rich organic molecules sublimate thermally from dust grains (e.g., \citealt{Nakano2003, Li2021, Houge2025}).} positions, which influence the distribution of elemental abundance ratios in the gas phase (such as C/O and N/O ratios) and complex organic molecules (e.g., \citealt{Eistrup2016, Eistrup2018, vantHoff2020carbon, Ohno2021, Notsu2022, Aikawa2024, Houge2025, Sellek2025}).
Moreover, the disk ionisation degree and ionisation rate $\zeta_{\mathrm{CR}}$, which affect gas-phase and grain-surface reactions, also have radial (and vertical) variations in disks and protostellar envelopes (e.g., \citealt{Ceccarelli2014, Cleeves2014, Cleeves2015, Aikawa2021, Seifert2021, Fujii2022, vantHoff2022, Cabedo2023, Long2024, Schwarz2026}).
Recently, \citet{Cabedo2023} reported the resolved map of $\zeta_{\mathrm{CR}}$ for the solar-type Class 0 protostar B335 using ALMA. They found an enhanced ionisation that increases at small envelope radii peaking at the central protostellar position (up to $\zeta_{\mathrm{CR}}\sim10^{-14}$ s$^{-1}$).
They posit that the extreme $\zeta_{\mathrm{CR}}$ can be attributed to the presence of cosmic-rays accelerated close to the protostar (e.g., \citealt{Feigelson2002, Turner2009, Rab2017}). 
We note that \citet{Ceccarelli2014} reported the enhanced ionisation rate of $>10^{-14}$ s$^{-1}$ for the whole envelope and $\sim10^{-12}$ s$^{-1}$ for the innermost envelope around 
the cluster of a few intermediate and low-mass protostars in the source OMC-2 FIR 4 with the total mass of 30$M_{\odot}$ (e.g., \citealt{Shimajiri2008, Lopez2013, Tobin2019}).
\citet{Bergin2024} suggested that such extreme cosmic-ray ionisation may help to generate the carbon isotope fractionation between molecules in younger disks, such as $^{12}$C-enriched methane and hydrocarbon ices  (see also \citealt{SakaiTakano2013}), and $^{13}$C-enriched CO \citep{Yoshida2022} and complex organic molecules (e.g., \citealt{Ichimura2024, Yamato2024, Busch2025, Hsieh2025, Jeong2025}).
Therefore, it is important to resolve the radial distribution of the ionisation degree and ionisation rate from the envelope to the innermost disk scale ($<100$ au) in Class 0/I objects.
\\ \\
In this paper, we report the results of molecular line observations of the disk and envelope (Keplerian disk $+$ pseudo disk, see Section \ref{sec:3-2}) around NGC 1333~IRAS~4C (hereafter IRAS~4C)\footnote{This object IRAS~4C is named as Per-emb-14 in some previous studies, such as \citet{Tobin2015} and \citet{Anderson2025}.} in the Perseus Molecular Cloud, 
as part of the ALMA Large Program Fifty AU STudy of the chemistry in the disk/envelope system of solar-like protostars (FAUST, see e.g., \citealt{Codella2021})\footnote{\url{http://faust-alma.riken.jp}}
The FAUST program
explores physical and chemical structures on a 50 au scale toward 13 nearby young protostellar sources with the same surface brightness sensitivity (e.g., \citealt{Bianchi2020, Codella2021, Codella2022, Codella2024, Okoda2021, Okoda2023, Imai2022, Ohashi2022, Vastel2022, Vastel2024, Cacciapuoti2023, Cacciapuoti2025, Evans2023, Mercimek2023, Mercimek2025, Chahine2024a, Chahine2024b, DeSimone2024, Maureira2024, Maureira2026, Podio2024, Sabatini2024, Sabatini2025, Zhang2024, Duran2025, Frediani2025, Hirota2025, Martinez-Henares2025, Oya2025, Cuello2026, Saury2026, Yang2026}).
Previous observations of IRAS~4C (ALMA/PEACHES,  \citealt{Zhang2018, Yang2021, Anderson2025}) have shown that 
the disk rotation axis of this source is largely tilted to the line of sight,
making it suitable for studying the molecular abundance distribution along the disk height (see also Section \ref{sec:2-1}). 
In this paper, we focus on the vertical and radial chemical structures in the disk and envelope.
The molecular lines analyzed in this paper are listed in Table \ref{Table:1}.
We analyzed molecular line emission of CCH, c-C$_{3}$H$_{2}$, H$_{2}$CO, and C$^{18}$O for vertical chemical layers and H$^{13}$CO$^{+}$, DCO$^{+}$, and C$^{18}$O for ionisation.
\\ \\
The paper is organized as follows. Section \ref{sec:2} introduces the previous studies on our target, and outlines the observational setup and our data reduction process. 
The results and discussion of our observations are described in Sections \ref{sec:3} and \ref{sec:4}, respectively.
In Section \ref{sec:5}, we provide our conclusions.
\section{Observations}\label{sec:2}
\subsection{Target Object}\label{sec:2-1}
Our target IRAS~4C is a low-mass protostar ($L_{\mathrm{bol}}=1.1L_{\odot}$; \citealt{Enoch2009, Sadavoy2014, Tobin2016})\footnote{Previous studies reported $L_{\mathrm{bol}}$ of this object as $0.7L_{\odot}$, but here we adopted the updated value based on the newly reported distance (293 pc)} located in the NGC 1333 star-forming region in the Perseus Molecular Cloud at a distance of 293 pc \citep{OrizLeon2018}\footnote{Previous studies adopted the value of 235 pc for the distance of this object \citep{Hirota2008, Zhang2018}. In this paper, we newly used the value of 293 pc which was reported based on data in Gaia DR2 \citep{OrizLeon2018}, and discussion in au units has been corrected using the new distance value.}.
It is classified as a Class 0 source based on its bolometric temperature of $T_{\mathrm{bol}}=31$ K (e.g., \citealt{Enoch2009, Sadavoy2014, Koumpia2016, Tobin2016}).
IRAS~4C lies 47$''$ northeast of IRAS 4A, and 54$''$ north-northeast of IRAS 4B.
IRAS~4C does not show hot corino activity although both IRAS~4A and 4B are hot
corino sources which are extremely rich in complex organic molecules (e.g., \citealt{Persson2012, Taquet2015, Lopez2017, Sahu2019, DeSimone2022, Chahine2024a, Chahine2024b} for IRAS~4A; \citealt{Sakai2006, Bottinelli2007, Persson2012, DeSimone2022} for IRAS~4B).
\citet{Koumpia2016} suggested that IRAS~4C seems to be in a different evolutionary state than IRAS~4A and 4B based on its lower luminosity and colder nature than the other two sources, but they also discussed that neither the younger nor the older age scenario seems to be fully consistent with the body of observational data on IRAS~4C.
The inclination $i$ between the disk plane and the line of sight is about $25^\circ$ \citep{Tobin2015, Segura-Cox2016}. In this paper, we define the inclination such that $i=0^\circ$ for an edge-on disk and $i=90^\circ$ for a face-on disk.
Note that this inclination value is estimated by the image-plane 2D Gaussian fitting of the dust continuum emission \citep{Segura-Cox2016} assuming geometrically thin disk, and thus it should be considered as lower limit.
Since IRAS 4C is not fully edge-on disk ($i=25^\circ$), in integrated intensity maps, even at the disk surface there are vertical contamination from the midplane.
If the disk radius is around 200 au (based on continuum in Figures \ref{Figure1} and \ref{Figure2}), emission from the midplane side would be mixed in by up to around $200$ au$\times \sin(25^\circ) \sim84 $ au, which would contaminate to the emission in the disk surface. However, such contamination will concentrate around the region close to the protostar only (small $R$, see details in Section \ref{sec:3-2}).
\\ \\
\citet{Zhang2018} observed this object in the ALMA/PEACHES survey and resolved the outflow cavity structure in the CCH and CS line emission with clear signatures of rotation with respect to the outflow axis.
The velocity pattern is highly symmetric with respect to the outflow axis, and the northern cavity wall is blue-shifted while the southern cavity wall is red-shifted.
While the CCH emission appears to mostly trace the outflow cavity structure, the CS emission traces the envelope.
The rotation is detected from about 150 au up to about 1750 au above the envelope/disk midplane. 
As the distance from the central source increases, the rotation velocity of the outflow decreases while the outflow width increases, which gives a flat specific angular momentum distribution along the outflow. 
From the CS and SO emission which trace the envelope and disk in this source, \citet{Zhang2018} roughly estimated the centrifugal barrier radius of $0.2''-0.6''$  ($\sim60-180$ au).
We note that this centrifugal barrier radius is the boundary between the infalling envelope and the rotating disk (e.g., \citealt{Sakai2014a, Oya2016}).
\\ \\
At the base of the outflow cavity, previous observations have found a decrement in the molecular emission (such as CCH, see \citealt{Zhang2018}) along the disk midplane.
However, due to the lack of spatial resolution and the limited number of observed molecules, it was unclear if this effect is due to a dark lane caused by optically thick dust emission or chemical structures (e.g., \citealt{Li2017, Podio2020, Lin2023}).
\subsection{Observational Setup and Data Reduction}\label{sec:2-2}
The observations of NGC 1333 IRAS~4C were conducted with ALMA (Cycle 6) between 2018 November and 2021 June
as part of the large program FAUST (2018.1.01205.L, PI: Satoshi Yamamoto).
The molecular lines analyzed in this paper (see Table \ref{Table:1}) are included in two frequency settings in Band 6; $216-235$ GHz (Setup 1: 1.3 mm) and $244-262$ GHz (Setup 2: 1.2 mm).
We used the 12 m array data from two different configurations (C43-6 and C43-3 for extended and compact configurations, respectively) and the 7 m array data of the Atacama Compact Array (ACA/Morita Array, Setup 2 only), combining these visibility data in the UV plane.
The ACA was employed to ensure a maximum
recoverable size of $8''.4$ ($=2460$ au, at the source distance).
We note that such maximum
recoverable size is reached by the 12m array for setup 1, therefore no ACA observation was performed for setup 1.
The field center is ($\alpha_{\mathrm{2000}}$, $\delta_{\mathrm{2000}}$)$=(03^{\mathrm{h}}29^{\mathrm{m}}13^{\mathrm{s}}.550$, $+31^{\circ}13'58''.110$), which is close to the protostellar position.
The spectral windows optimized for targeting specific molecular line observations 
were set to a bandwidth of 58.59 MHz and a resolution of 141 kHz.
The channel width corresponds to a velocity resolution of $\sim0.19$ km s$^{-1}$ and $\sim0.17$ km s$^{-1}$ for Setup 1 and Setup 2, respectively.
The spectral windows used for broad continuum coverage in the observations have a bandwidth of 1875 MHz and a channel width of 0.488 MHz (Setup 1) and 0.977 MHz (Setup 2).
This channel width corresponds to a velocity resolution of $\sim1.25$ km s$^{-1}$ and $\sim1.38$ km s$^{-1}$ for Setup 1 and Setup 2,
respectively. 
\\ \\
The Common Astronomy Software Applications (CASA) package \citep{McMullin2007, CASA2022} was used for data reduction and imaging.
The data were calibrated using a modified version of the ALMA calibration pipeline in CASA v5.6.1-8, with an additional in-house calibration routine to correct for the system temperature and spectral line data normalization\footnote{\url{https://help.almascience.org/index.php?/Knowledgebase/Article/ View/419}}. 
Self-calibration was carried out using line-free continuum emission for each configuration. Two rounds of phase-only self-calibration were performed using solution intervals of 6 seconds (for the more compact configurations) or 18 seconds (for the extended configuration). This was followed by amplitude self-calibration using per-scan solutions and {\tt solnorm=True} in the CASA task {\tt gaincal}. Finally, once the source model was as complete as possible, offsets between spectral windows were corrected using a single amplitude solution per observation using {\tt solnorm=False}. The complex gain corrections derived from the self-calibration were then applied to all channels in the data, and the continuum model derived from the self-calibration was subtracted from the data to produce continuum-subtracted line data. A similar self-calibration technique was also used to align both amplitudes and phases across the multiple configurations. The overall improvement in the dynamic range of the continuum images is more than a factor of two.
The uncertainty in the absolute flux density scale is approximately 10\%.
\\ \\
Maps of the spectral line emission were obtained by CLEANing the dirty images with a Briggs robustness parameter of 0.5.
The synthesized beams of the molecular line data are summarized in Table \ref{Table:2}.
\begin{table*}
\caption{List of observed lines$^{\mathrm{a}}$}              
\label{Table:1}      
\centering                                      
\begin{tabular}{c c c c c c c }          
\hline\hline                        
Molecule & Transition & Frequency & $E_{l}$ & $E_{u}$ & $A_{ul}$ & $n_{\mathrm{cr}}$ \\
&& (GHz) & (K) & (K) & (s$^{-1}$) & (cm$^{-3}$)  \\ 
\hline                                   
C$^{18}$O&$J=2-1$&219.5603541& 5.269 &15.806&$6.01\times10^{-7}$&$1\times10^{4}$\\  
CCH&$N$=3-2, $J$=7/2-5/2, $F$=4-3&262.0042600&12.575&25.149&$5.31\times10^{-5}$&$8\times10^{6}$\\
        &$N$=3-2, $J$=7/2-5/2, $F$=3-2&262.0064820&12.574&25.148&$5.10\times10^{-5}$&$8\times10^{6}$\\
c-C$_{3}$H$_{2}$&$6_{0,6}-5_{1,5}$&217.8221480&28.154&38.607&$5.40\times10^{-4}$&$5\times10^{7}$\\
                             &$6_{1,6}-5_{0,5}$&217.8221480&28.154&38.607&$5.40\times10^{-4}$&$5\times10^{7}$\\
H$_{2}$CO&$3_{0,3}-2_{0,2}$&218.2221920&10.483&20.956&$2.82\times10^{-4}$&$6\times10^{5}$\\
H$^{13}$CO$^{+}$&$J=3-2$&260.2553390&12.491&24.981&$1.34\times10^{-3}$&$2\times10^{6}$\\
DCO$^{+}$&$J=3-2$&216.1125822&10.372&20.744&$2.38\times10^{-3}$&$2\times10^{6}$\\

\hline                                             
\end{tabular}
\tablecomments{$^{\mathrm{a}}$ Line parameters are taken the from Cologne Database for Molecular Spectroscopy \citep{Endres2016}. 
$E_{l}$ and $E_{u}$ are the energies of the lower and upper levels, respectively. $A_{ul}$ is the Einstein A coefficient. 
$n_{\mathrm{cr}}$ is the critical density for $T_{\mathrm{ex}}=50$ K \citep{Chandra2000, Schoier2005, vanderTak2020}.
}
%
\end{table*}
\section{Results}\label{sec:3}
\subsection{Dust Continuum Emission}\label{sec:3-1}
\begin{figure*}[htbp]
\begin{center}
\vspace{9cm}
\includegraphics[scale=0.65]{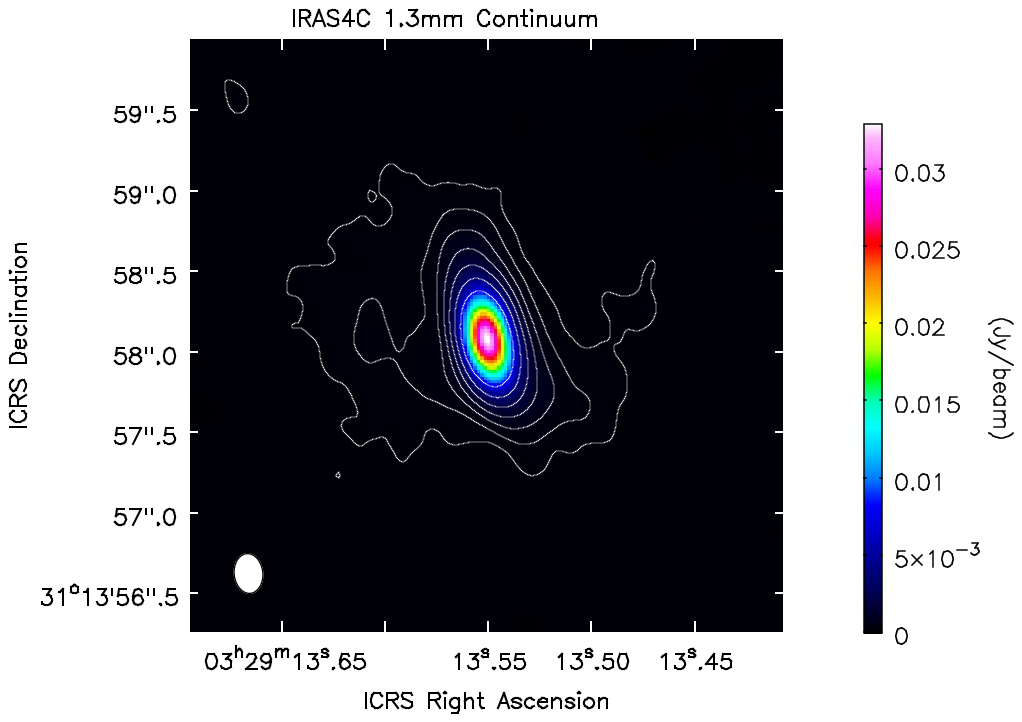}
\end{center}
\vspace{-9.5cm}
\caption{
Image of the 1.3 mm dust continuum.
The peak position is ($\alpha_{\mathrm{ICRS}}$, $\delta_{\mathrm{ICRS}}$)$=(03^{\mathrm{h}}29^{\mathrm{m}}13^{\mathrm{s}}.551$, $+31^{\circ}13'58''.08$).
The rms noise level ($\sigma$) of this continuum map is 20 $\mu$Jy beam$^{-1}$.
The white contour levels start at 5$\sigma$ and double in value up to 320$\sigma$. A white ellipse represents the beam size.
The continuum emission is elongated in the north-south direction. The direction of this elongation is perpendicular to the outflow direction \citep{Zhang2018}.
\vspace{0.2cm}
}\label{Figure1}
\end{figure*}

Figure \ref{Figure1} shows the 1.3 mm continuum map, whose beam size is $0\farcs247\times0\farcs178$ (P.A. $=5.^{\circ}61$).
We adopt the 1.3 mm continuum map in the present study as it has a higher angular resolution than the 1.2 mm continuum map.
The rms noise level ($\sigma$) of this continuum map is 20 $\mu$Jy/beam. 
The peak position determined by a 2D Gaussian fit algorithm performed in the image and provided by CASA in the image is ($\alpha_{\mathrm{ICRS}}$, $\delta_{\mathrm{ICRS}}$)$=(03^{\mathrm{h}}29^{\mathrm{m}}13^{\mathrm{s}}.551$, $+31^{\circ}13'58''.08$).
The integrated flux density and peak intensity are $63.75\pm0.79$ mJy and $32.24\pm0.27$ mJy/beam, respectively.
The peak intensity corresponds to a brightness temperature of $17.6\pm0.1$ K.
The source image size de-convolved from beam is $0\farcs332\pm0\farcs006$$\times$$0\farcs110\pm0\farcs003$ with a P.A. of $14.^{\circ}93\pm0.^{\circ}58$.
This corresponds to a linear size of ($97.3\pm1.8$) au $\times$ ($32.2\pm0.9$) au, which is consistent with previous estimates \citep{Tobin2015}.
The continuum emission is elongated in the north-south direction,
which might trace the outflow launching region \citep{Zhang2018}.
This elongation of the aspect ratio of the continuum emission is consistent with the fact that 
the disk rotation axis of this source is largely tilted to the line of sight
\citep{Tobin2015}.
In addition, there is faint ($\sim5-10\sigma$) continuum emission extended toward the outflow direction. 
\begin{table*}
\caption{rms and beam sizes for observed lines$^{\mathrm{b}}$}              
\label{Table:2}      
\centering                                      
\begin{tabular}{c c c c}          
\hline\hline                        
Molecule & Transition & $\sigma$ & Synthesized Beam \\
&& (mJy beam$^{-1}$) & \\ 
\hline                                   
C$^{18}$O&$J=2-1$&1.5&$0.''26\times0.''19$ (P.A. $=5.^{\circ}5$)\\
CCH&$N$=3-2, $J$=7/2-5/2, $F$=4-3&1.5&$0.''26\times0.''15$ (P.A. $=4.^{\circ}6$)\\
        &$N$=3-2, $J$=7/2-5/2, $F$=3-2&1.3&$0.''26\times0.''15$ (P.A. $=4.^{\circ}6$)\\
c-C$_{3}$H$_{2}$&$6_{0,6}-5_{1,5}$&0.9&$0.''27\times0.''20$ (P.A. $=-1.^{\circ}8$)\\
                             &$6_{1,6}-5_{0,5}$&0.9&$0.''27\times0.''20$ (P.A. $=-1.^{\circ}8$)\\
H$_{2}$CO&$3_{0,3}-2_{0,2}$&0.7&$0.''27\times0.''19$ (P.A. $=6.^{\circ}3$)\\
H$^{13}$CO$^{+}$&$J=3-2$&1.3&$0.''26\times0.''15$ (P.A. $=4.^{\circ}5$)\\
DCO$^{+}$&$J=3-2$&0.8&$0.''27\times0.''21$ (P.A. $=-1.^{\circ}4$)\\

\hline                                             
\end{tabular}
\tablecomments{$^{\mathrm{b}}$ 
$\sigma$ is the rms noise level.
The beam size for each line is obtained from the observations.
}
%
\end{table*}
\subsection{The Vertical Layered Structures}\label{sec:3-2}
\begin{figure*}[htbp]
\begin{center}
\vspace{-1cm}
\includegraphics[scale=0.3]{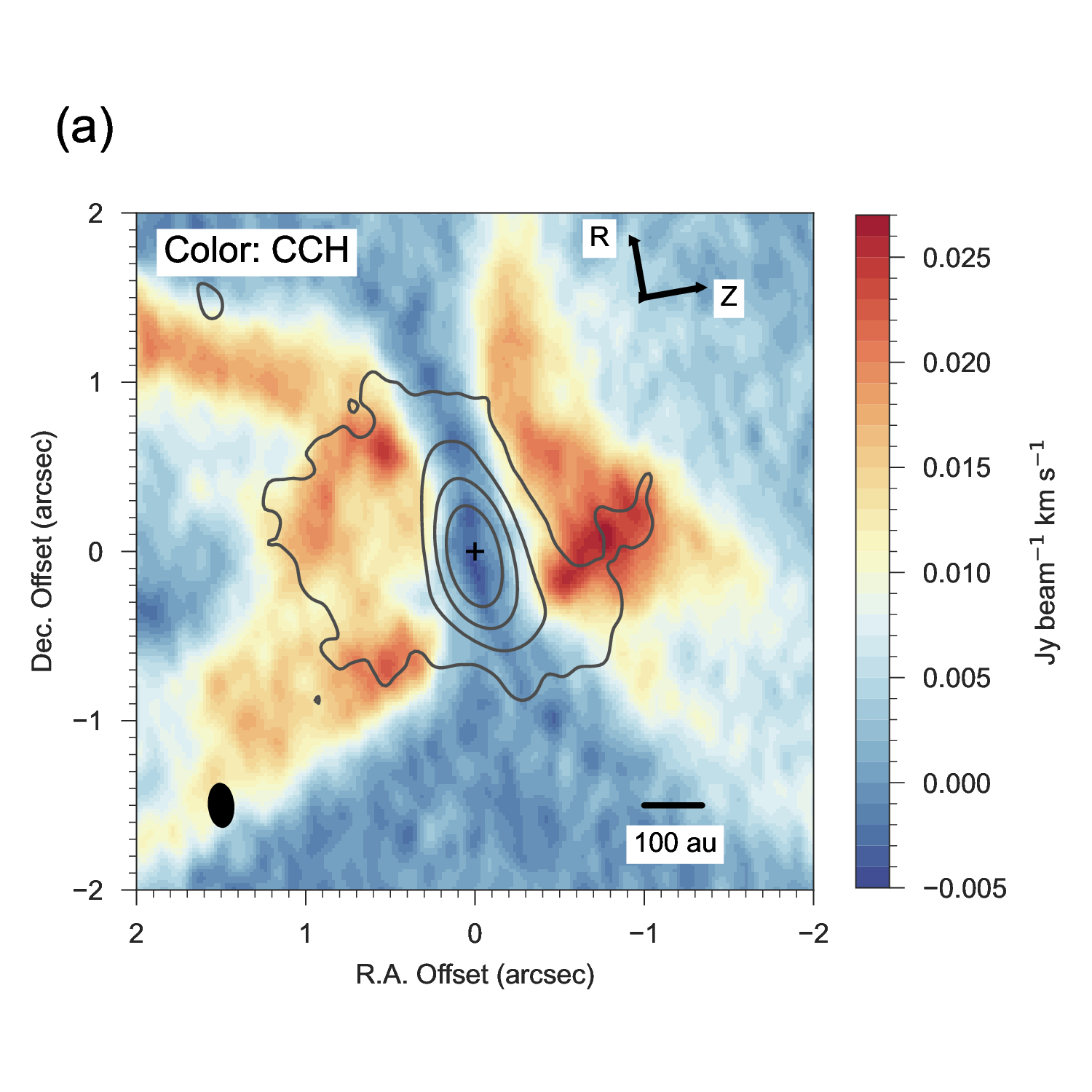}
\includegraphics[scale=0.3]{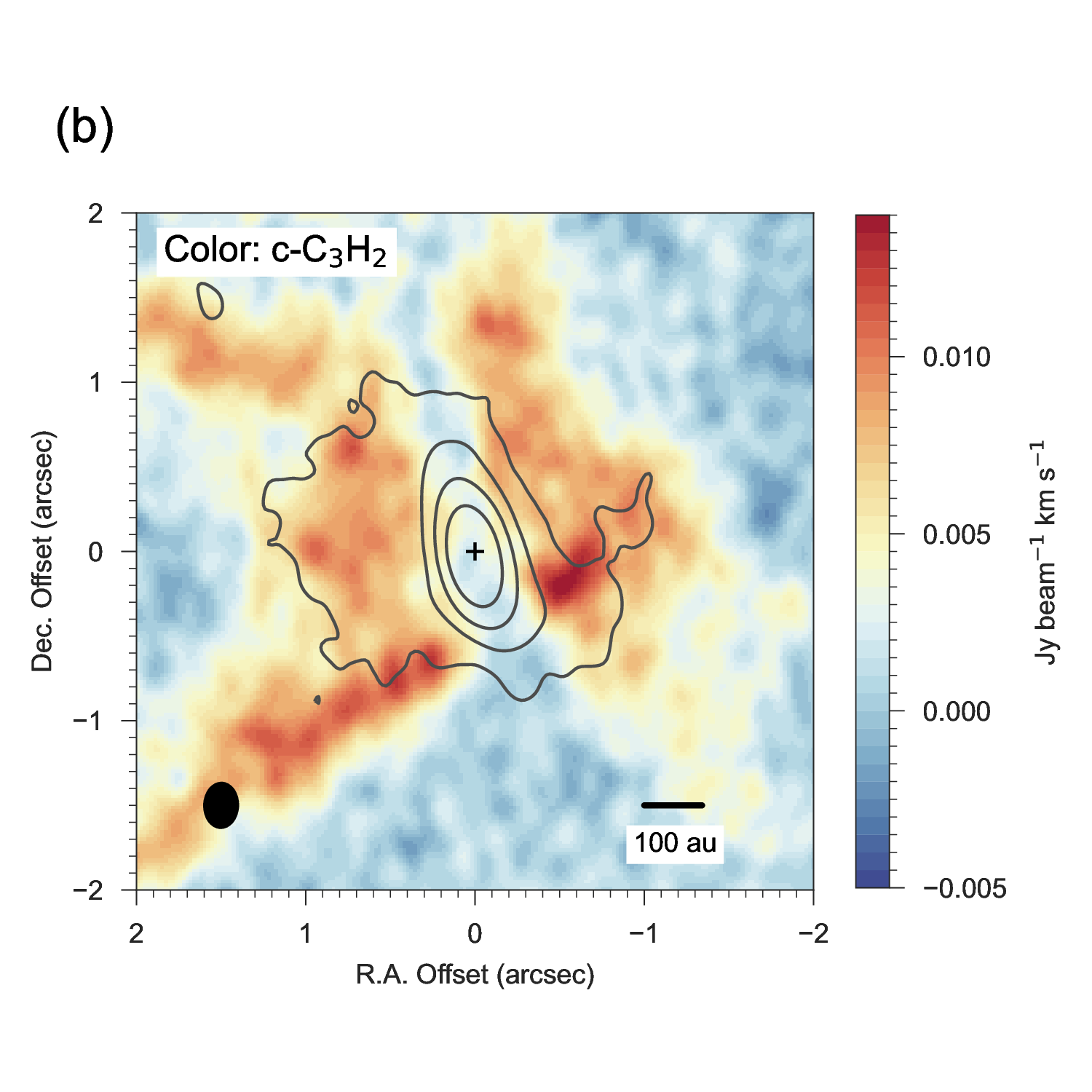}
\vspace{-1cm}
\includegraphics[scale=0.3]{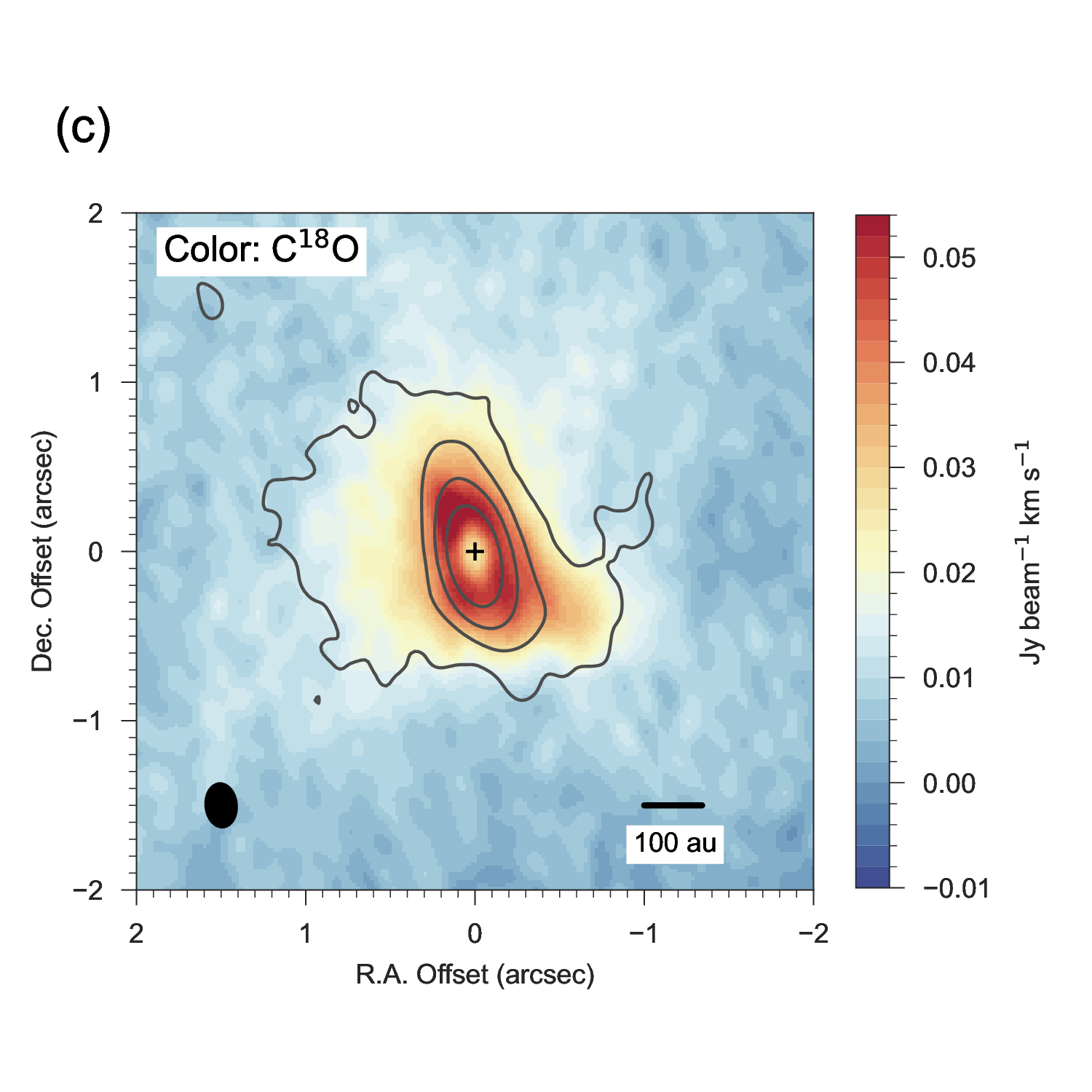}
\includegraphics[scale=0.3]{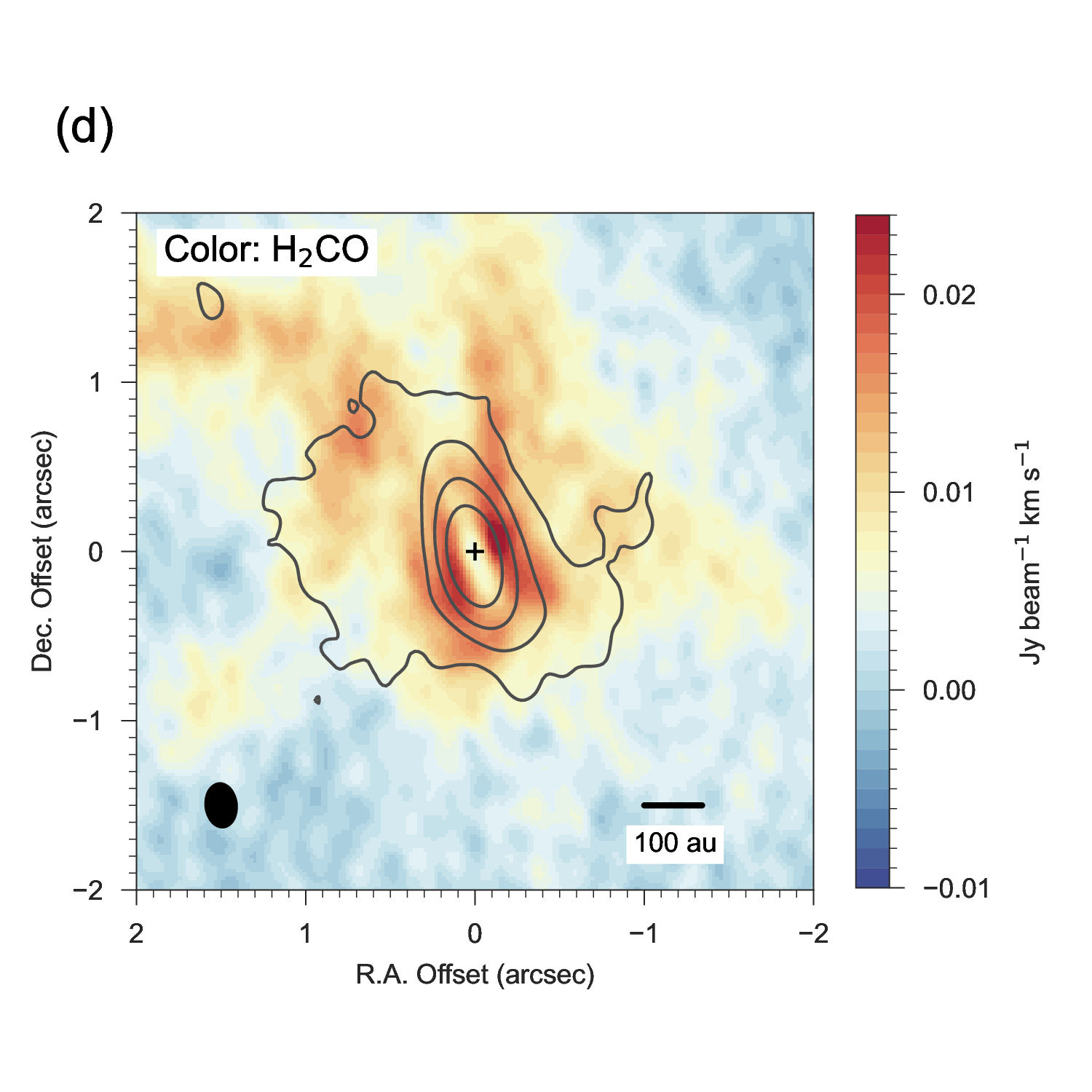}
\vspace{-1cm}
\includegraphics[scale=0.3]{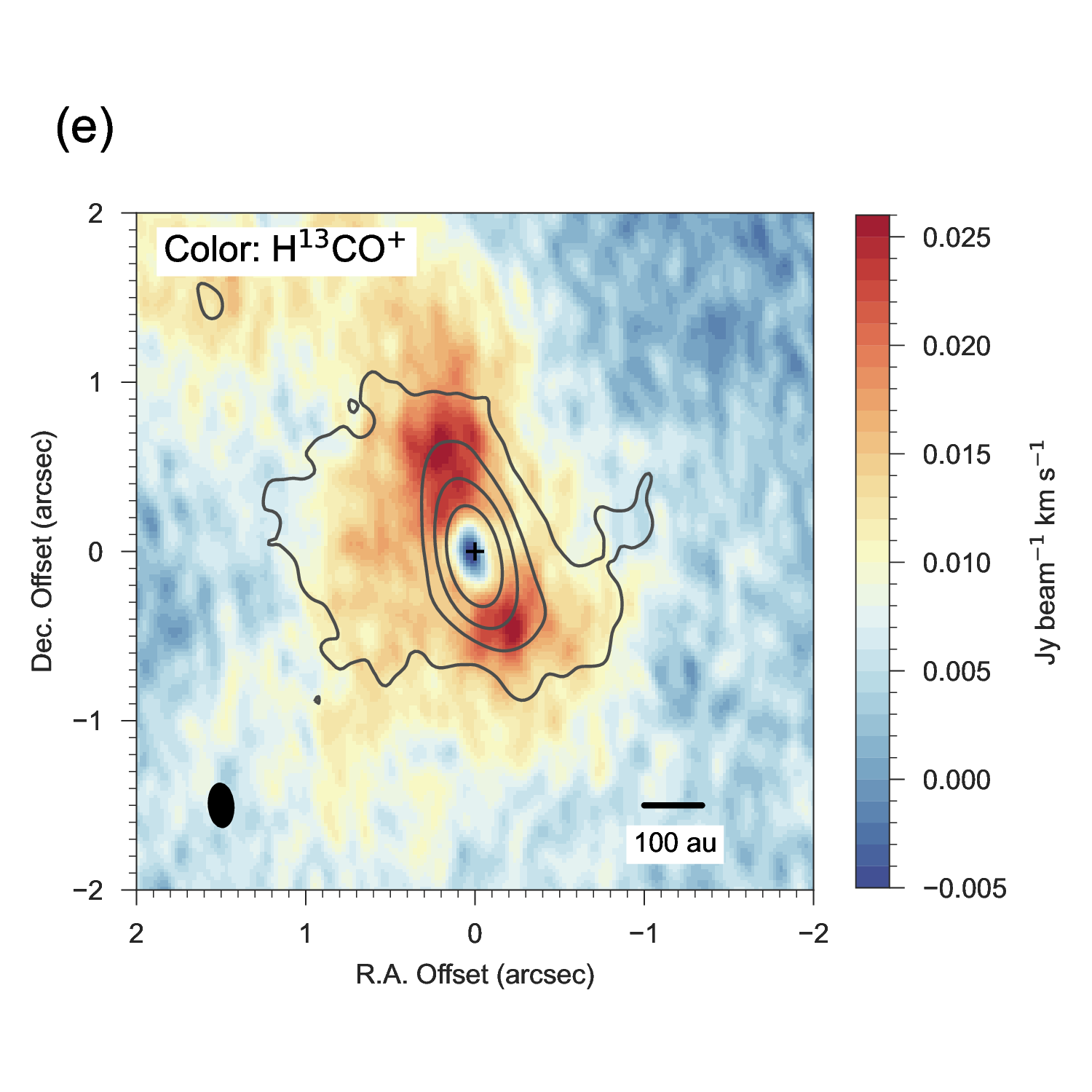}
\includegraphics[scale=0.3]{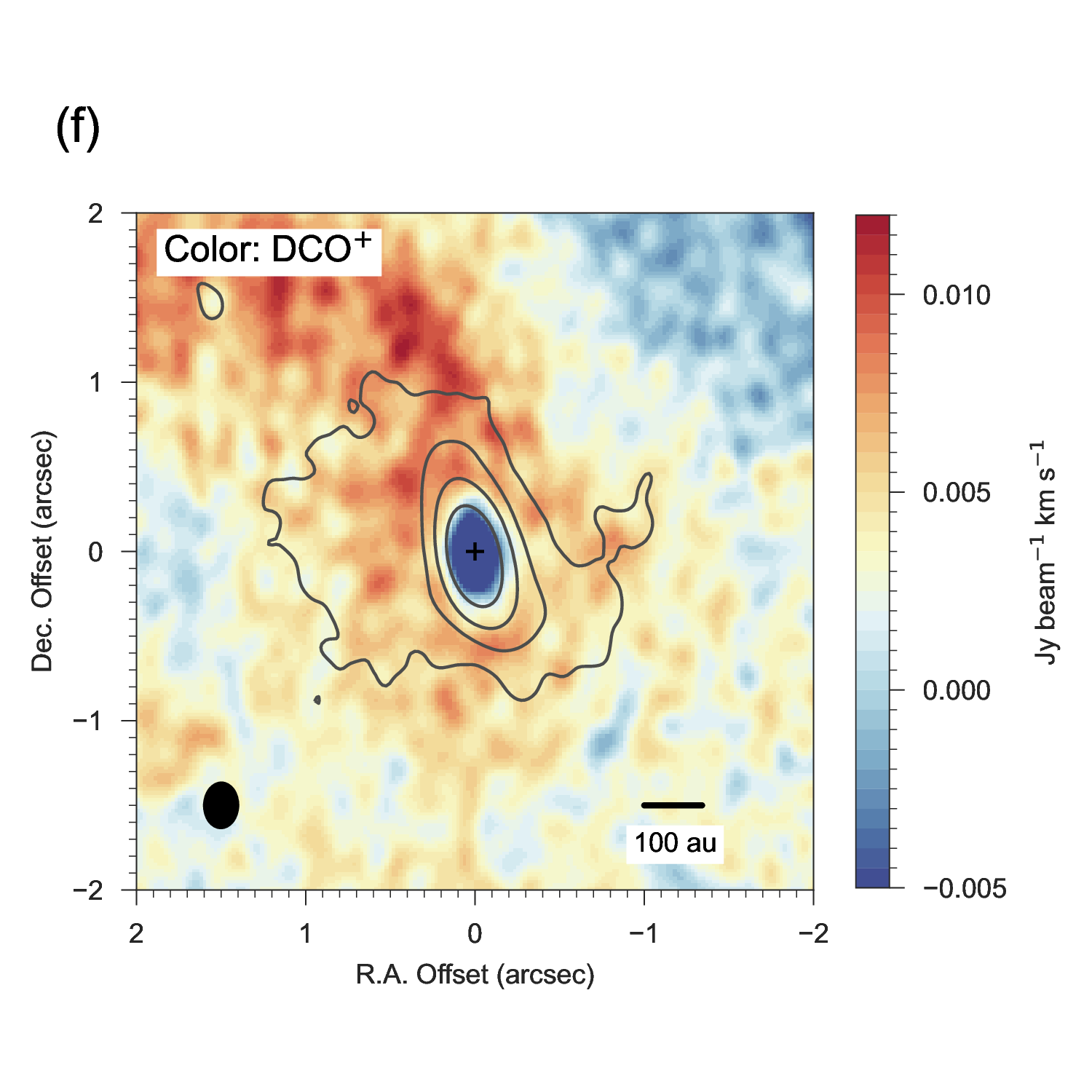}
\end{center}
\vspace{0.5cm}
\caption{
Integrated intensity maps (moment 0 maps, color) of CCH ($N$=3-2, $J$=7/2-5/2, $F$=4-3, top left), c-C$_{3}$H$_{2}$ ($6_{0,6}-5_{1,5}$, top right), C$^{18}$O ($J=2-1$, middle left), H$_{2}$CO ($3_{0,3}-2_{0,2}$, middle right), H$^{13}$CO$^{+}$ ($J=3-2$, bottom left), and DCO$^{+}$ ($J=3-2$, bottom right).
The 1.3 mm continuum image is overploted by black contours (see also Figure \ref{Figure1}).
For the continuum emission, the rms noise level ($\sigma$) of the continuum is 20 $\mu$Jy beam$^{-1}$, and the contour levels are 5$\sigma$, 20$\sigma$, 80$\sigma$ and 320$\sigma$.
Here, we define the $R$ and $Z$ axes in the disk major axis direction (the disk/envelope direction) and minor axis direction (the outflow direction), respectively, where $Z$ is positive in NW direction and $R$ is positive in NE direction (see Figure \ref{Figure2} (a)).
The rms noise level for line emission is $\sim1$ mJy beam$^{-1}$ km s$^{-1}$.
The beam size for each line is shown in the bottom left corner of each panel, and the cross in each panel represents the field center, which is close to the source position.
\vspace{0.2cm}
}\label{Figure2}
\end{figure*}
\begin{figure*}[htbp]
\begin{center}
\vspace{-1cm}
\includegraphics[scale=0.3]{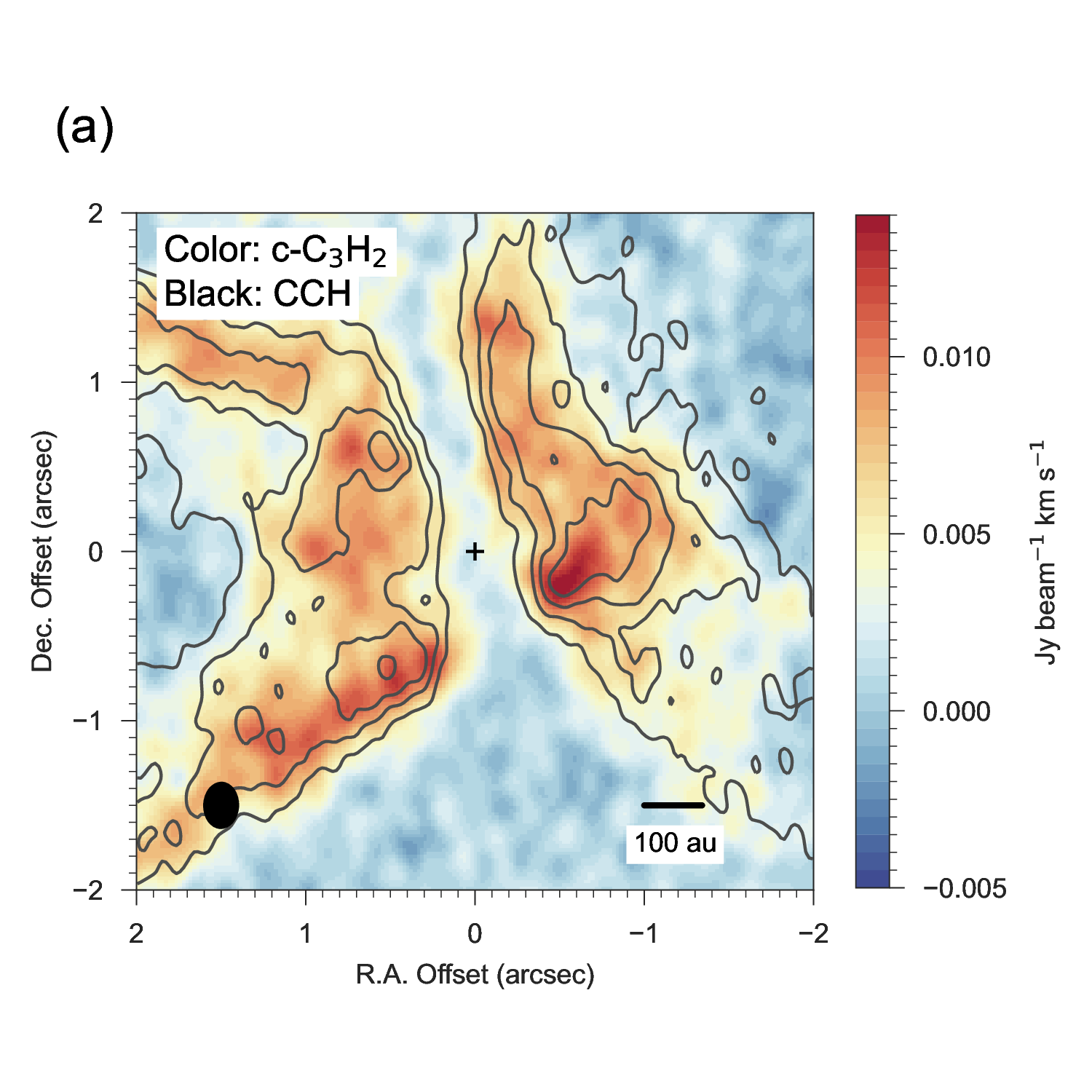}
\includegraphics[scale=0.3]{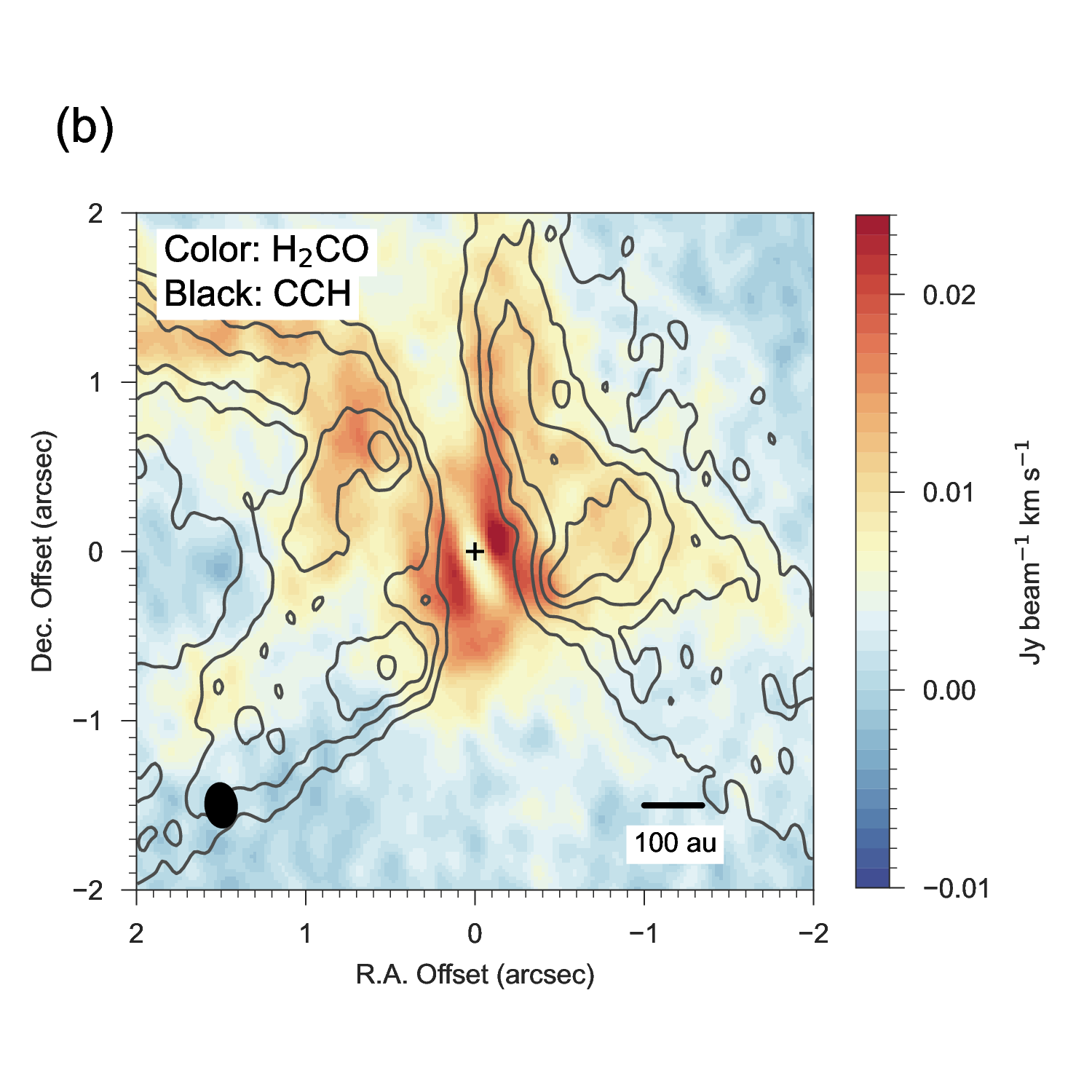}
\vspace{-1cm}
\includegraphics[scale=0.3]{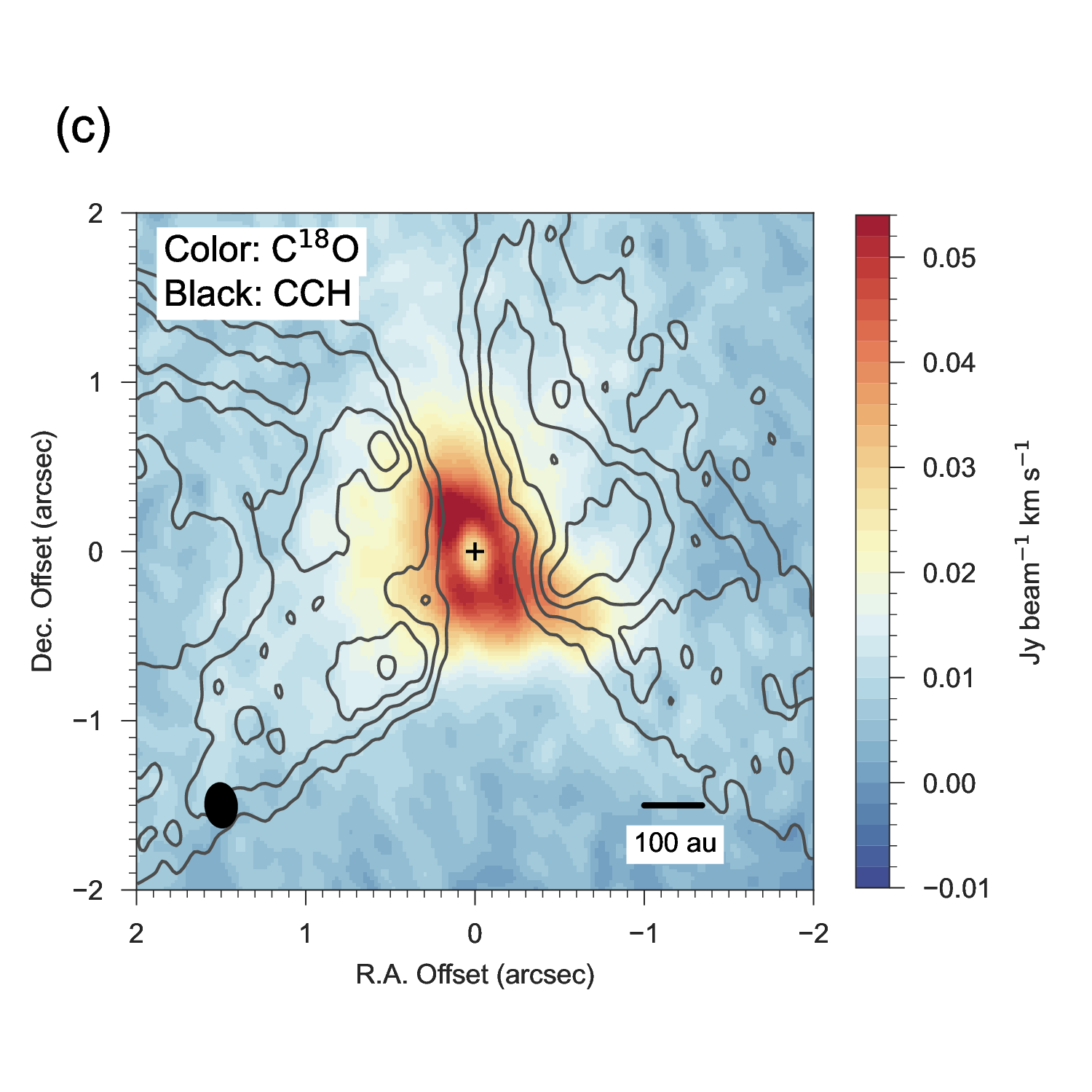}
\includegraphics[scale=0.3]{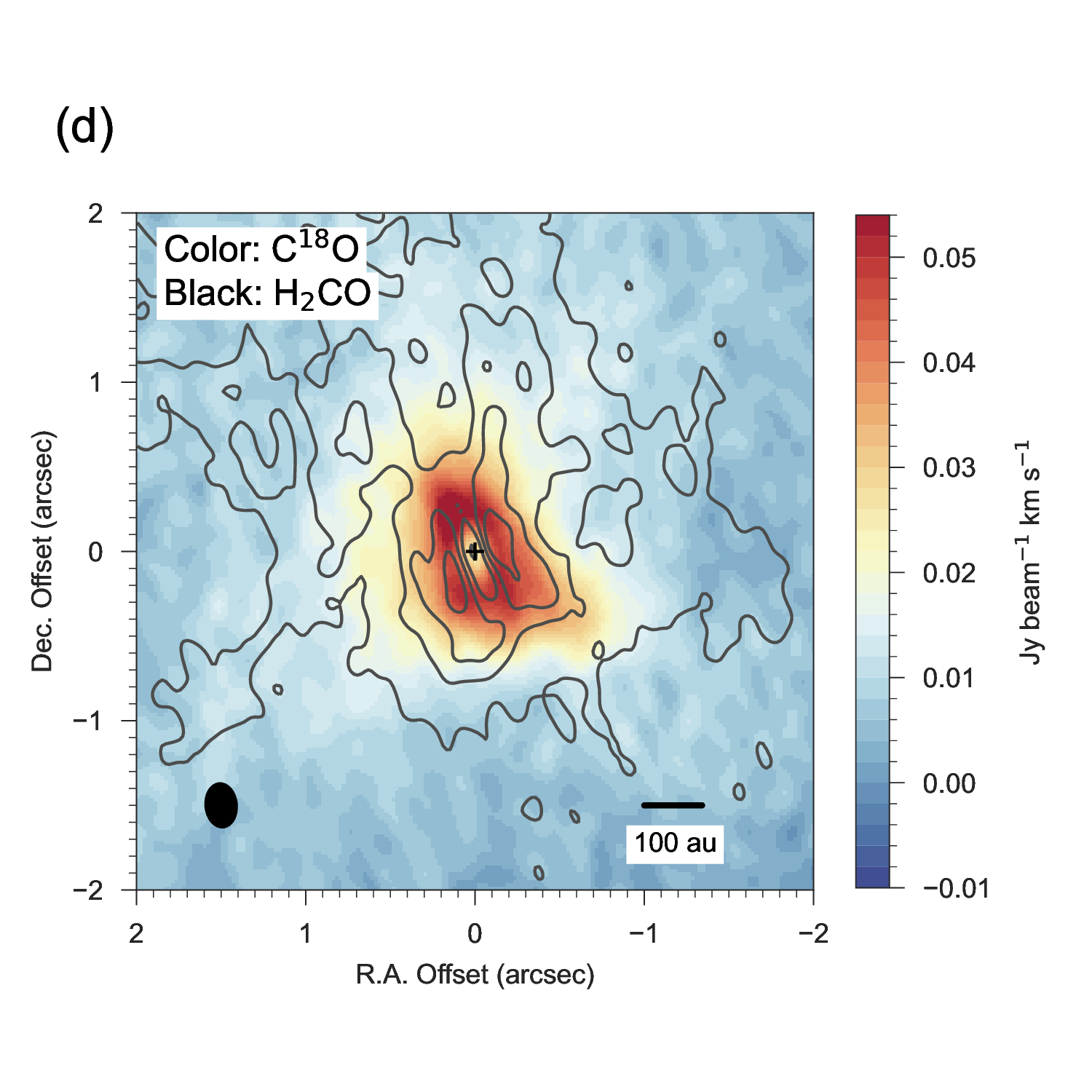}
\vspace{-1cm}
\includegraphics[scale=0.3]{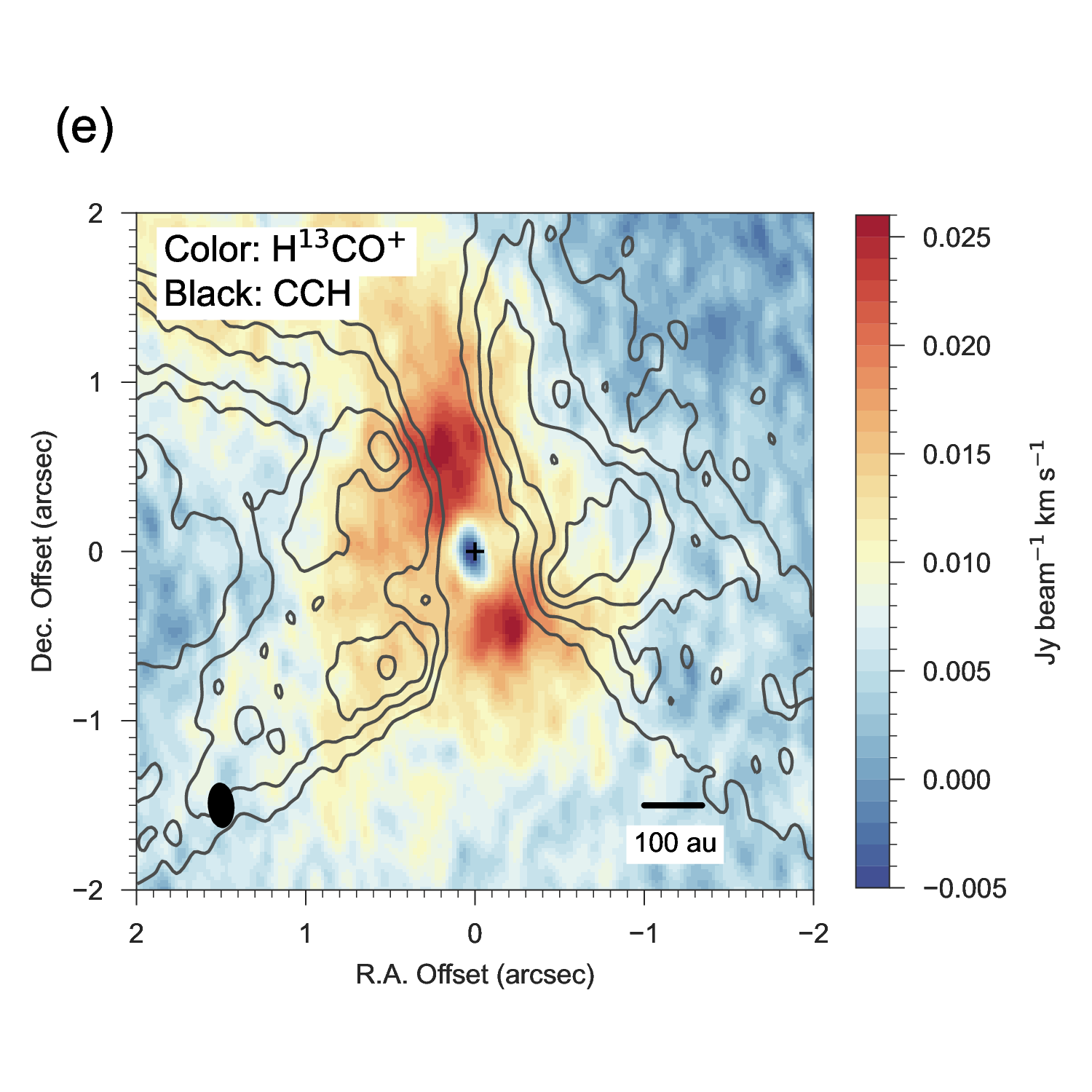}
\includegraphics[scale=0.3]{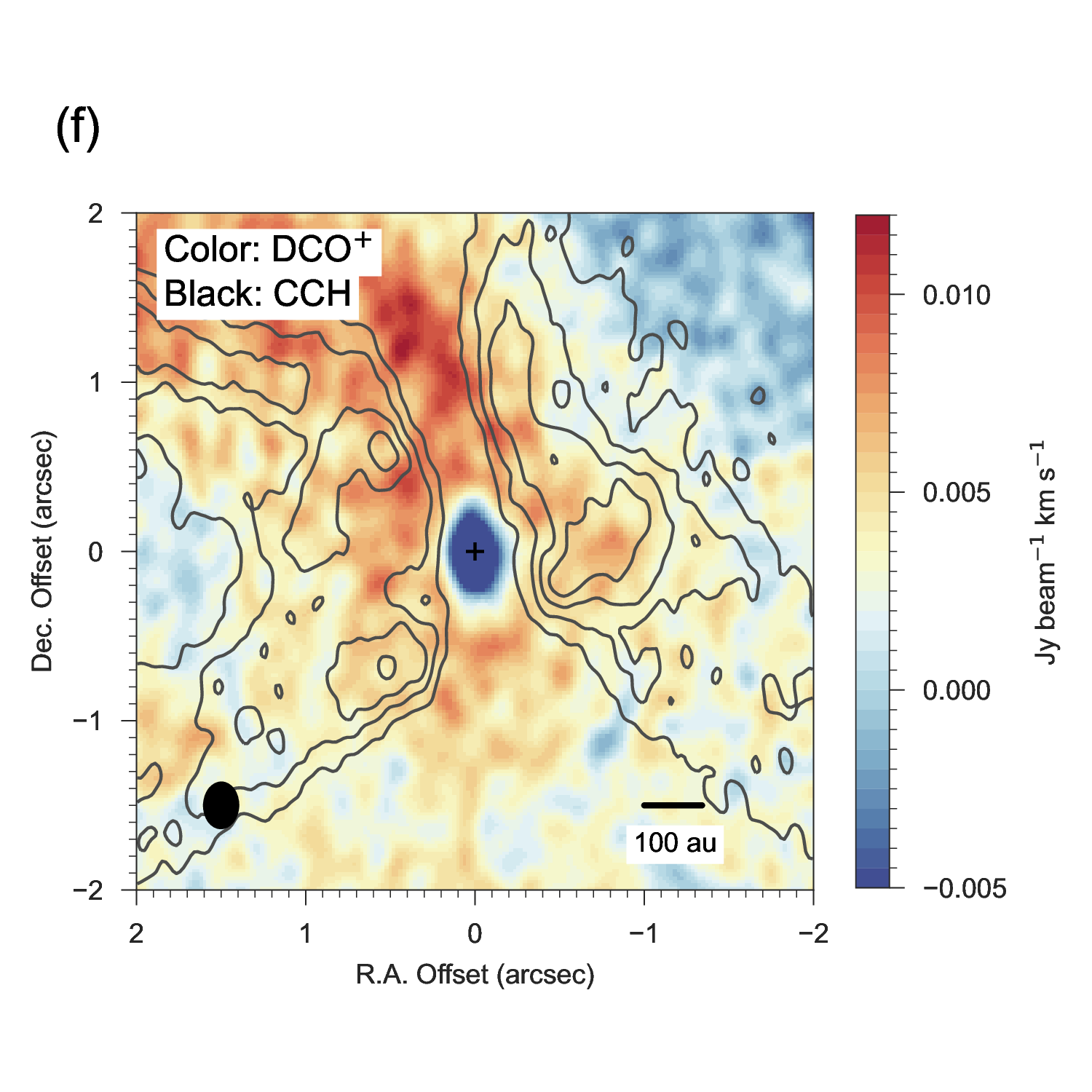}
\end{center}
\caption{
Integrated intensity maps (moment 0 maps) of c-C$_{3}$H$_{2}$ (top left), H$_{2}$CO (top right),
C$^{18}$O (middle left and right), H$^{13}$CO$^{+}$ (bottom left), DCO$^{+}$ (bottom right) (color) overlaid on the H$_{2}$CO emission (middle right) or CCH emission (other plots) (black contours; see also Figure \ref{Figure2}).
The rms noise level for line emission is $\sim1$ mJy beam$^{-1}$ km s$^{-1}$.
The contour levels of the CCH and H$_{2}$CO line emission are 5$\sigma$, 10$\sigma$, 15$\sigma$, and 20$\sigma$.
\vspace{0.2cm}
}\label{Figure3}
\end{figure*}
\begin{figure*}[htbp]
\begin{center}
\vspace{2cm}
\includegraphics[scale=0.65]{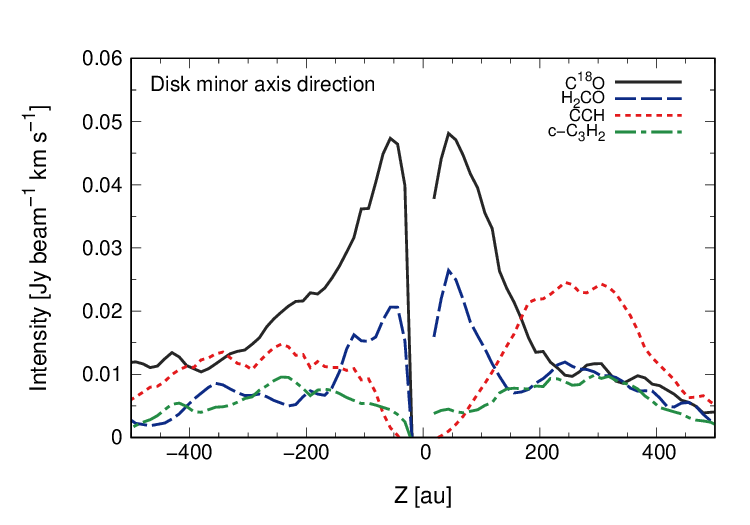}
\includegraphics[scale=0.65]{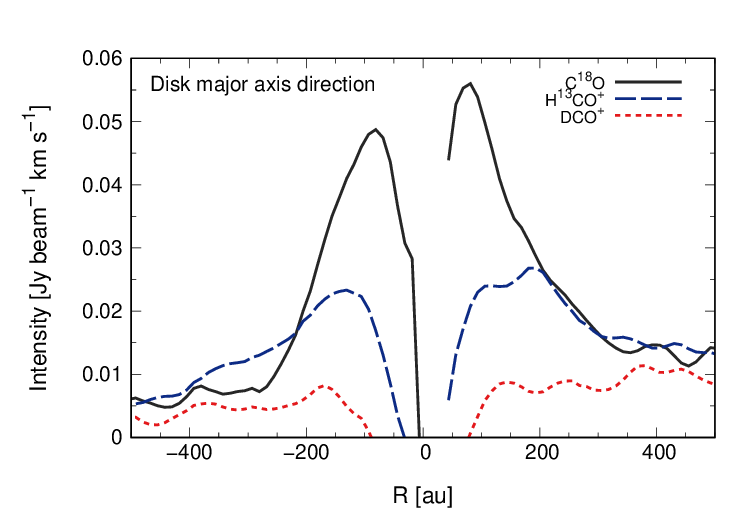}
%
\end{center}
\caption{
Left Panel: Radial profiles of the integrated intensities of C$^{18}$O, H$_{2}$CO, CCH, and c-C$_{3}$H$_{2}$ lines in the vertical (disk minor axis) direction.
Right Panel: Radial profiles of the integrated intensities of C$^{18}$O, H$^{13}$CO$^{+}$, and DCO$^{+}$ lines in the radial (disk major axis) direction.
We define the $R$ and $Z$ axes in the disk major and minor axis directions, respectively, where $Z$ is positive in NW direction and $R$ is positive in NE direction (see Figure \ref{Figure2}).
\vspace{0.2cm}
}\label{Figure4}
\end{figure*}

Figures \ref{Figure2} (a) and (b) show the integrated intensity (moment 0, between $4-12$ km s$^{-1}$) maps of CCH ($N$=3-2, $J$=7/2-5/2, $F$=4-3) and c-C$_{3}$H$_{2}$ ($6_{0,6}-5_{1,5}$), respectively.
The black contours show the 1.3 mm continuum emission intensity (Figure \ref{Figure1}).
Here, we define $R$ and $Z$ axis in the disk major axis direction (the disk/envelope direction) and minor axis direction (the outflow direction), respectively, where $Z$ is positive in NW direction and $R$ is positive in NE direction (see Figure \ref{Figure2} (a)).
We note that in this paper ``disk'' does not necessarily refer to the Keplerian disk only, and it cannot be distinguished from a flattened envelope (also known as a pseudo disk).
Figures \ref{Figure2} (a) and (b) show the dark lane of apparently negative (CCH) and almost
zero (c-C$_{3}$H$_{2}$) intensity, respectively, around the continuum emission peak ($Z\lesssim0."3$).
Emission from the molecules seems to trace the upper surface of the disk/envelope structures and outflow cavity walls.
\\ \\
We note that the CCH and and c-C$_{3}$H$_{2}$ emission extend toward the outflow direction (perpendicular to the disk/envelope direction)
from a position slightly offset from the continuum peak, and the offsets are larger along the northern sides of the cavity ($\sim1.0-1.5''$) than on the southern sides ($\sim0.7''$).
In addition, the length of the disk-like structure along the major axis of the continuum estimated from the CCH and c-C$_{3}$H$_{2}$ line emission is also longer toward the north ($\sim1.0-1.5''$) than toward the south ($\sim0.7''$), although the continuum emission ($>5\sigma$) does not show such clear asymmetry (see Figure \ref{Figure1}).
The distribution of the CCH line emission is consistent with the results of the previous ALMA observations. \citet{Zhang2018} analyzed the velocity structure of the CCH line emission toward the outflow direction carefully. 
They described from the moment 1 map that toward the outflow direction, CCH emission has constant velocity at least $Z\lesssim0.5"$.
We note that the apparent starting radii of CCH extended emission in the $Z$ direction ($R\sim0-0."1$) is not the same as the outflow launching radius ($R\lesssim0.''05$) derived from angular momentum measurements by \citet{Zhang2018}.
In this paper we mainly focus on the disk/envelope component.
\\ \\
Figures \ref{Figure2} (c) and (d) show the integrated intensity maps of C$^{18}$O ($J=2-1$) and H$_{2}$CO ($3_{0,3}-2_{0,2}$), respectively.
Unlike the CCH and c-C$_{3}$H$_{2}$ line emission, the C$^{18}$O line emission is concentrated around the central source and it exists along the midplane of the disk/envelope structure.
The spatial extent is similar to the continuum emission ($R\sim300$ au), although the line intensity decreases around the continuum peak position ($R\lesssim50$ au).
The H$_{2}$CO line emission appears to trace mainly the warm molecular layer ($\sim50-150$ au from the central source), i.e. between the line emitting regions of C$^{18}$O and CCH/c-C$_{3}$H$_{2}$. It is also located along the outflow cavity walls.
The length of the H$_{2}$CO line emission along the disk-envelope direction is also longer toward the north ($\sim1.0-1.5''$) than toward the south ($\sim0.7''$), similar to CCH and c-C$_{3}$H$_{2}$ line emission lengths.
\citet{Evans2023} and \citet{Martinez-Henares2025} discussed the similar asymmetric structures in H$_{2}$CO line emission of the Class 0/I protobinary system [BHB2007] 11.
\\ \\
We note that the H$_2$CO emitting layer has a slightly parabolic X-shape (especially in the side with $Z>0$), and the $Z$-coordinate of the emitting layer increases as $R$-coordinate increases (see Figure \ref{Figure2} (d)).
This result might suggest that contamination from emission originating in the disk midplane is less significant than emission from the disk surface layer, since the radius $R$ of the contaminating emission is expected to be largest in the region at $Z\sim0$ (see also Section Figure \ref{sec:2-1}). We note that the parabolic X-shape structure of H$_{2}$CO line emission was also confirmed by previous ALMA observations in edge-on disks, such as the disk around the Class I protostar IRAS 04302+2247 (e.g., \citealt{Podio2020}).
\\ \\
Figures \ref{Figure3} (a), (b), and (c) show the integrated intensity maps of c-C$_{3}$H$_{2}$, H$_{2}$CO, and C$^{18}$O overlaid with the CCH emission, which focus on the disk/envelope component.
The line emitting regions of CCH and c-C$_{3}$H$_{2}$ are similar to each other. 
The left panel of Figure \ref{Figure4} shows the profiles of the integrated intensities of C$^{18}$O, H$_{2}$CO, CCH, and  c-C$_{3}$H$_{2}$ lines in the vertical (disk minor axis, $Z$) direction passing through the protostar position. The distances of the line intensity peaks from the central source are $\sim50$ au for C$^{18}$O, $\sim50-150$ au for H$_{2}$CO, and $\sim150-300$ au for CCH and c-C$_{3}$H$_{2}$.
\\ \\
In Figures \ref{Figure2} (a), (b), (c), and (d), while the ``hole'' seen in the C$^{18}$O emission map could originate from high dust optical depth, the dark lane found in CCH and c-C$_{3}$H$_{2}$ emission map is unlikely to be solely due to the high optical depth.
We focus on the individual channel maps after the continuum subtraction, i.e., (Line brightness temperature) - (Dust brightness temperature).
For the C$^{18}$O and H$_{2}$CO, the fact that there is a dip in the emission coincident with the dust continuum but it does not go negative except at the continuum peak means that the C$^{18}$O and H$_{2}$CO emission has a brightness temperature similar to, or greater than, that of the dust.
It is therefore more natural to consider that this dark lane seen in CCH and c-C$_{3}$H$_{2}$ maps is due to the scarcity of the chemical
species (See Section \ref{sec:4-1} for details).
Indeed, the dark lane continues to the outer part of the disk/envelope system, $\sim1''$ ($\sim300$ au) away from the protostar in the disk/envelope direction, where the dust continuum emission is very weak.
We note that if the emission from CCH was obscured by the dust, we would not expect to have C$^{18}$O at the same region.
\\ \\
We note that the reason why both C$^{18}$O and H$_{2}$CO have very weak emission in the central region where we have the peak continuum emission could be (1) because the molecules freeze onto the dust grains or (2) because the continuum emission is optically thick thus absorbing the line emission. 
The second hypothesis is more likely because we would see no emission if the C$^{18}$O molecules freeze onto dust grains.
In addition, the first hypothesis can be denied because the line emission have negative components in Position-Velocity diagrams (Figure \ref{Figure6}).
\subsection{The Radial Profiles of C$^{18}$O, H$^{13}$CO$^{+}$, and DCO$^{+}$ Line Emission}\label{sec:3-3}
\begin{figure*}[htbp]
\begin{center}
\vspace{1cm}
%
\includegraphics[scale=0.35]{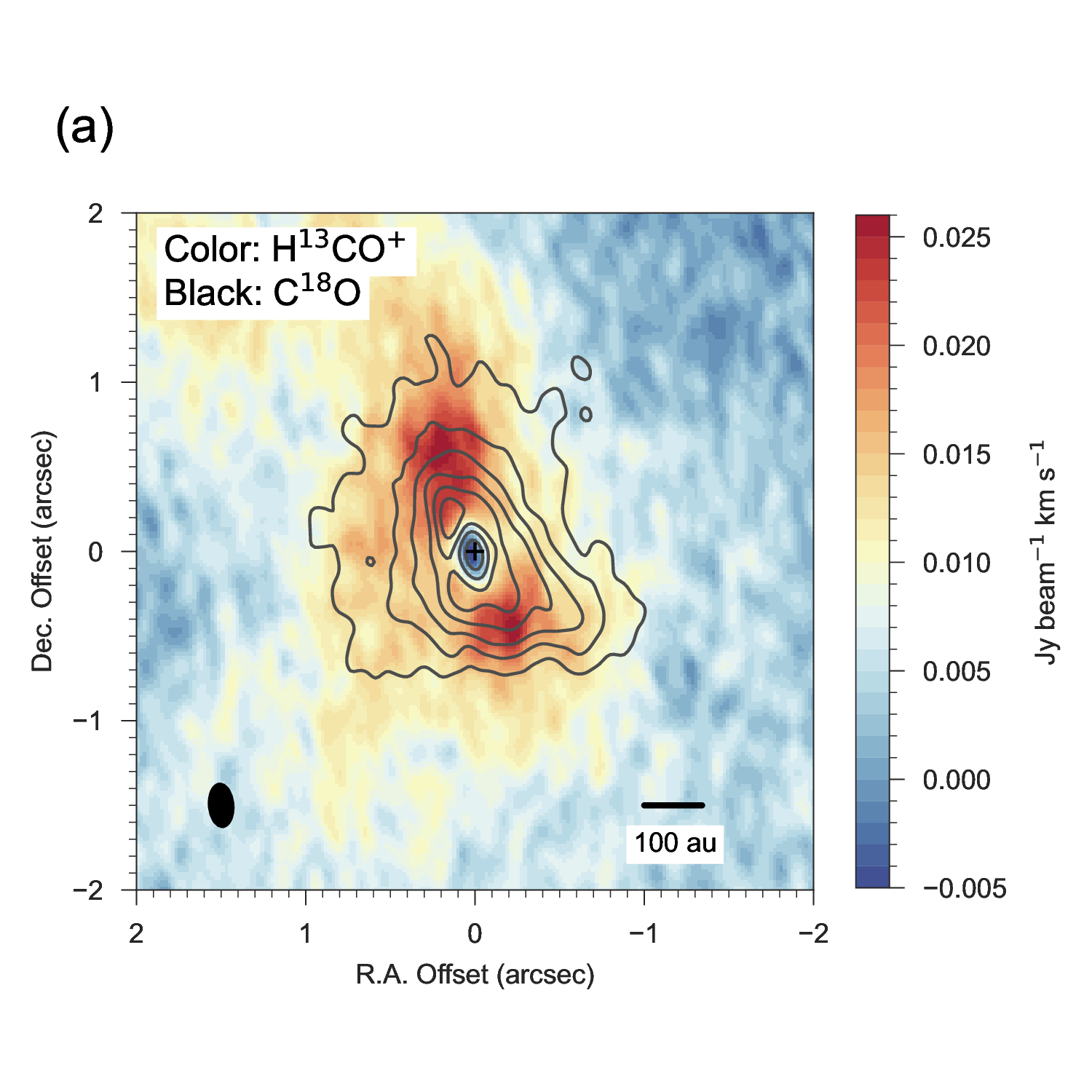}
\includegraphics[scale=0.35]{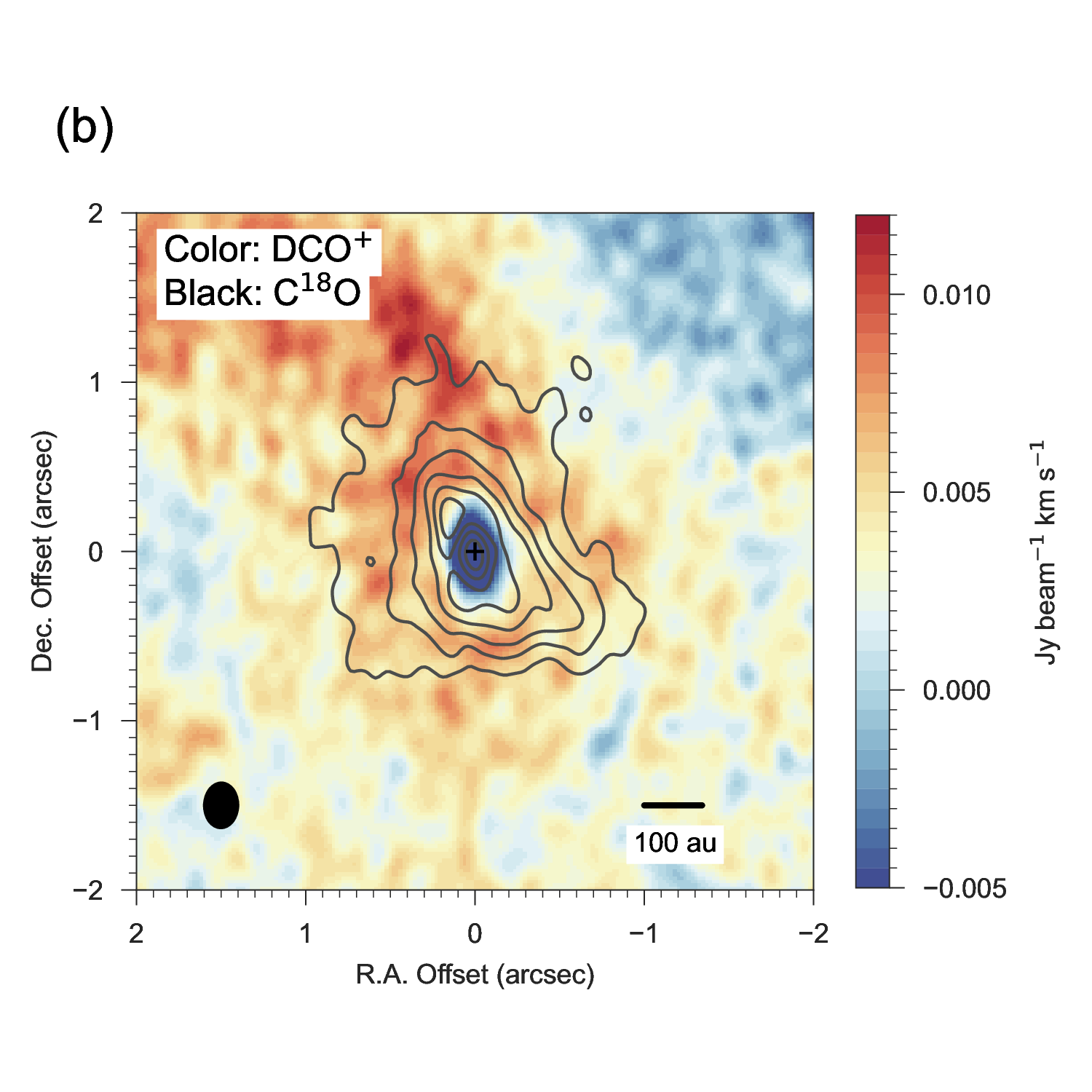}
%
\end{center}
\vspace{-1cm}
\caption{
Integrated intensity map (moment 0 map) of H$^{13}$CO$^{+}$ (left) and DCO$^{+}$ (right) overlaid on the C$^{18}$O emission (black contours; see also Figure \ref{Figure2}).
The rms noise level for line emission is $\sim1$ mJy beam$^{-1}$ km s$^{-1}$.
The contour levels of the C$^{18}$O line emission are 10$\sigma$, 15$\sigma$, 20$\sigma$, 25$\sigma$, 30$\sigma$, and 35$\sigma$.
\vspace{0.2cm}
}\label{Figure5}
\end{figure*}

Figures \ref{Figure2} (e) and (f) show the integrated intensity maps of H$^{13}$CO$^{+}$ ($J=3-2$), and DCO$^{+}$ ($J=3-2$) lines overlaid on the dust continuum emission contours.
Figures \ref{Figure3} (e) and (f) show the integrated intensity maps of H$^{13}$CO$^{+}$ ($J=3-2$), and DCO$^{+}$ ($J=3-2$) overlaid with the CCH line emission.
Figure \ref{Figure5} shows the integrated intensity maps of H$^{13}$CO$^{+}$ ($J=3-2$), and DCO$^{+}$ ($J=3-2$) overlaid with the C$^{18}$O line emission.
The H$^{13}$CO$^{+}$ line emission mainly exists along the midplane of the disk/envelope structure.
The intensity peaks of the H$^{13}$CO$^{+}$ line ($\sim0.''5$ from the central source) are located outside the peaks of the continuum emission and C$^{18}$O line emission.
\\  \\
The DCO$^{+}$ line emission mainly exists along the midplane of the disk/envelope structure, although there is weak emission from DCO$^{+}$ which is located at the outflow cavity walls.
The DCO$^{+}$ line emission from northern side ($>0.01$ Jy beam$^{-1}$ km s$^{-1}$) is stronger than that from southern side ($\sim0.008$ Jy beam$^{-1}$ km s$^{-1}$ at most).
In addition, the intensity peak of the DCO$^{+}$ line ($>1.''0$ from the central source) is located outside the C$^{18}$O and H$^{13}$CO$^{+}$ line emission.
\\ \\
The right panel of Figure \ref{Figure4} shows the radial profiles of the integrated intensities of C$^{18}$O, H$^{13}$CO$^{+}$, and DCO$^{+}$ lines in the radial (disk major axis, $R$) direction.
The distances of the line intensity peaks from the central source are $\sim50-100$ au for C$^{18}$O, $\sim100-200$ au for H$^{13}$CO$^{+}$, and $\gtrsim300$ au for DCO$^{+}$.
\begin{figure*}[htbp]
\begin{center}
\vspace{4cm}
\includegraphics[scale=0.58]{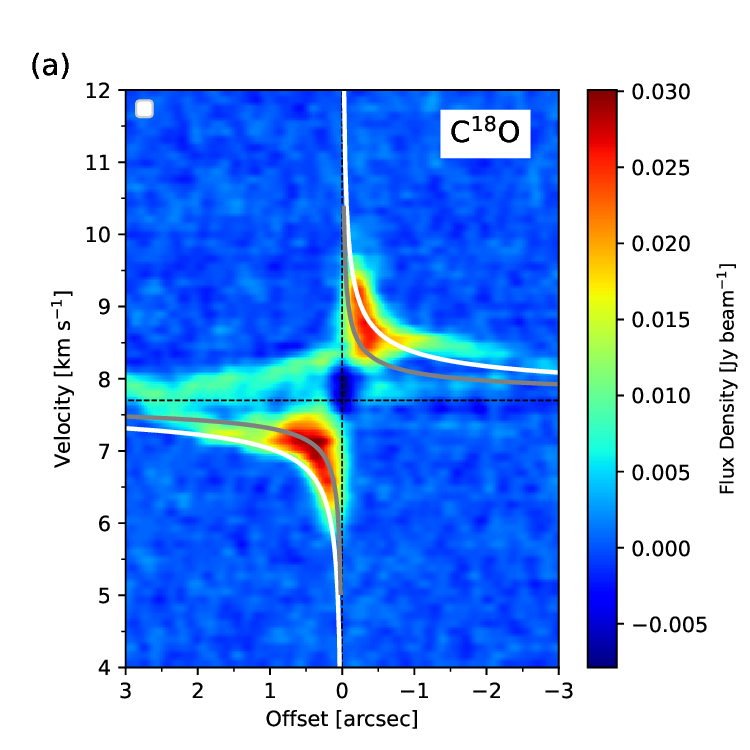}
\includegraphics[scale=0.58]{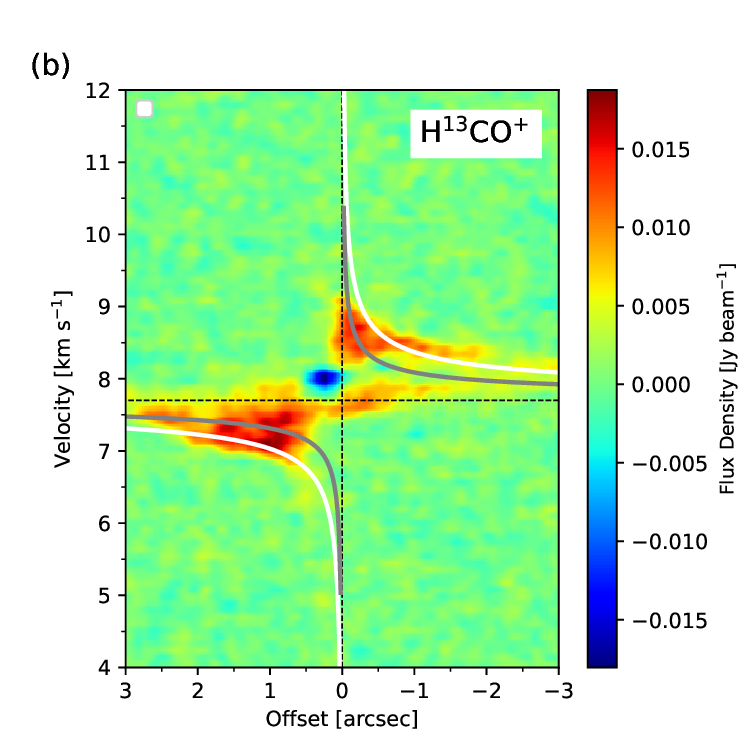}
\includegraphics[scale=0.58]{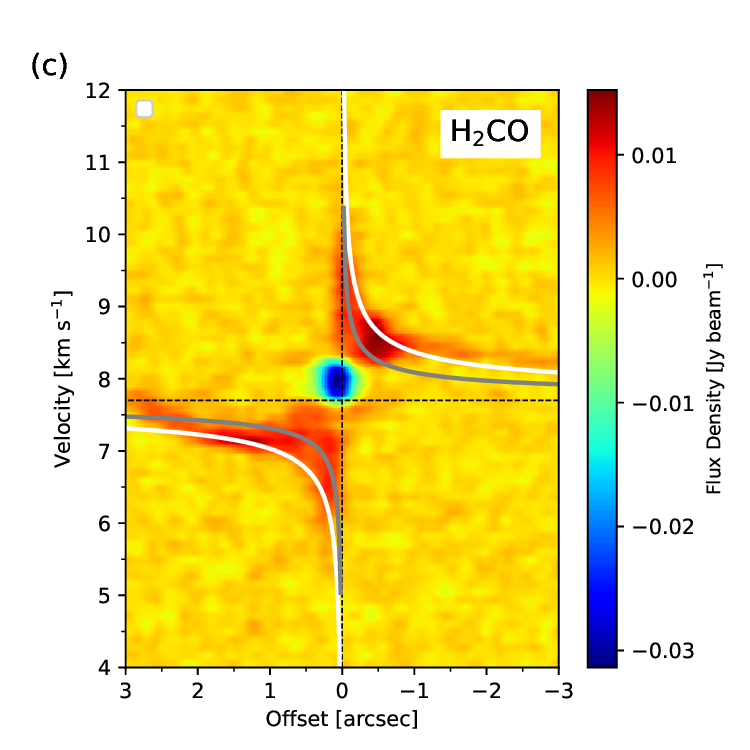}
\includegraphics[scale=0.58]{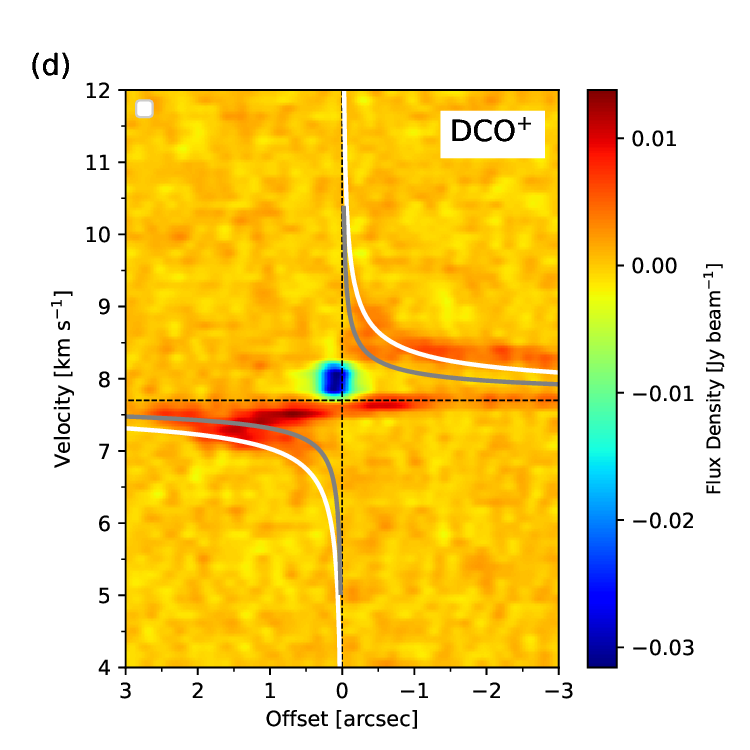}
%
\end{center}
\vspace{-2cm}
\caption{
Position-Velocity diagrams of C$^{18}$O, H$^{13}$CO$^{+}$, H$_{2}$CO, and DCO$^{+}$
line emission along cuts perpendicular to the outflow axis and passing through the continuum peak, i.e., along the major axis of the continuum emission ($R$ direction, see Figure \ref{Figure2}).
The white and gray curves correspond to rotation curves assuming Keplerian rotation.
In calculating the rotation curves, we assume the distance of the source as 293 pc, the central star mass as 0.18M$_{\mathrm{\odot}}$ (based on previous estimates in \citealt{OrizLeon2018, Zhang2018}, white curves) and 0.06M$_{\mathrm{\odot}}$ (gray curves), the inclination angle $i$ as $25^\circ$ \citep{Tobin2015, Segura-Cox2016} (here we adopted inclination between the disk plane and the line of sight), and the source systemic velocity as 7.7 km s$^{-1}$.
The white squares on the upper left corners show the pixel sizes to make these PV diagrams.
}\label{Figure6}
\end{figure*}
\\ \\
Finally, Figure \ref{Figure6} shows the position-velocity (PV) diagrams of C$^{18}$O, H$^{13}$CO$^{+}$, H$_{2}$CO, and DCO$^{+}$
line emission along the cuts perpendicular to the outflow axis and passing through the continuum peak, i.e., along the major axis of the continuum emission and therefore along the disk ($R$ direction). We set the averaging widths as 0.1$''$, so that we extract the emission component around the midplane only.
The white and gray curves correspond to the rotation profile assuming Keplerian rotation.
In calculating the rotation curves, we assume the distance of the source as 293 pc, the central star mass as 0.18~M$_{\mathrm{\odot}}$ (based on previous estimates in \citealt{OrizLeon2018, Zhang2018}, white curves) and 0.06~M$_{\mathrm{\odot}}$ (gray curves), the disk inclination angle $i$ as $25^\circ$ \citep{Tobin2015, Segura-Cox2016}, and the source systemic velocity as 7.7 $\mathrm{km}~\mathrm{s}^{-1}$.
Here, we adopted inclination between the disk plane and the line of sight, thus we assumed completely edge-on as $0^\circ$ instead of $90^\circ$ throughout this paper.
We note that the rotation curves with lower central star mass (0.06~M$_{\mathrm{\odot}}$) well fit the PV diagrams of C$^{18}$O and H$^{13}$CO$^{+}$ than those with 0.18~M$_{\mathrm{\odot}}$, suggesting that the central star mass might be smaller than the previous estimates.
We note that fitting infalling envelope models would also be useful to investigate the velocity profiles in detail (e.g., \citealt{Oya2022}).
The detailed kinematic analyses of the disk/envelope system will be described in a separate FAUST publication, combined with the outflow emission analyses.
\\ \\
In Figure \ref{Figure6}, the emission peaks of the C$^{18}$O line are located at $\lesssim0.5''$ and $\gtrsim1$ km s$^{-1}$ away from the source systemic velocity position ($\sim7.7$ km s$^{-1}$, \citealt{Higuchi2018, Zhang2018}) in second and fourth quadrant, whereas those of H$^{13}$CO$^{+}$ and DCO$^{+}$ are located $\sim1.0''$ and $<1$ km s$^{-1}$ away.
We note that the differences between H$^{13}$CO$^{+}$ and DCO$^{+}$ are also clear in the PV diagrams. There are second emission peaks at $\sim1.5$ km s$^{-1}$ observed within the inner regions in H$^{13}$CO$^{+}$, which are not seen in the DCO$^{+}$ emission.
\subsection{Derivation of Column Densities and Optical Depth}\label{sec:3-4}
Figure \ref{Figure7} shows the distribution of column densities for C$^{18}$O, H$^{13}$CO$^{+}$, DCO$^{+}$, and H$_{2}$.
We calculated the column densities $N_{i}$ except for H$_{2}$ based on the following equation which \citet{Persson2012} and \citet{Cabedo2023} adopted:
\begin{equation}\label{eqNi}
\begin{split}
N_{i} = &\frac{8\pi}{{\lambda}^{3}A_{ul}}\frac{1}{B_{\nu}(T_{\mathrm{ex}})-B_{\nu}(T_{\mathrm{bg}})}\\ &\times \frac{1}{1-\exp(-h\nu/k_{\mathrm{B}}T_{\mathrm{ex}})}  \frac{Q_{\mathrm{rot}}}{g_{u}\exp(-E_{l}/k_{\mathrm{B}}T_{\mathrm{ex}})}\\ &\times \int I_{\nu} \mathrm{d}v,
\end{split}
\end{equation}
where ${\lambda}$ and $\nu$ are the wavelength and frequency of the transition, $A_{ul}$ is the Einstein A coefficient, 
$k_{\mathrm{B}}$ is the Boltzmann constant, $g_{u}$ is the upper state degeneracy, $Q_{\mathrm{rot}}$ is the partition function at the temperature $T=T_{\mathrm{ex}}$, 
$E_{l}$ is the energy of the lower level,
$B_{\nu}(T)$ is the Planck function for the background ($T_{\mathrm{bg}}=2.7$ K) and for the excitation temperature of the line ($T_{\mathrm{ex}}$), and $\int I_{\nu} \mathrm{d}v$ is the integrated intensity.  This equation assumes optically thin emission.
First of all, in Figures \ref{Figure7} and \ref{Figure8} we assume $T_{\mathrm{ex}}=50$ K throughout the disk/envelope system, since we mainly focus on the inner region where 
the SO line emission is detected on the scale of 1$''$ (300 au) radius from the central protostar \citep{Zhang2018}. The sublimation temperature of SO in disk condition is around 50 K (e.g., \citealt{Sakai2014a}).
We have confirmed that the column density values increase / decrease only by a factor of around two by changing $T_{\mathrm{ex}}$ to 20 K /100 K, respectively.
\\ \\
The bolometric luminosity $L_{\mathrm{bol}}$ is $1.1L_{\odot}$ for IRAS 4C (See Section \ref{sec:2-1}).
Thus, assuming that the dust temperature in the midplane $T_{d}(R)$
is determined by the balance of heating from the central star radiation and cooling by radiation of optically thick dust, we can estimate $T_{d}(R)$ by the following equation (e.g., \citealt{Ceccarelli2000}),
\begin{equation}
T_{d}(R)=49 \left(\frac{L_{\mathrm{bol}}}{L_{\odot}} \right)^{1/4} \left(\frac{R}{300 \mathrm{au}}\right)^{-1/2} K,
\end{equation}
where we assume that $T_{d}(R)=50$ K at $L_{\mathrm{bol}}$$=1.1L_{\odot}$ and $R=300$ au (on the scale of 1$''$) based on the spatial extent of the detected SO line emission \citep{Zhang2018}.
In this case, the disk radii where $T(R)<20$ K and $T(R)>100$ K are around $R\gtrsim1.9\times10^{3}$ au and around $R\lesssim76$ au.
In the innermost region the $T_{d}(R)$ would be much larger because of the mass accretion heating (like the disk around Class I protostar R CrA IRS7B-a, \citealt{Takakuwa2024}).
We note that \citet{Tobin2020} described the empirical equation of the average dust temperature for the whole disk as $T_{d}=43(L_{\mathrm{bol}}/L_{\odot})^{0.25}$. Based on this empirical equation, the averaged dust temperature $T_{d}$ is estimated to be 44 K.
\\ \\
The peak column densities are $\sim10^{13}$ cm$^{-2}$ for H$^{13}$CO$^{+}$, $\sim2\times10^{12}$ cm$^{-2}$ for DCO$^{+}$, and $\sim(3-4)\times10^{16}$ cm$^{-2}$ for C$^{18}$O.
The peaks of the H$^{13}$CO$^{+}$ column density ($\sim0.''5$ ($\sim150$ au) from the central source) are located outside the peaks of the continuum emission and C$^{18}$O column density (see also Figure \ref{Figure5}).
The DCO$^{+}$ column density in the northern side ($\sim(1-2)\times10^{12}$ cm$^{-2}$) are larger than those on the southern side ($\lesssim5\times10^{11}$ cm$^{-2}$).
In addition, the peaks of the DCO$^{+}$ column density ($>1.''0$ from the central source) are located outside the C$^{18}$O and H$^{13}$CO$^{+}$ column density peaks.
\\ \\
Here we note that the derivation of column densities for the molecules is dependent on the assumption that the emission is optically thin \citep{vantHoff2018_L1527, Sharma2025}.
Figure \ref{FigureA1_tau} in Appendix \ref{Appendix-A} shows the line optical depth $\tau$ for H$^{13}$CO$^{+}$, DCO$^{+}$, and C$^{18}$O.
We calculated the line optical depth $\tau$ based on the following equation which \citet{vantHoff2018} adopted: 
\begin{equation}
\begin{split}
\tau = &\frac{{\lambda}^{3}}{8\pi} \frac{A_{ul}}{\Delta V} \frac{g_{u}N_{i}}{Q_{\mathrm{rot}}}\exp\left(-\frac{E_{u}}{k_{\mathrm{B}}T_{\mathrm{ex}}}\right)\left(1-\exp\left(\frac{-h\nu}{k_{\mathrm{B}}T_{\mathrm{ex}}}\right)\right),
\end{split}
\end{equation}
where $N_{i}$ is the column densities calculated in equation (\ref{eqNi}) (see also Figure \ref{Figure7}) and we assume $T_{\mathrm{ex}}=50$ K and line velocity width $\Delta V=0.17$ km s$^{-1}$ throughout the disk/envelope system. 
According to Figure \ref{FigureA1_tau} in Appendix \ref{Appendix-A}, the peak positions of optical depth are coincident with those of column densities, and 
the peak optical depth are $\sim0.7$ for DCO$^{+}$, $\sim1.5$ for H$^{13}$CO$^{+}$, and $\sim3.5$ for C$^{18}$O.
The H$^{13}$CO$^{+}$ line emission is marginally optically thick ($\tau\sim1-1.5$) at $r\sim100-200$ au, and C$^{18}$O is optically thick $\tau\gtrsim2$ in $r\lesssim150$ au.
Thus, the column densities of C$^{18}$O might be underestimated compared to those of H$^{13}$CO$^{+}$.
We have confirmed that the values of $\tau$ increase / decrease only by a factor of around two by changing $T_{\mathrm{ex}}$ to 20 K /100 K, respectively. 
\\ \\
The column densities of H$_{2}$ are calculated from the dust continuum emission using the following equation: 
\begin{equation}
\begin{split}
N_{\mathrm{H}_{2}} = - \frac{1}{\mu_{\mathrm{H}_{2}}m_{\mathrm{H}}\kappa_{\nu}} \ln \left[1-\frac{S_{\nu}^{\mathrm{beam}}}{\Omega_{\mathrm{beam}}B_{\nu}(T_{d})} \right],
\end{split}
\end{equation}
where $\mu_{\mathrm{H}_{2}}=2.8$, $m_{\mathrm{H}}$ is the atomic hydrogen mass, $\Omega_{\mathrm{beam}}$ is the beam solid angle, $B_{\nu}(T_{d})$ is the Planck function at the dust temperature $T_{d}$ (here we assume $T_{\mathrm{ex}}=T_{d}=50$ K), 
$S_{\nu}^{\mathrm{beam}}$ is the dust continuum emission flux density per beam (see Figure \ref{Figure1}),
and $\kappa_{\nu}$ is the dust mass opacity per gas mass at the frequency $\nu$. For the dust opacity we adopt a power-law fit given by the following equation,
\begin{equation}
\kappa_{\nu} = \frac{\kappa_{0}}{\xi_{d}} \left(\frac{\nu}{\nu_{0}} \right)^{\beta}, 
\end{equation}
with a dust mass absorption coefficient $\kappa_{0}=1.6$ cm$^{2}$ g$^{-1}$ at $\lambda_{0}=1.3$ mm, following \citet{Ossenkopf1994} and \citet{Cabedo2023}.
We assume a standard gas-to-dust ratio $\xi_{d}=100$, 
and a dust emissivity exponent $\beta=1.7$ ($=$ the value of interstellar average).
\citet{Cacciapuoti2025} reported the values of $\beta$ in protostellar envelopes for the whole samples of FAUST observations, and they found $\beta>1.4$ for IRAS 4C.
We note that some previous (sub)millimeter observations of protoplanetary disks have reported much smaller values of $\beta$ (such as $<0$), which may be explained by dust self-scattering \citep{Li2017}.
The resulting H$_{2}$ column densities are $3\times10^{23}$ cm$^{-2}$ within 0.5$"$ from the central protostar.
\section{Discussion}\label{sec:4}
\subsection{The Vertical Chemical Layered Structures}\label{sec:4-1}
\begin{figure*}[htbp]
\begin{center}
%
\includegraphics[scale=0.35]{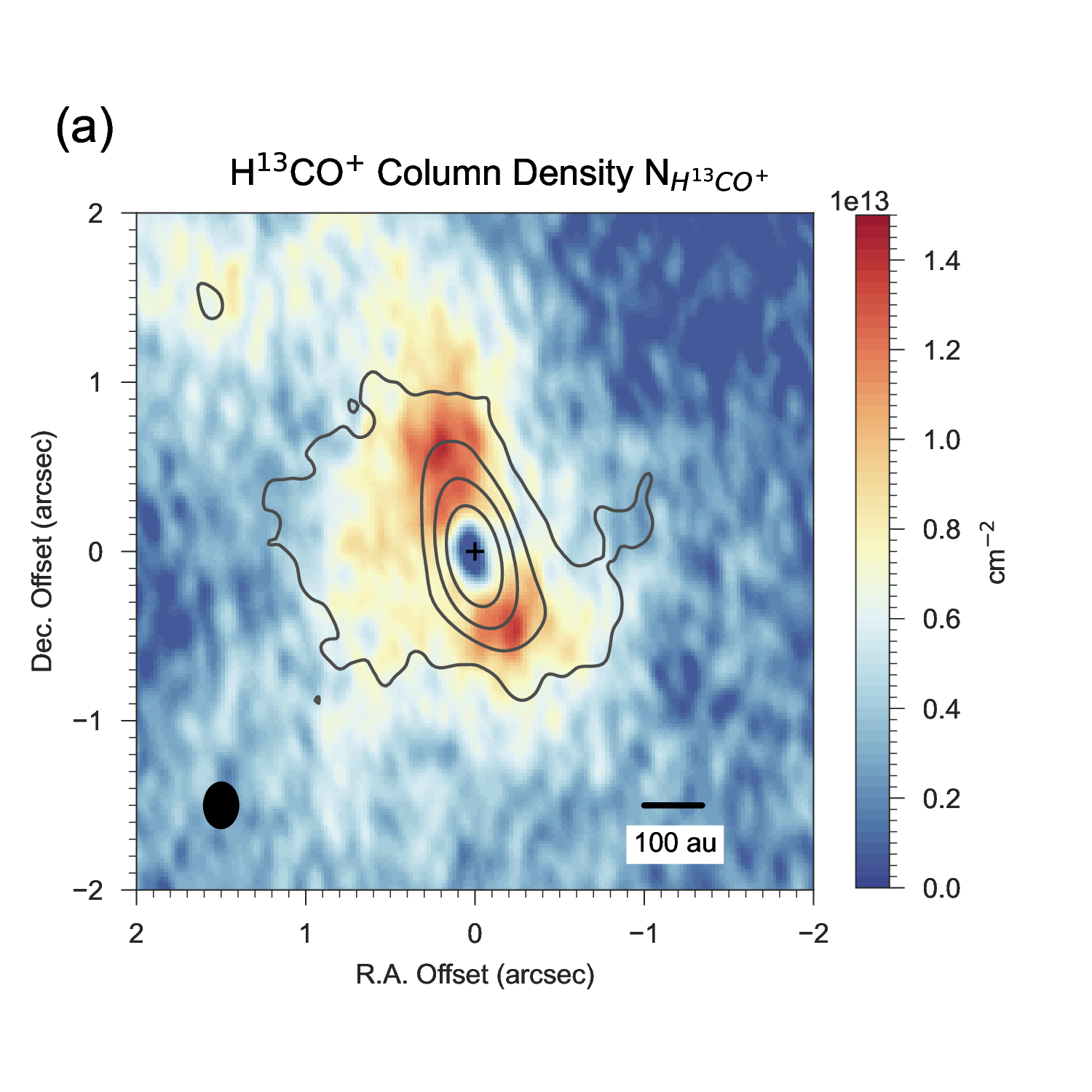}
\includegraphics[scale=0.35]{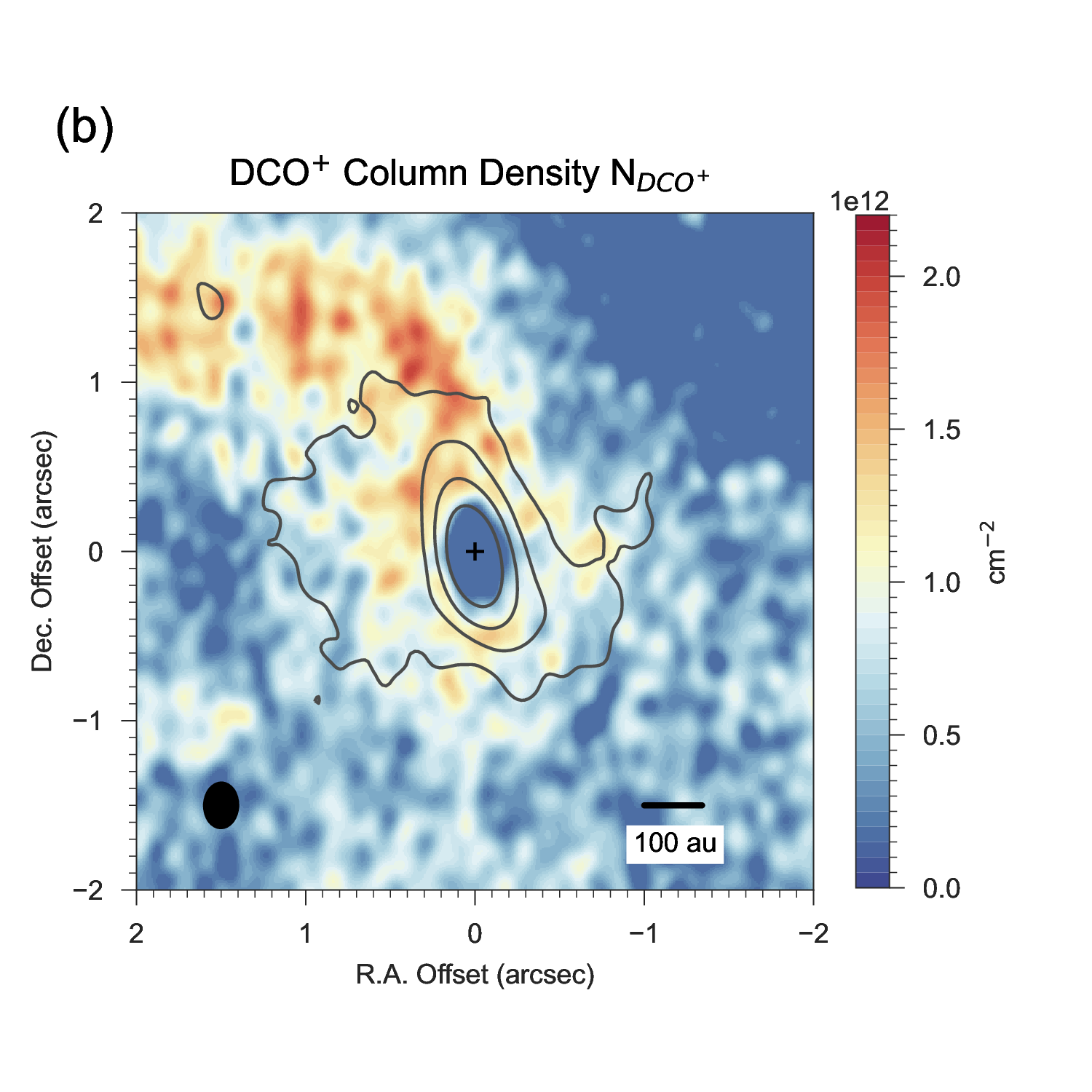}
\includegraphics[scale=0.35]{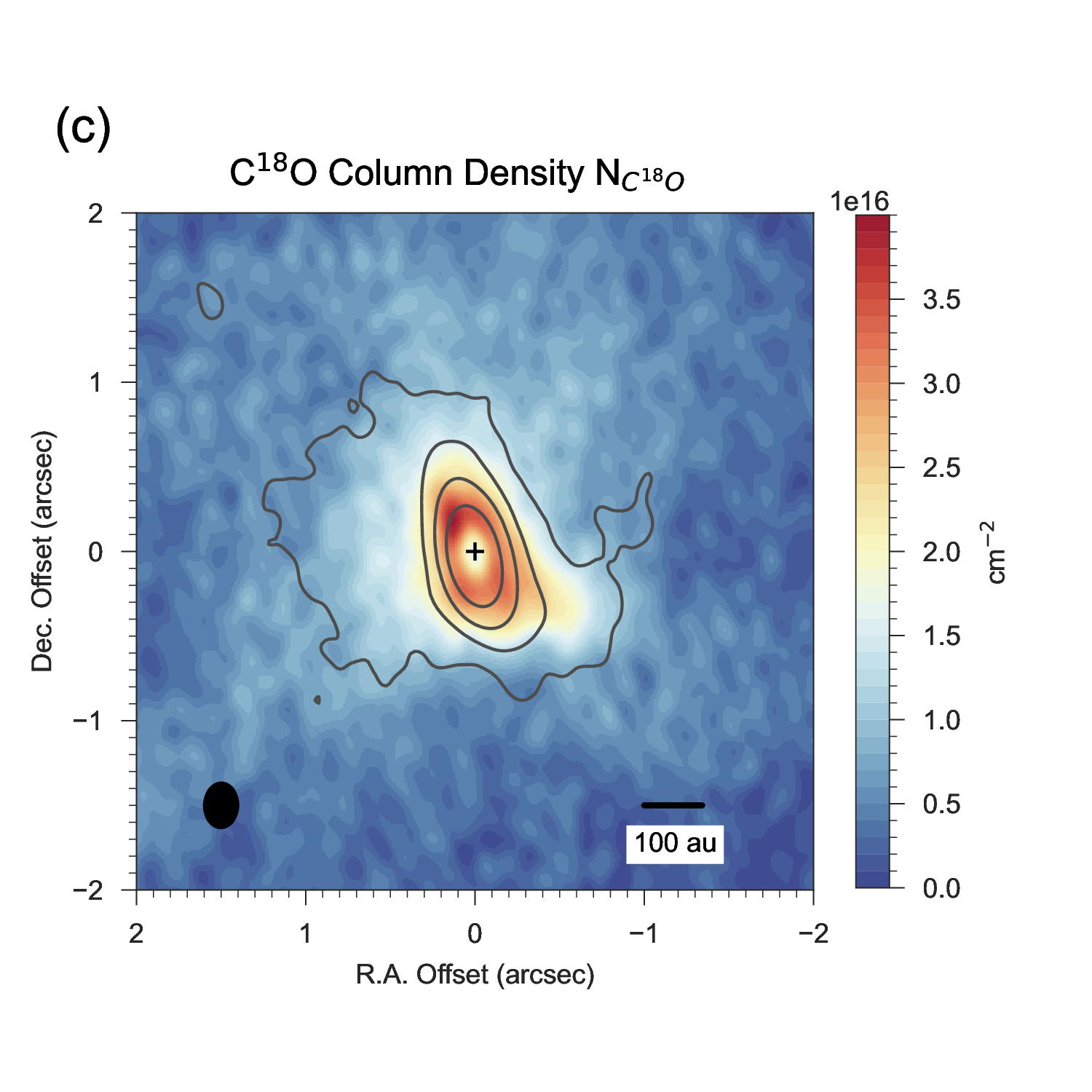}
\includegraphics[scale=0.35]{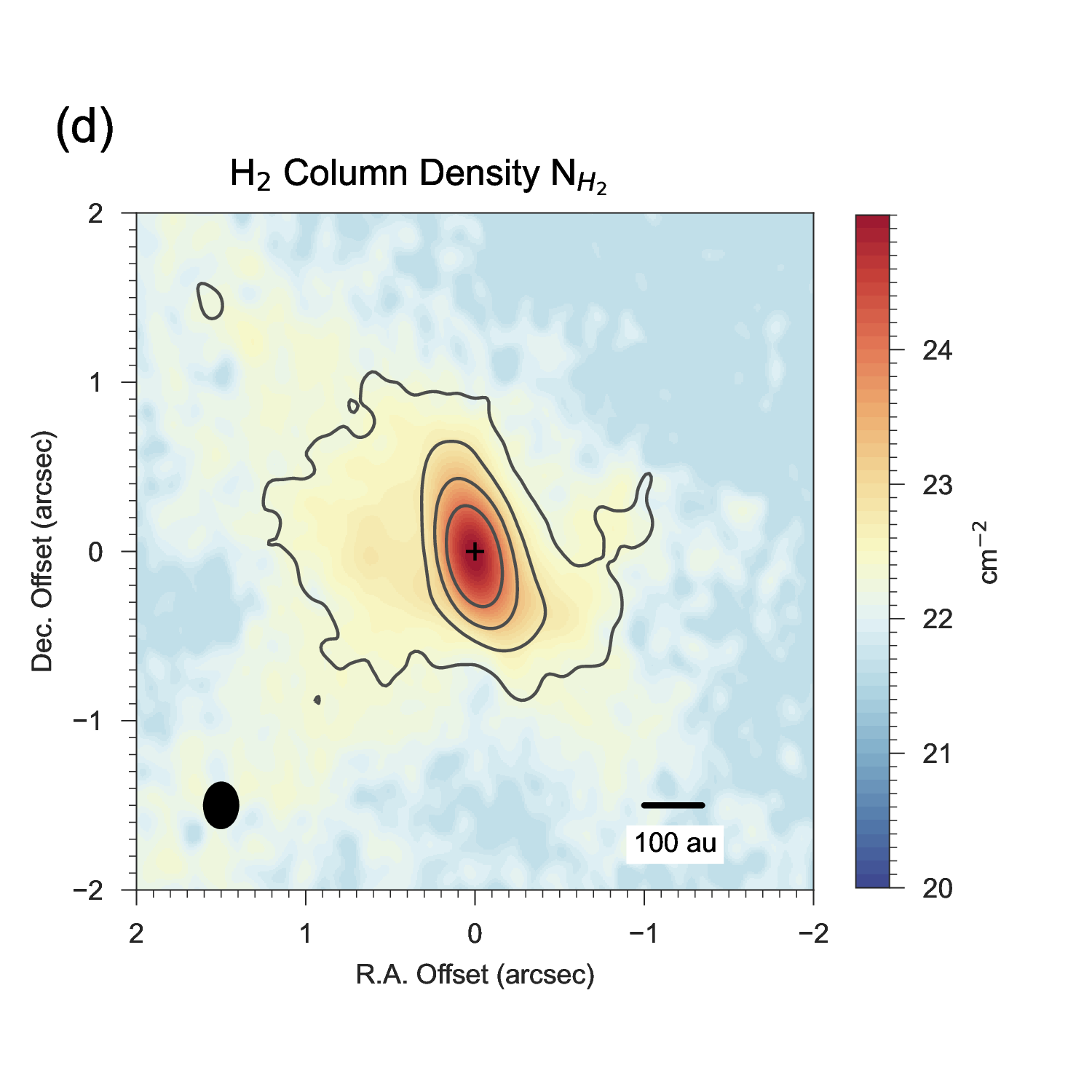}
\end{center}
\caption{
Distribution of Column Densities for H$^{13}$CO$^{+}$ (a), DCO$^{+}$ (b), C$^{18}$O (c), and H$_{2}$ (d).
Note that column densities of H$_{2}$ (bottom right) are shown using a logarithm scale.
The 1.3 mm continuum image is overploted by black contours (see also Figure \ref{Figure1}).
For the continuum emission, the rms noise level ($\sigma$) of the continuum is 20 $\mu$Jy beam$^{-1}$, and the contour levels are 5$\sigma$, 20$\sigma$, 80$\sigma$ and 320$\sigma$.
In calculating the column densities, we assume $T_{\mathrm{ex}}=50$ K throughout the disk/envelope system.
In calculating these panels, we clipped the values of integrated intensities of each molecule with 1$\sigma$ ($=$1 mJy beam$^{-1}$ km s$^{-1}$) and the value of continuum intensity with 1$\sigma$ ($=$20 $\mu$Jy beam$^{-1}$).
\vspace{0.2cm}
}\label{Figure7}
\end{figure*}
\begin{figure*}[htbp]
\begin{center}
\includegraphics[scale=0.35]{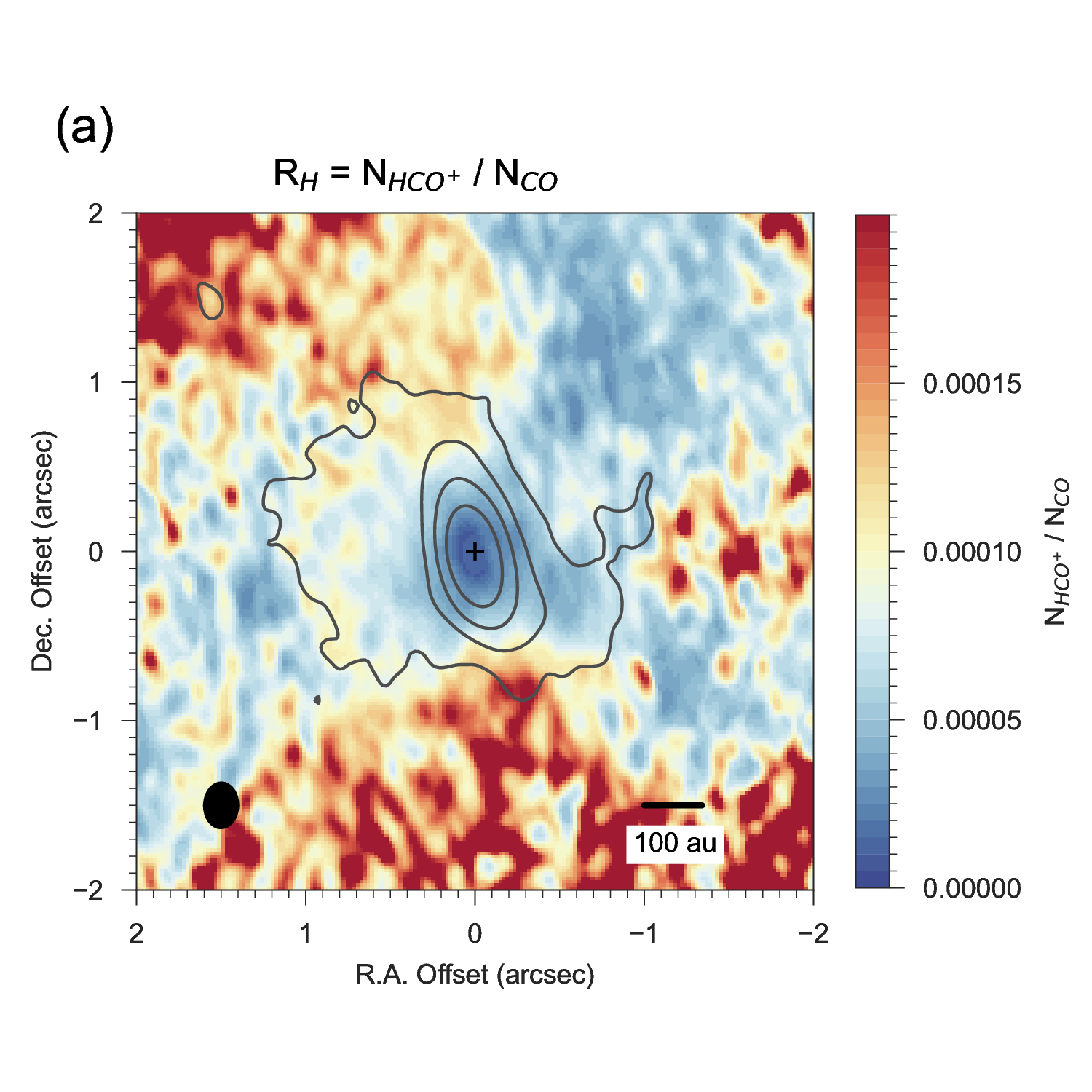}
\includegraphics[scale=0.35]{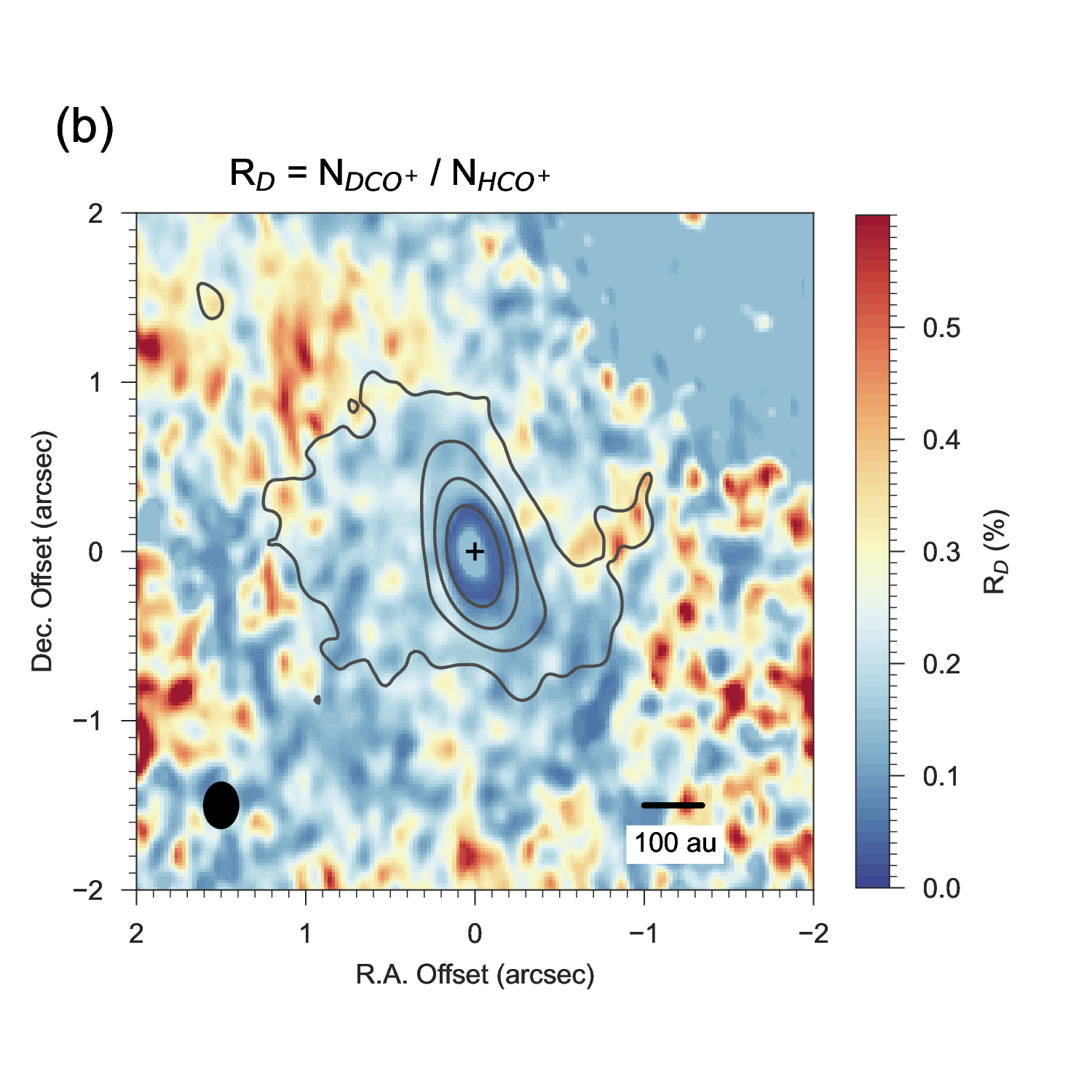}
\includegraphics[scale=0.35]{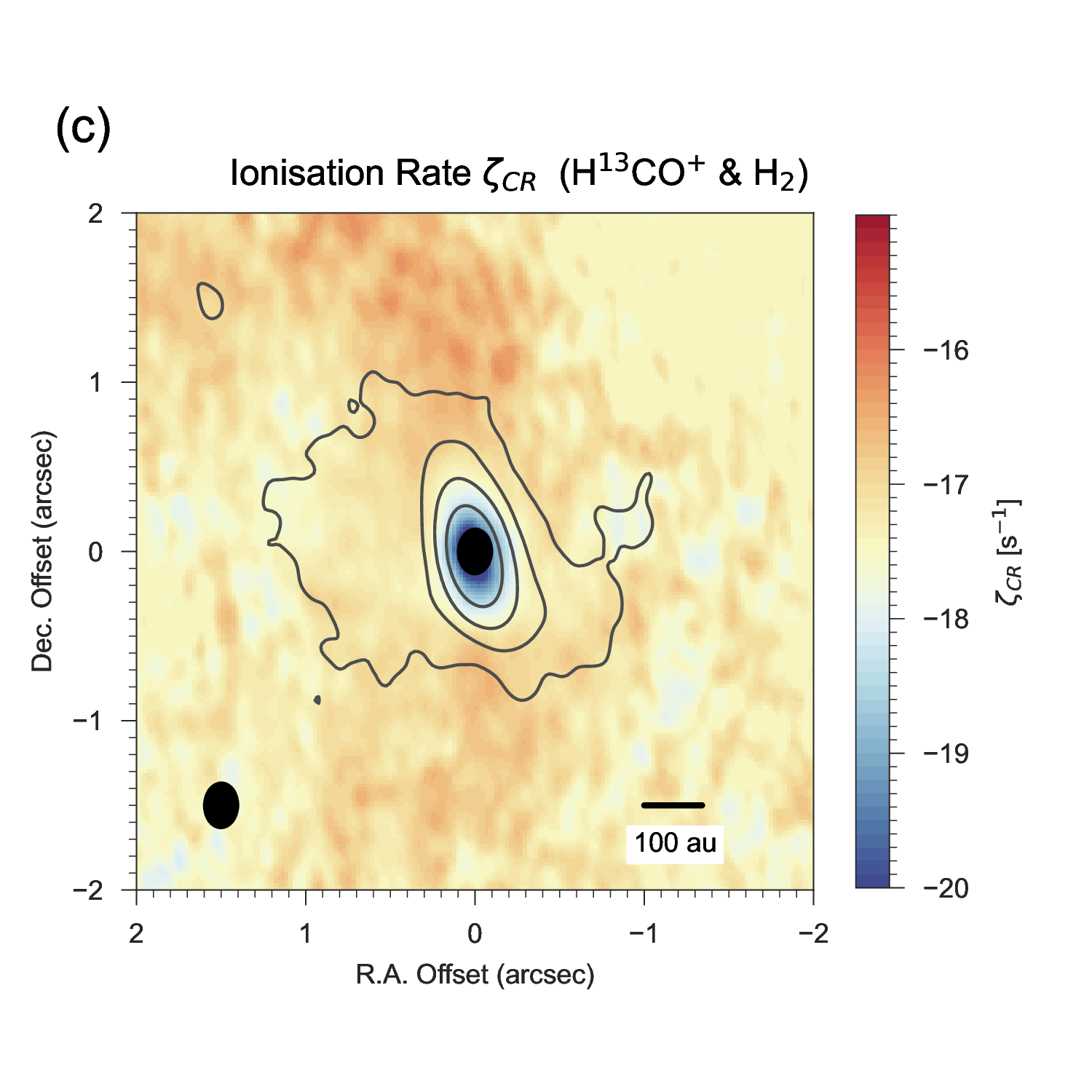}
\includegraphics[scale=0.35]{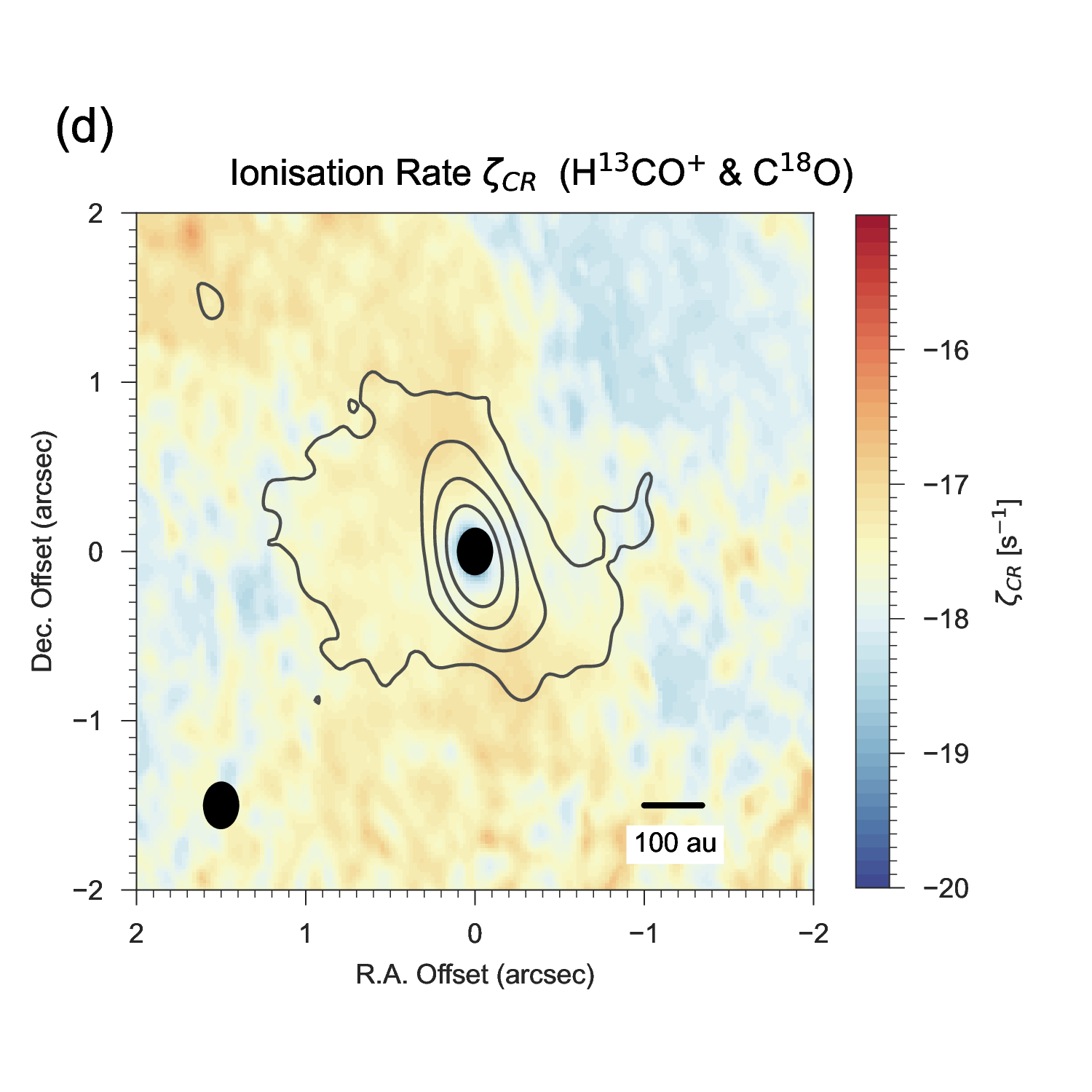}
\end{center}
\caption{
(a): The distribution of the column density ratio $R_{\mathrm{H}}=N_{\mathrm{HCO}^{\mathrm{+}}}/N_{\mathrm{CO}}$ (CO ionisation degree).
(b): The distribution of the deuteration fraction $R_{\mathrm{D}}$ calculated from column density ratio between DCO$^{+}$ and HCO$^{+}$.
(c) The distribution of the ionisation rate $\zeta_{\mathrm{CR}}$ estimated from H$^{13}$CO$^{+}$ and H$_{2}$.
(d) The distribution of the ionisation rate $\zeta_{\mathrm{CR}}$ estimated from H$^{13}$CO$^{+}$ and C$^{18}$O.
In calculating these panels, we clipped the values of integrated intensities of each molecule with 1$\sigma$ ($=$1 mJy beam$^{-1}$ km s$^{-1}$) and the value of continuum intensity with 1$\sigma$ ($=$20 $\mu$Jy beam$^{-1}$).
In addition, we also clipped the values of column densities for H$^{13}$CO$^{+}$ and DCO$^{+}$ with $2.0\times10^{12}$ cm$^{-2}$ and $2.0\times10^{11}$ cm$^{-2}$, respectively, to cut the plots which we do not discuss about ionisation because of its low accuracy.
We assume $T_{\mathrm{ex}}=50$ K throughout the disk/envelope system.
In addition, we assume $L$ (=depth of the structure) as $6\times10^{4}$ au based on \citet{Pineda2024} (see also Figure \ref{FigureB3}).
In (c) and (d), we cut the innermost region with significantly higher dust optical depth by the black filled ellipse, whose size and shape are the same as the beam of the dust continuum emission (see also Figures \ref{FigureB1}-\ref{FigureB3}).
\vspace{0.2cm}
}\label{Figure8}
\end{figure*}
\begin{figure}[htbp]
\begin{center}
\vspace{1cm}
\hspace{-2cm}
\includegraphics[scale=0.7]{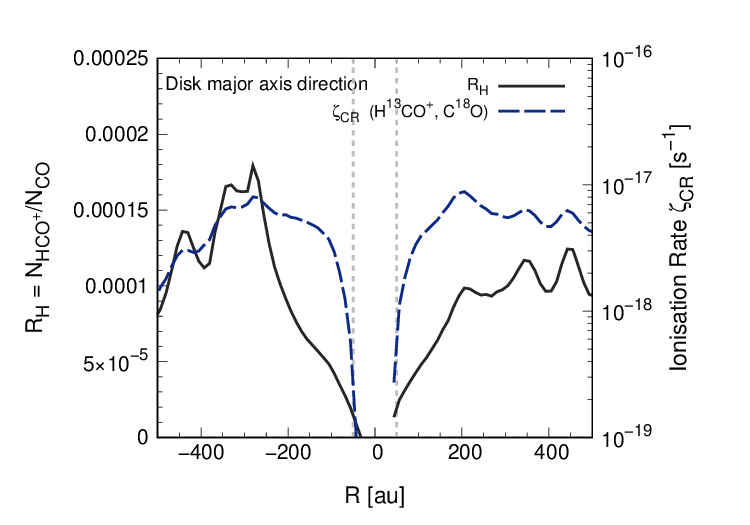}
\end{center}
\caption{Radial profiles of $R_{\mathrm{H}}$=$N_{\mathrm{HCO}^{+}}$/$N_{\mathrm{CO}}$ (black solid line) and the ionisation rate $\zeta_{\mathrm{CR}}$ (blue dashed line) estimated from the H$^{13}$CO$^{+}$ and H$_{2}$ with $L=6\times10^{4}$ au in the $R$ direction (disk major axis direction, see Figure \ref{Figure2}).
The gray dotted lines show the radius of $R=\pm50$ au, We include these dotted lines since it is not appropriate to discuss the ionisation rates at $R<50$ au.
\vspace{0.2cm}
}\label{Figure9}
\end{figure}
As we reported in Section \ref{sec:3-2}, 
it was found that the line emission of C$^{18}$O, H$_{2}$CO, c-C$_{3}$H$_{2}$, and CCH show a layered distribution in this order, outward and starting closest to the disk midplane.
Two possibilities can be considered as the origin of this layered structure. One is the excitation effect of molecular line emission (that is, molecular lines with higher critical densities trace the regions closer to the disk midplane), and the other is the varying abundance of molecules in each layer, also referred to as chemical stratification.
\\ \\
The C$^{18}$O line, which is detected most strongly in the region closest to the midplane and has the lowest critical density ($1\times10^{4}$ cm$^{-3}$ for $T_{\mathrm{ex}}=50$ K, \citealt{Chandra2000, Schoier2005}) among the observed species (see e.g., \citealt{Schoier2005, Shirley2015}), suggests that C$^{18}$O is abundant near the midplane ($Z\sim50$ au). 
When we compare the C$^{18}$O map with the dust continuum map, the C$^{18}$O emission appears slightly weaker along the midplane ($Z\sim0$), which is considered to be in the highest density condition (It is easily recognized by the contour map in Figure \ref{Figure2} and vertical intensity profiles in Figure \ref{Figure4}). 
Since we observed only one transition, we cannot rule out the possibility that the intensity of the spectral line changes due to temperature variations.
In addition, since the emission components of C$^{18}$O line at around 8 km s$^{-1}$ on the Position-Velocity diagram (see Figure \ref{Figure6}) are negative, we cannot remove the possibility that highly optically thick dust in the foreground obscured the information of the exact disk midplane.
However, it is also natural to consider that C$^{18}$O line emission is depleted because the regions along the exact disk midplane ($Z\sim0$) have the highest density and lower temperature conditions.  
Such phenomena have indeed been reported in some other young disks, such as L1489 IRS and IRAS 04302+2247 (e.g., \citealt{Pinte2018, Yamato2023}). Note that the column density map in Figure \ref{Figure7} was created under the assumption of a constant temperature ($T_{\mathrm{ex}}=50$ K).   
\\ \\
Intensity decreasing towards the disk height direction as approaching the disk midplane becomes much clearer in the H$_{2}$CO line emission map. Recognizing the much higher critical density of the H$_{2}$CO line ($6\times10^{5}$ cm$^{-3}$ for $T_{\mathrm{ex}}=50$ K, \citealt{Shirley2015}) compared to the C$^{18}$O line, with similar upper state energies (see Table \ref{Table:1}), the difference seen between the C$^{18}$O line and the H$_{2}$CO line maps could originate from a different distribution for these molecules.  Since we do not know the temperature distribution of the disk structure in this object, it is difficult to make a firm argument. 
However, considering that H$_{2}$CO has a higher sublimation temperature ($\sim40-70$K, see e.g., \citealt{Ferrero2020}) than CO ($\sim20$K), it is possible that H$_{2}$CO is more depleted than CO in the disk midplane where its temperature is lower than that in the warm molecular layer of the disk.
The left panel in Figure \ref{Figure4} shows that in the region within 200 au from the disk midplane, emissions from both C$^{18}$O and H$_{2}$CO become significantly stronger than at $Z\sim0$. These components might reflect the distribution of the molecular gas in the disk.
\\ \\
The observed line of c-C$_{3}$H$_{2}$ has a higher critical density and excitation than that of CCH (the critical densities for $T_{\mathrm{ex}}=50$ K are $5\times10^{7}$ cm$^{-3}$ for c-C$_{3}$H$_{2}$ and $8\times10^{6}$ cm$^{-3}$ for CCH, \citealt{Chandra2000, Schoier2005, Spielfiedel2012}), but their distributions do not differ significantly (see Figure \ref{Figure3}). Therefore, it is reasonable to consider that 
these molecules are inherently abundant in this layer, rather than being bright due to excitation effects.
Considering that the outermost layer is exposed to irradiation from the protostar and even radiation from outside the system, it is reasonable to interpret that photochemistry occurs in this layer. These molecules are indeed representative of photochemistry, as also reported with spatially resolved imaging in photodissociation regions (e.g., \citealt{Cuadrado2015, Guzman2015}).
\\ \\
Overall, the layered structure found in the C$^{18}$O, H$_{2}$CO, c-C$_{3}$H$_{2}$, and CCH lines appear to originate from the different abundance of molecules in each layer, which can be referred to as a chemical layered structure.
This chemical layered structure is found previously in Class I-II edge-on disks where gas accretion from the outer envelope is almost completed (e.g., \citealt{Podio2020, Oberg2023}).
On the other hand, in IRAS~4C, the H$^{13}$CO$^{+}$ and DCO$^{+}$ line emission seems to trace not layered structure but disk/envelope system, as we described in Section \ref{sec:3-3} and Figure \ref{Figure6}.
\\ \\
Based on thermo-chemical models, protoplanetary disks consist of three chemical layers in the vertical direction (e.g., \citealt{Aikawa2002, Aikawa2024, Woitke2009, Dutrey2014, Walsh2010, Walsh2012, Notsu2017, Oberg2023}).
The first layer is the hot surface layer, directly irradiated by the stellar radiation and thus warm, and molecules are photodissociated into atoms and radicals by (inter-)stellar UV and X-ray radiation. Thus, this disk surface is a photon-dominated region (PDR).
CCH and c-C$_{3}$H$_{2}$ become abundant in this region, especially in conditions where C/O$\gtrsim1$ (e.g., \citealt{Bergin2016, Cleeves2018, Bosman2021}).
The second layer is the warm ($\gtrsim20$ K) molecular layer, where UV radiation is moderately shielded and molecules are in the gas phase and gas-phase chemistry is active.
Ion-molecule reactions are triggered by X-rays from the central stars and cosmic-rays.
The third layer is the freeze-out layer, that is, the cold outer disk midplane where molecules freeze out onto dust grains (beyond the snowlines).
Even in the outer cold midplane, reactions in the gas phase and within ices can be triggered by high-energy cosmic-rays, X-rays, and decay of radioactive nuclei (e.g., $^{26}$Al).
Various complex organic molecules, such as CH$_{3}$OH, only form on the icy mantles of dust grains, whereas unsaturated molecules such as carbon chain molecules can form by gas-phase reactions in the warm molecular and hot surface layers.
\\ \\
Here we compare our observed images in IRAS~4C with the vertical chemical layered structures obtained in other objects.
\citet{vantHoff2023} reported the results of high-spatial resolution ($\sim24$ au) observations of molecular line emission towards the Class 0 protostar L1527 in ALMA Large Program eDisk and also discussed the presence of similar chemical layered structures between C$^{18}$O and H$_{2}$CO.
They showed that C$^{18}$O, H$_{2}$CO, and SO trace the disk and inner envelope (see also \citealt{Zhang2024} for SO in L1527), with H$_{2}$CO emission arising predominantly in the disk surface layers.
In contrast, they discussed that c-C$_{3}$H$_{2}$ is absent in the disk and its emission originates solely in the envelope (see also \citealt{vantHoff2026}), which is consistent with previous ALMA observations of different transitions \citep{Sakai2014a} and the CCH distribution \citep{Sakai2014b}.
We consider the possible reasons why CCH and c-C$_{3}$H$_{2}$ are absent in the surface layer of the disk in L1527, unlike IRAS~4C, that in L1527 (1) The vertical density gradient is steep and the PDR layer is geometrically thin, (2) As exemplified by the presence of SO emission \citep{Zhang2024}, oxygen is supplied to the gas phase, creating unfavorable conditions for formation of carbon-chain molecules \citep{Bergin2016, Anderson2025} (3) The temperature is high and CCH $+$ H$_{2}$ (with energy barrier) reaction proceeds.
\\ \\
Recently, spatially resolved observations of some Class I-II edge-on ($=$small inclination angle) disks have been obtained with CO and simple molecules (e.g., \citealt{Dutrey2017, Dutrey2025, Louvet2018, Podio2020, Teague2020, vantHoff2020}).
\citet{Podio2020} reported from their observations for the edge-on disk around the Class I source IRAS 04302+2247 (the butterfly star) by ALMA-DOT program (ALMA chemical survey of Disk-Outflow sources in Taurus, see e.g., \citealt{Garufi2020}) that they could resolve, the vertical structure of a disk with multiple molecular tracers (see also Section \ref{sec:1} and \citealt{vantHoff2020, Lin2023}). 
They showed that molecular emission is vertically stratified, with $^{12}$CO observed at larger disk heights compared to both CS and H$_{2}$CO, which is nearly co-spatial.
They found that the height of molecular emitting regions scaled by radius ($z/r$) are $\sim 0.41-0.45$ for $^{12}$CO and $\sim 0.21-0.28$ for H$_{2}$CO at $r\sim100-200$ au.
\citet{vantHoff2020} also showed the vertically stratified structures in this disk from their ALMA observations that C$^{17}$O line emission originates in the disk midplane and H$_{2}$CO line emission traces more surface layers. They reported the presence of CO freeze-out in the outermost region ($\gtrsim100$ au) of the disk midplane. 
\citet{Lin2023} also confirmed these chemical layered structures in the same disk using ALMA eDisk observations.
\\ \\
At first glance, the chemical layer structures observed in Class I/II disks and those of the Class 0 IRAS~4C disk look similar, but also have some differences as well.
First of all, the spatial extent of the IRAS~4C emission is $r\sim300$ au along the midplane of the disk and envelope structures.
The H$^{13}$CO$^{+}$ and DCO$^{+}$ line emission traces the infalling envelope structure from the outer envelope to the disk-forming region of IRAS~4C, which is clearly different from the Class I/II disks. It is difficult to clarify with only single source observations how continuous infall motion affects the formation of the chemically layered structure.
Furthermore, the radiation in the disk and envelope surface in the Class 0 sources are dominated by accretion luminosity and protostellar outflow shocks, and not a stellar photosphere like Class I/II sources.
Because the disk$+$envelope (Keplerian disk $+$ pseudo disk) in IRAS4C is less dense than Keplerian disk in Class I/II sources, UV radiation from outflow shocks is transmitted to more deeper region in the vertical direction, and the ``surface layer'' might be geometrically thicker. To clarify the difference of the temperature structures in the disk vertical direction between IRAS4C and L1527 \citep{vantHoff2023, Zhang2024}, where CCH and c-C$_{3}$H$_{2}$ are absent in the disk surface layer, it would be useful to analyze the distribution of multiple line transitions of SO and SO$_{2}$ gas, which are formed by SO and SO$_{2}$ ice sublimation where gas is heated by UV radiation, or by H$_{2}$O photodissociation \citep{vanGelder2021}.
\subsection{The Radial Variations of the Disk Ionisation Degree and Ionisation Rates}\label{sec:4-2}
Disk ionisation is thought to be a fundamental parameter in the physical and chemical evolution of protostellar envelopes and disks. The non-ideal MHD (magnetohydrodynamic) effects, which depend on ionisation degree, play a major role in angular momentum transfer and mass accretion (e.g., \citealt{Suzuki2009, Tomida2015, Tsukamoto2015, Kuffmeier2020, Nishio2025, Zhao2026}).
Recently, \citet{Kuffmeier2020} showed from their non-ideal MHD simulations of a collapsing protostar that a higher ionisation rate leads to stronger magnetic braking, and hence to the formation of smaller disks.
In particular, they speculate that the protostellar clusters in the ONC and CrA cause a higher cosmic-ray ionisation rate and thereby smaller disks in these regions than in regions such as Taurus, Ophiuchus, Chamaeleon I, or Lupus (see also \citealt{Tobin2020}). 
In addition, ionisation also triggers ion-molecule reactions which, for example, convert CO to other species, such as CO$_{2}$, CH$_{4}$, and CH$_{3}$OH (e.g., \citealt{Furuya2014, Bosman2018, Eistrup2016, Schwarz2018, Notsu2022, Sellek2025}).
There are several possible ionisation sources in disks and envelopes: X-rays and gamma-ray emission from the central star, cosmic-rays and stellar energetic particles, and the decay of short-lived radioactive nuclei (SLRs) (e.g., \citealt{Umebayashi1981, Umebayashi2009, Turner2009, Cleeves2013, Cleeves2014, Cleeves2015, Cleeves2017, Rab2017, Rab2018, Padovani2018, Padovani2020, Notsu2021, Fujii2022, Waggoner2022, Kimura2023, Long2024, Long2026, Washinoue2024, Trapman2025, Brunn2026, Desch2026, Schwarz2026}).
The typical cosmic-ray (CR) ionisation rate $\zeta_{\mathrm{CR}}$ in nearby star-forming regions is $\sim10^{-16}-10^{-17}$ s$^{-1}$ (e.g., \citealt{Redaelli2021, Redaelli2025, Sabatini2023, Pineda2024}), and it is attenuated in the denser locations such as the disk midplane \citep{Umebayashi1981, Fujii2022}, with the attenuation length of cosmic-rays much larger than for X-rays.
\\ \\
\citet{Aikawa2021} analyzed HCO$^{+}$, H$^{13}$CO$^{+}$, N$_{2}$H$^{+}$, and N$_{2}$D$^{+}$, line emission from five Class II disks obtained by the ALMA Large Program ``Molecules with ALMA at Planet-forming Scales (MAPS)'' (e.g., \citealt{Oberg2021MAPS}). They reported that the ionisation degree is low in the inner dense gas, especially within the CO snowline, and the midplane ionisation rate $\zeta_{\mathrm{CR}}$ is $\gtrsim10^{-18}$ s$^{-1}$ for the disks around the Class II objects IM Lup, AS 209, and HD 163296.
\citet{Seifert2021} reported from independent ALMA observations of H$^{13}$CO$^{+}$ and N$_{2}$H$^{+}$ line emission for the IM Lup disk that the distribution of line emission is best reproduced using a radially increasing cosmic-ray gradient, with low cosmic-ray ionisation in the inner disk ($\zeta_{\mathrm{CR}}\lesssim10^{-20}$ s$^{-1}$), high cosmic-ray ionisation in the outer disk ($\zeta_{\mathrm{CR}}\gtrsim10^{-17}$ s$^{-1}$), and a transition at $\sim80-100$ au.
In addition, \citet{Long2024} reported from their ALMA observations of HCO$^{+}$, H$^{13}$CO$^{+}$, and N$_{2}$H$^{+}$ line emission for the Class II DM Tau disk that the cosmic-ray ionisation rate is modestly reduced ($\zeta_{\mathrm{CR}}\gtrsim10^{-18}$ s$^{-1}$) and there is a hard X-ray spectrum produced by flares.
They also discussed that their best-fit model under-produced emission in the inner disk, suggesting that there may be an additional mechanism enhancing ionisation in DM Tau's inner disk, such as stellar energetic particles (e.g., \citealt{Turner2009, Rab2017, Brunn2026}).
\cite{Trapman2025} estimated the radially averaged disk midplane cosmic-ray ionisation rates for 10 disks in ALMA Large Program exoALMA samples (e.g., \citealt{Teague2025}), by assuming the kinematically derived disk gas mass and refitting the N$_{2}$H$^{+}$ and CO line emission spectra. They 
reported a wide range of midplane ionisation rates, from $\zeta_{\mathrm{CR}}\sim10^{-17}$ s$^{-1}$ for AS 209 and RXJ 1852.3-3700 to $\zeta_{\mathrm{CR}}\sim3\times10^{-20}$ s$^{-1}$ for Elias 2-27. They also reported the median cosmic-ray ionisation rate of $1.5\times10^{-18}$ s$^{-1}$, although the uncertainty of ionisation rates for individual sources is large, typically $\sim$1.5 dex, suggesting that N$_{2}$H$^{+}$ alone only provides a moderate constraint on the disk ionisation rate.
\\ \\
\citet{vantHoff2022} discussed from the HCO$^{+}$ and H$^{13}$CO$^{+}$ line emission obtained by ALMA that the cosmic-ray ionisation rate has a canonical value of $\sim10^{-17}$ s$^{-1}$ in the inner envelope and may be attenuated to $\sim10^{-18}$ s$^{-1}$  in the disk around L1527.
\citet{Ceccarelli2014} reported the enhanced ionisation rate of $>10^{-14}$ s$^{-1}$ for the whole envelope and $\sim10^{-12}$ s$^{-1}$ for the innermost envelope around 
the cluster of a few intermediate and low-mass protostars in the source OMC-2 FIR 4 with a total mass of 30$M_{\odot}$ (e.g., \citealt{Shimajiri2008, Lopez2013, Tobin2019})
from their $Herschel$/HIFI observations of HCO$^{+}$ and N$_{2}$H$^{+}$, suggesting the presence of stellar energetic particles whose energies are $>10$ MeV.
Recently, \citet{Cabedo2023} reported the first resolved map of the cosmic-ray ionisation rate at $10^{2}-10^{3}$ au scales in a solar-type Class 0 protostar, and found an enhanced ionisation that increases at small envelope radii, reaching values up to the cosmic-ray ionisation rate of $\sim10^{-14}$ s$^{-1}$ at a few hundred astronomical units (au) from the central Class 0 protostellar object B335.
They showed that the extremely high cosmic-ray ionisation rate can be attributed to the presence of cosmic-rays accelerated close to the protostar (e.g., \citealt{Turner2009, Rab2017, Brunn2026}). 
It is therefore important to resolve the radial distribution of the ionisation degree and ionisation rate from the envelope to inner disk scales ($<100$ au) in Class 0/I objects.
\\ \\
In this study, we estimate the distribution of ionisation degree and ionisation rates around IRAS~4C from the observed C$^{18}$O, H$^{13}$CO$^{+}$, and DCO$^{+}$ line emission.
Figure \ref{Figure8} (a) shows the distribution of the column density ratio $R_{\mathrm{H}}=N_{\mathrm{HCO}^{\mathrm{+}}}/N_{\mathrm{CO}}$ (CO ionisation degree).
To estimate $N_{\mathrm{HCO}^{\mathrm{+}}}$ from $N_{\mathrm{H}^{13}\mathrm{CO}^{\mathrm{+}}}$ and $N_{\mathrm{CO}}$ from $N_{\mathrm{C}^{18}\mathrm{O}}$, we assume the $^{12}$C/$^{13}$C ratio as 69 and the $^{16}$O/$^{18}$O ratio as 560 \citep{Wilson1999, Jorgensen2010}.
We note that the carbon isotope fractionation in cold envelope with higher ionisation rate ($\gtrsim5\times10^{-17}$ s$^{-1}$) would decrease $^{12}$C/$^{13}$C ratios in CO and HCO$^{+}$ by a factor of $2-3$ (e.g., \citealt{Bergin2024}). Thus, the values of $N_{\mathrm{H}\mathrm{CO}^{\mathrm{+}}}$ might be overestimated by a factor of $2-3$.
%
In Figure \ref{Figure8} (a), $R_{\mathrm{H}}$ is 
$\sim10^{-5}-10^{-4}$ at $R\sim$0.2$''$-0.7$''$ ($R\sim$50-200 au) and $(1-2)\times10^{-4}$ at $R\gtrsim$0.7'' ($R\gtrsim$200 au) from the central protostar, respectively (see also Figure \ref{Figure9}).
\citet{Aikawa2021} estimated that $R_{\mathrm{H}}$ is $\sim10^{-5}-10^{-4}$ at $R\gtrsim100$ au, and $\lesssim10^{-5}$ at $R\lesssim100$ au in Class II disks.
Thus, both Class II disks and the IRAS~4C disk/envelope system have the radial gradients of $R_{\mathrm{H}}$, although the values in the same radii are different.
\\ \\
We note that according to Figure \ref{FigureA1_tau} in Appendix \ref{Appendix-A}, the H$^{13}$CO$^{+}$ line emission is marginally optically thick ($\tau\sim1-1.5$) at $r\sim100-200$ au, and C$^{18}$O is optically thick $\tau\gtrsim2$ in $r\lesssim150$ au. Thus, $R_{\mathrm{H}}$ might be overestimated by a factor of around 2-3 at $r\lesssim150$ au since the column densities of C$^{18}$O might be underestimated compared to those of H$^{13}$CO$^{+}$.
\\ \\
Based on \citet{vantHoff2018}, by assuming HCO$^{+}$ is formed by the reaction between CO and H$_{3}^{+}$ and $N_{e^{-}}=N_{\mathrm{HCO}^{+}}$, the ionisation rate $\zeta_{\mathrm{CR}}$ is estimated by the following equation,
\begin{equation}
\zeta_{\mathrm{CR}}= \frac{N_{\mathrm{HCO}^{+}}^{2}}{N_{\mathrm{H_{2}}} L} k_{1},
\end{equation}
where $L$ is the depth of the structure, $k_{1}$ is the reaction rate of HCO$^{+}$ and $e^{-}$ and $k_{1}=2.40\times10^{-7}\times(T_{\mathrm{ex}}/300)^{-0.69}$ cm$^{3}$ s$^{-1}$ ($=8.3\times10^{-7}$ cm$^{3}$ s$^{-1}$ at $T_{\mathrm{ex}}=50$ K, \citealt{Brian1990}).
We note that this assumption of the relation between HCO$^{+}$ and CO is basically reasonable outside the water snowline, whereas HCO$^{+}$ molecules are significantly destroyed by water gas within the water snowline \citep{vantHoff2018, Notsu2021}. Thus, within the water snowline, $\zeta_{\mathrm{CR}}$ could be underestimated.
We assumed that the size of the NGC 1333 cloud and the structure of IRAS~4C are the same, and assumed $L=6\times10^{4}$ au based on the assumption by \citet{Pineda2024} for the NGC 1333 cloud. 
Here we note that if we assume a much smaller value of $L$ (=size of the structure), the ionisation rates increase in inverse proportion to $L$ (see below).
Figure \ref{Figure8} (c) shows the distribution of $\zeta_{\mathrm{CR}}$ estimated from $N_{\mathrm{H}^{13}\mathrm{CO}^{\mathrm{+}}}$ and $N_{\mathrm{H}_{2}}$.
In addition, Figure \ref{Figure8} (d) shows the distribution of $\zeta_{\mathrm{CR}}$ estimated from $N_{\mathrm{H}^{13}\mathrm{CO}^{\mathrm{+}}}$ and $N_{\mathrm{C}^{18}\mathrm{O}}$, by assuming the $^{16}$O/$^{18}$O ratio as 560 and $N_{\mathrm{CO}}$/$N_{\mathrm{H}_{2}}$$=10^{-4}$.
In Figures \ref{Figure8} (c) and (d), we mask the innermost region with significantly higher dust optical depth by the black-filled ellipse, whose size and shape are the same as the beam of the dust continuum emission.
We assumed that CO depletion has not efficiently happened in the Class 0 object IRAS~4C, based on previous observations (e.g., \citealt{Zhang2020}).
Although the actual values of $\zeta_{\mathrm{CR}}$ are different by a factor of 2-3 between these two maps,
$\zeta_{\mathrm{CR}}$ is larger in the outer region and smaller in the inner region in both panels.
\\ \\
As described in Section \ref{sec:3-1}, the dust optical depth is larger in the inner region of the disk/envelope system, especially within 50 au of the protostar.
We note that the value ``50 au'' has an uncertainty, since it is similar to the beam size of our observations (see Table \ref{Table:2}).
Considering the high dust optical depth at $R<50$ au and beam sizes of our observations ($\sim0.''2$, see Table \ref{Table:2}), in this innermost region at $R<50$ au, 
it is not appropriate to discuss the ionisation rates by comparing the distribution of C$^{18}$O, H$^{13}$CO$^{+}$, and DCO$^{+}$ line emission (see Figure \ref{Figure4}).
Thus, in Figures \ref{Figure8} and \ref{Figure9} we consider only $R>50$ au.
Figure \ref{Figure9} shows the radial profile of
$R_{\mathrm{H}}$=$N_{\mathrm{HCO}^{+}}$/$N_{\mathrm{CO}}$ and the ionisation rate estimated from the H$^{13}$CO$^{+}$ and H$_{2}$ with $L=6\times10^{4}$ au along the radial (disk major axis) direction.
We note that in deriving the CO ionisation degree, we assume that both C$^{18}$O and H$^{13}$CO$^{+}$ line emission trace the same region along the line of sight.
Focusing on the region with $R\sim50-200$ au (see Figures \ref{Figure4} and \ref{Figure9}), we can see that the relationship between the line emission intensities of the C$^{18}$O line and ion molecules (H$^{13}$CO$^{+}$ and DCO$^{+}$) is different from that at $R>200$ au, and thus we can also see that the $R_{\mathrm{H}}$ is changing around $R\sim200$ au ( 
$R_{\mathrm{H}}\sim10^{-5}-10^{-4}$ at $50<R<200$ au and $R_{\mathrm{H}}\sim(1-2)\times10^{-4}$ at $R>200$ au, see Figures \ref{Figure8} (a) and \ref{Figure9}).
We consider that the gas densities traced by the CO emission line increase within $R\sim200$ au. 
It might suggest that the disk within $\sim200$ au is Keplerian disk whereas the disk outside $\sim200$ au is pseudo disk where infall motion is dominant.
In Figure \ref{Figure6}, the C$^{18}$O line emission on the PV diagram increases at $\lesssim0.7''$ ($\lesssim200$ au), suggesting a significant increases in gas density due to the presence of a Keplerian disk. However, further detailed analyses with infalling envelope models (e.g., \citealt{Oya2022}) are required to constrain the Keplerian disk radius more precisely.
\\ \\
In Figures \ref{Figure8} and \ref{Figure9}, the absolute values of ionisation rates $\zeta_{\mathrm{CR}}$ in the IRAS~4C disk/envelope system are estimated to be $\sim10^{-17}-10^{-16}$ s$^{-1}$ at $R>100$ au and $\lesssim10^{-18}$ s$^{-1}$ at $50<R<100$ au.
\citet{Pineda2024} reported the maps of ionisation rates in the larger scale of NGC~1333 star-forming region, and the ionisation rates of the clouds around IRAS~4C ($\sim$ a few 10$^{3}$ au) is $\sim10^{-16}$ s$^{-1}$, which is consistent with the value for the outer envelope of IRAS~4C in Figure \ref{Figure8}.
We have confirmed that the absolute value of ionisation rates for IRAS~4C varies within a factor of $\sim$2 by changing $T_{\mathrm{ex}}$ by 20K and 100K (see Figures \ref{FigureB1} and \ref{FigureB2} in Appendix \ref{Appendix-B}).
We consider that the disk formation may have increased the local gas density and decreased the ionisation rate in the inner dense midplane regions of the disk/envelope system.
In addition, apart from B335 \citep{Cabedo2023}, the additional cosmic-rays acceleration close to the protostar would not be effective for the disk ionisation at the regions with $\gtrsim50-100$ au of IRAS~4C.
\\ \\ 
However, for the estimated values of ionisation rates in IRAS~4C, it has an uncertainty up to orders of magnitude based on the assumptions used to estimate the ionisation fraction \citep{Sabatini2023}. 
The assumption of $N_{e^{-}}=N_{\mathrm{HCO}^{+}}$ and the value of $L$ we adopted also brings uncertainty.
If we assume a much smaller value of $L$ (=depth of the structure), the ionisation rates increase in inverse proportion to $L$.
We do not know the exact value of the depth of IRAS~4C structure, but if we assume $L$ as around a few $\times10^{3}$ au, the ionisation rates increase by a factor of a few $\times10$.
Figure \ref{FigureB3} in Appendix \ref{Appendix-B} shows the distributions of the ionisation rate by assuming $L=1000$ au, which is the size of relatively smaller protostellar envelopes (such as in B335, \citealt{Cabedo2023}).
In Figure \ref{FigureB3}, the values of ionisation rates are estimated to be $5\times10^{-16}-10^{-14}$ s$^{-1}$ at $R>100$ au and $\lesssim10^{-16}$ s$^{-1}$ at $50<R<100$ au.
\\ \\
Recent studies (e.g., \citealt{Bovino2020, Sabatini2020, Sabatini2023, Redaelli2024, Redaelli2025, Socci2024, Latrille2025}) proposed a new analytical approach to estimate the ionisation rates by using H$_{2}$D$^{+}$ lines as the main observational constraint to derive the amount of H$_{3}^{+}$.
Future comparison of the line emission of multiple ionisation tracers (such as $\mathrm{HCO}^{+}$, H$^{13}$CO$^{+}$, N$_{2}$H$^{+}$, H$_{2}$D$^{+}$, and N$_{2}$D$^{+}$) with more detailed disk models including chemical reaction networks and temperature structures will help to clarify the distribution of ionisation rates in the IRAS~4C disk-envelope system (see e.g., \citealt{Luo2024}).
\\ \\
The top right panel in Figure \ref{Figure8} shows the deuteration fraction $R_{\mathrm{D}}$=$N_{\mathrm{DCO}^{\mathrm{+}}}$/$N_{\mathrm{HCO}^{\mathrm{+}}}$.
According to this Figure, $R_{\mathrm{D}}$ is $\sim0.4-0.5\%$ at $R\gtrsim0.5$'' ($R\gtrsim150$ au) and $\lesssim0.1\%$ at $R\lesssim0.5$'' ($R\lesssim150$ au).
These values of $R_{\mathrm{D}}$ are around one order of magnitude smaller than those at the $\sim1000$ au scale envelope around B335 \citep{Cabedo2023} and prestellar cores \citep{Caselli1998}.
%
%
We try to estimate $\zeta_{\mathrm{CR}}$ from DCO$^{+}$, C$^{18}$O, and H$^{13}$CO$^{+}$ line emission based on the method of \citet{Caselli1998} and \citet{Cabedo2023}, but the estimated value of $\zeta_{\mathrm{CR}}$ is unreasonably high ($\sim10^{-12}-10^{-11}$ s$^{-1}$).
Thus, apart from the lower density molecular clouds and prestellar cores \citep{Caselli1998, Fontani2017}, $R_{\mathrm{D}}$=$N_{\mathrm{DCO}^{\mathrm{+}}}$/$N_{\mathrm{HCO}^{\mathrm{+}}}$ would not be an appropriate probe of the ionisation degree in higher density disks and envelopes around protostars.
The observations of other deuterated molecules (such as H$_{2}$D$^{+}$, D$_{2}$H$^{+}$, N$_{2}$D$^{+}$, and DCN) would be helpful to estimate the ionisation rates correctly (e.g., \citealt{Caselli2012, Favre2017, Bovino2020, Sabatini2023, Socci2024, Redaelli2024, Redaelli2025}), although the deuteration chemistry of H$_{3}$$^{+}$ (especially H$_{2}$D$^{+}$/ D$_{2}$H$^{+}$ ratio) is not fully understood \citep{Pagani2024}.
We note that \citet{Koumpia2016} reported from their single-dish JCMT\footnote{James Clerk Maxwell Telescope} observations that the deuteration fraction estimated from DCO$^{+}$ and HCO$^{+}$ emission lines is $\sim12\pm2\%$ toward IRAS 4A, $\sim3.5\pm1\%$ toward IRAS 4B, and an upper limit of $\sim20\%$ toward IRAS~4C.
They also pointed out that the outer envelope of IRAS~4C should show the higher degree of deuteration since it is the coldest among the sources and thus the enhancement of H$_{2}$D$^{+}$ abundance contributes to the enhancement of other deuterated molecules.
\\ \\
\section{Conclusion}\label{sec:5}
We report the results of molecular line observations of the disk and envelope (Keplerian disk $+$ pseudo disk) around the Class 0 low-mass protostar NGC 1333 IRAS~4C in the Perseus Molecular Cloud, as part of the ALMA Large Program FAUST.
Previous observations (ALMA/PEACHES) have shown that 
the disk rotation axis of this source is largely tilted to the line of sight,
making it possible for studying the molecular abundance distribution along the disk height.
We found that the molecules generated by photochemical reactions such as CCH and c-C$_{3}$H$_{2}$ are distributed in the low-density upper layers of the disk/envelope system, while the molecules such as C$^{18}$O and H$_{2}$CO are distributed in the high density regions around the protostar, indicating a chemically layered structure in the disk vertical direction. 
The chemical layered structure is similar to those known for Class I/II objects, but also had some differences. 
In addition, we compared the differences of chemical layered structure between IRAS~4C and another Class 0 object L1527, and the difference may suggest that compared with L1527, the surface layer of the disk in IRAS~4C are more active conditions for formation of carbon-chain molecules such as CCH and c-C$_{3}$H$_{2}$ and/or the PDR layer is not geometrically thin.
Moreover, the distribution of ionisation degree and ionisation rate in the radial direction of the disk/envelope system is estimated from the comparison of H$^{13}$CO$^{+}$ and C$^{18}$O line emission.
The $R_{\mathrm{H}} (=N_{\mathrm{HCO}^{\mathrm{+}}}/N_{\mathrm{CO}}$) is 
$\sim10^{-5}-10^{-4}$ at $R\sim$0.2''-0.7'' ($R\sim$50-200 au) and $(1-2)\times10^{-4}$ at $r\gtrsim$0.7'' ($\gtrsim$200 au) from the central protostar, respectively.
The estimated values of the ionisation rate $\zeta_{\mathrm{CR}}$ are $\sim10^{-17}-10^{-16}$ s$^{-1}$ at $R>100$ au, and $\lesssim10^{-18}$ s$^{-1}$ at $50<R<100$ au, although the absolute values of ionisation rates have uncertainty (up to orders of magnitude) based on the assumptions invoked to estimate the ionisation fraction and the depth of the structure.
Disk formation may have increased the local gas density and decreased the ionisation degree and the ionisation rate in the inner dense midplane regions of the disk/envelope system.
\\ \\
\clearpage
This paper makes use of the following ALMA data set: ADS/JAO.ALMA\# 2018.1.01205.L (FAUST, PI: Satoshi Yamamoto).
ALMA is a partnership of the ESO (representing its member states), the NSF (USA) and NINS (Japan), together with the NRC (Canada) and the NSC and ASIAA (Taiwan), in cooperation with the Republic of Chile. The Joint ALMA Observatory is operated by the ESO, the AUI/NRAO, and the NAOJ. The National Radio Astronomy Observatory is a facility of the National Science Foundation operated under cooperative agreement by Associated Universities, Inc. The authors thank the ALMA staffs for their excellent support.
S.N.~is grateful for support from RIKEN Special Postdoctoral Researcher Program (Fellowships), 
Start-up Research Grant for The University of Tokyo Excellent Young Researcher 2024, 
Grants-in-Aid for JSPS (Japan Society for the Promotion of Science) Fellows Grant Number JP23KJ0329, and MEXT/JSPS Grants-in-Aid for Scientific Research (KAKENHI) Grant Numbers JP20K22376, JP23K13155, JP24K00674, and JP26K00755.
N.S. is supported by MEXT/JSPS Grants-in-Aid for Scientific Research (KAKENHI) Grant Numbers JP20H05844, JP20H05845, JP20H00182, and JP25H00676, and a pioneering project in RIKEN (Evolution of Matter in the Universe). 
Y.A. is supported by MEXT/JSPS Grants-in-Aid for Scientific Research (KAKENHI) Grant Number JP20H05847 and JP24K00674.
Y.Z. acknowledges the support from the Yangyang Development Fund.
H.B.L. is supported by the National Science and Technology Council (NSTC) of Taiwan (Grant Nos. 111-2112-M-110-022-MY3, 113-2112-M-110-022-MY3).
C.C., L.P., and E.B acknowledge financial support under the National Recovery and Resilience Plan (NRRP), Mission 4, Component 2, Investment 1.1, Call for tender No. 104 published on 2.2.2022 by the Italian Ministry of University and Research (MUR), funded by the European Union - NextGenerationEU-Project Title 2022JC2Y93 Chemical Origins: linking the fossil composition of the Solar System with the chemistry of protoplanetary disks - CUP J53D23001600006 - Grant Assignment Decree No. 962 adopted on 30.06.2023 by the Italian Ministry of University and Research (MUR). 
E.B. acknowledges financial support from the Italian Ministry of University and Research (MUR) under the Italian Science Fund (FIS 2 Call - Ministerial Decree No. 1236 of 1 August 2023).
\\ \\
Cl.Co., L.P., E.B., N.B., and G.S. also acknowledge the PRIN-MUR 2020 BEYOND-2p (Astrochemistry beyond the second period elements, Prot. 2020AFB3FX), the project ASI-Astrobiologia 2023 MIGLIORA (Modeling Chemical Complexity, F83C23000800005), the INAF-GO 2023 fundings PROTO-SKA (Exploiting ALMA data to study planet forming disks: preparing the advent of SKA, C13C23000770005), the INAF-MiniGrant 2022 ``Chemical Origins'' (PI: L. Podio), and the INAF-MiniGrant 2023 TRIESTE (``TRacing the chemIcal hEritage of our originS: from proTostars to planEts''; PI: G. Sabatini).
S.B.C. was supported by the NASA Planetary Science Division Internal Scientist Funding Program through the Fundamental Laboratory Research work package (FLaRe). 
ALS, CeCe, and CV acknowledge support by the Thematic Action ``Physique et Chimie du Milieu Interstellaire'' (PCMI) of INSU Programme National ``Astro'', with contributions from CNRS Physique \& CNRS Chimie, CEA, and CNES.
N.C. was supported by the funding from the European Research Council (ERC) under the 
European Union Horizon Europe programme (grant agreement No. 101042275, project Stellar-MADE).
D.J.\ is supported by NRC Canada and by an NSERC Discovery Grant.
S.V and M.B. acknowledge the support from the European Research Council (ERC) Advanced Grant MOPPEX 833460.
F.Me. acknowledges funding from the European Research Council (ERC) under the European Union's Horizon Europe research and innovation program (grant agreement No. 101053020, project Dust2Planets).
L.L. acknowledges the support of DGAPA/UNAM PAPIIT grants IN108324 and IN112820, CONAHCyT grant CF-263356 and
SECIHTI grant CBF-2025-I-109.
%
\appendix
\restartappendixnumbering
\section{Distribution of line optical depth}\label{Appendix-A}
\begin{figure*}[htbp]
\begin{center}
\vspace{1cm}
%
\includegraphics[scale=0.35]{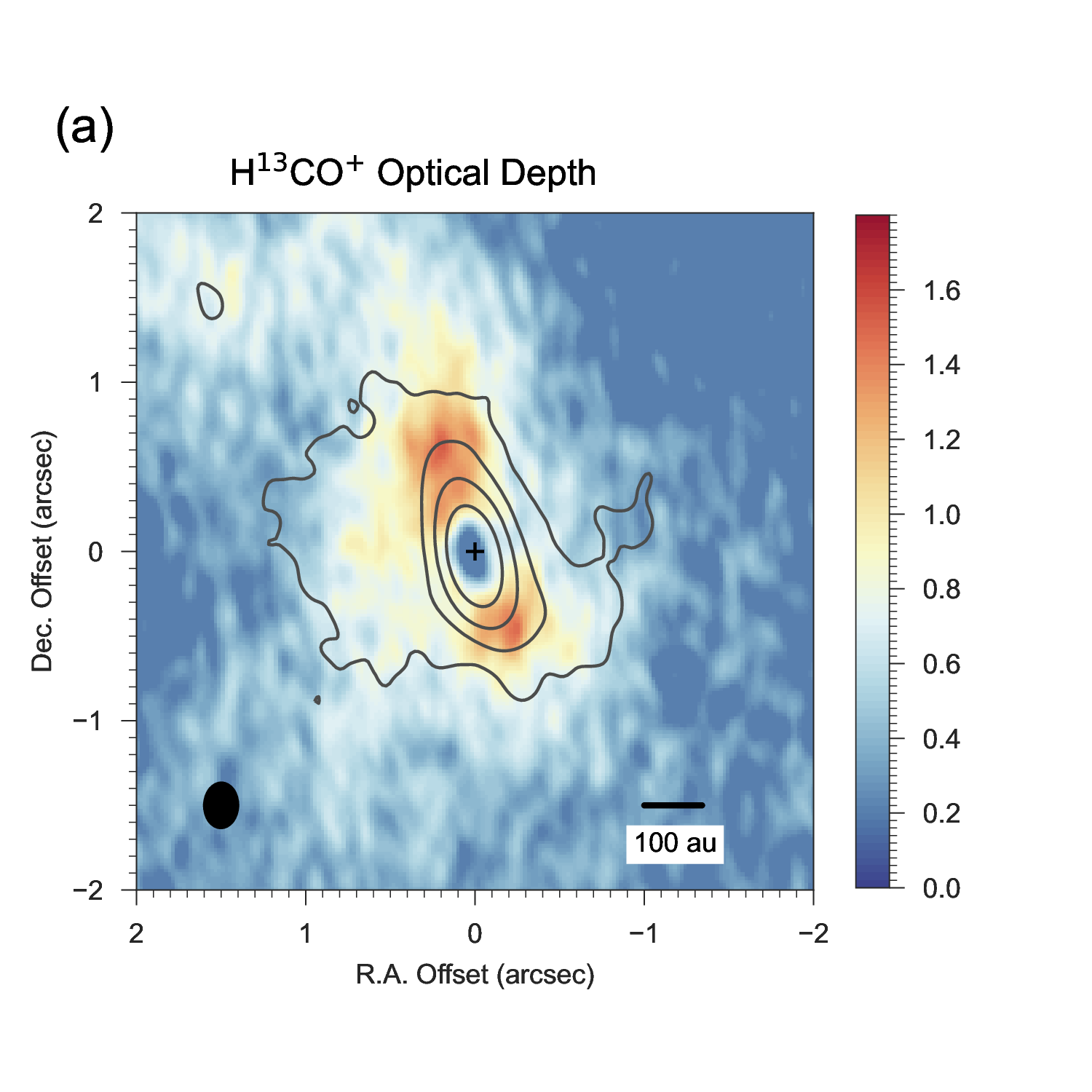}
\includegraphics[scale=0.35]{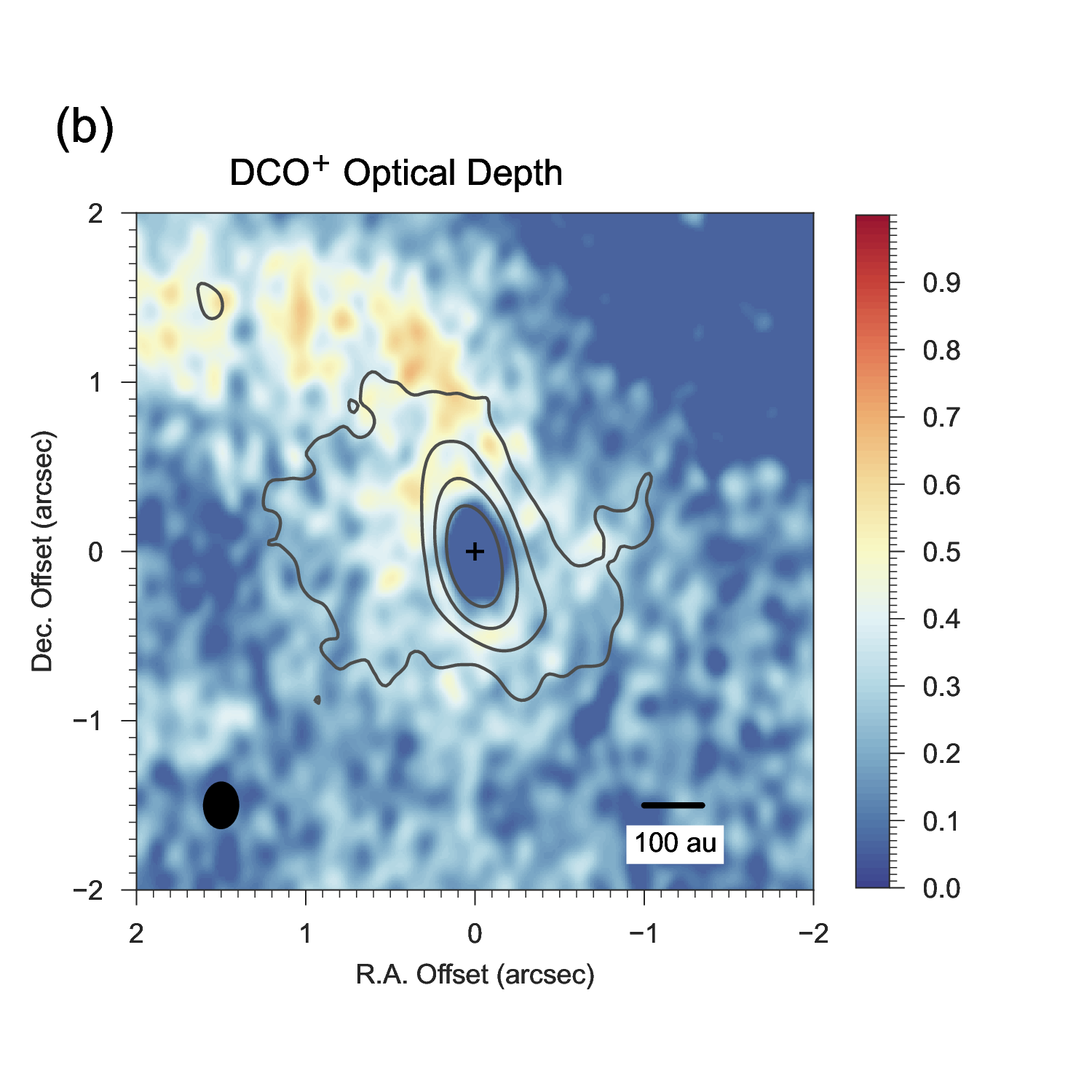}
\includegraphics[scale=0.35]{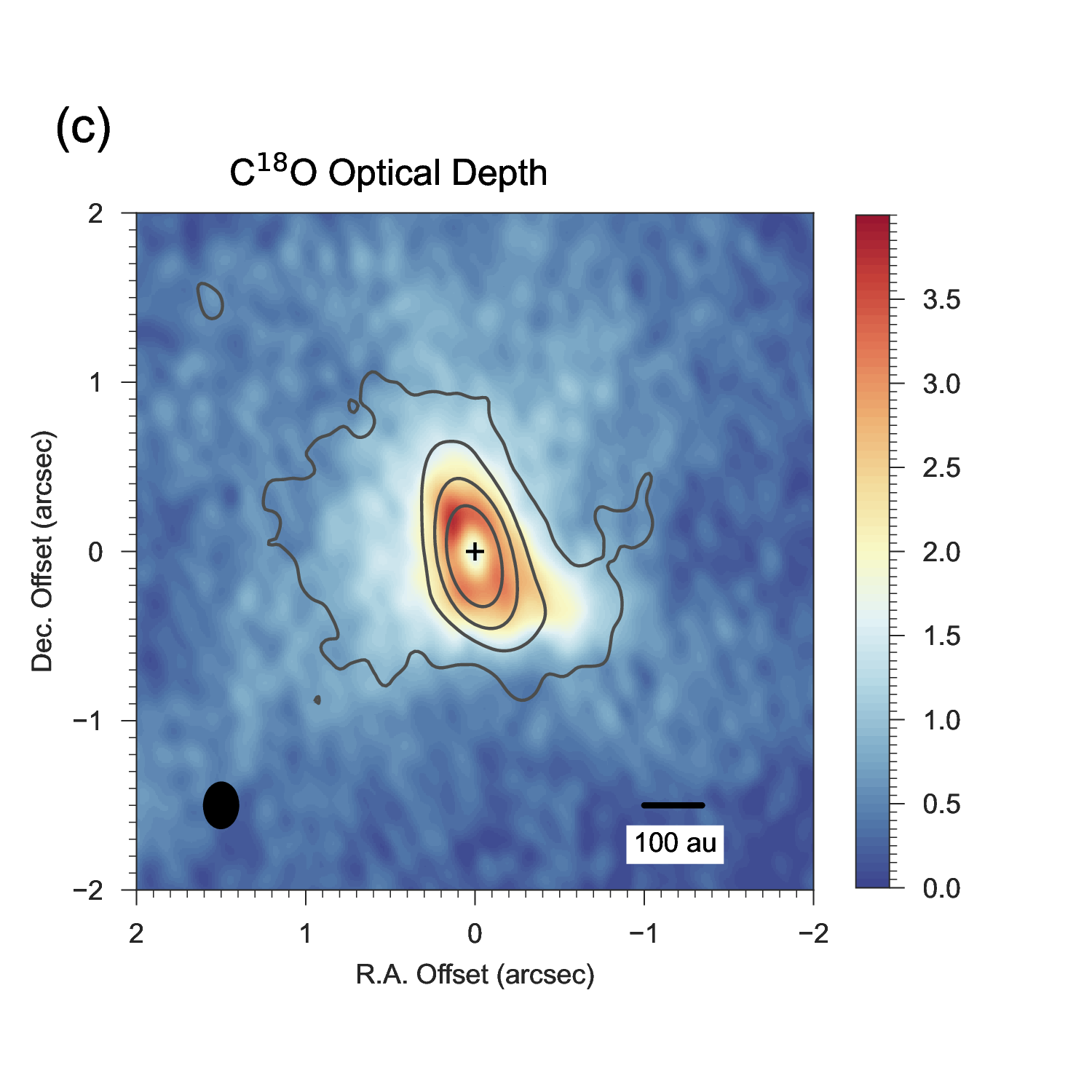}

\end{center}
\caption{
Distribution of line optical depth $\tau$ for H$^{13}$CO$^{+}$ (a), DCO$^{+}$ (b), and C$^{18}$O (c).
The 1.3 mm continuum image is overploted by black contours (see also Figure \ref{Figure1}).
For the continuum emission, the rms noise level ($\sigma$) of the continuum is 20 $\mu$Jy beam$^{-1}$, and the contour levels are 5$\sigma$, 20$\sigma$, 80$\sigma$ and 320$\sigma$.
In calculating the line optical depth, we use the values of column densities shown in Figure \ref{Figure7}, and we assume $T_{\mathrm{ex}}=50$ K and line velocity width $\Delta V=0.17$ km s$^{-1}$ throughout the disk/envelope system (see details in Section \ref{sec:3-4}).
\vspace{0.2cm}
}\label{FigureA1_tau}
\end{figure*}

Figure \ref{FigureA1_tau} shows the distribution of line optical depth $\tau$ for H$^{13}$CO$^{+}$ (a), DCO$^{+}$ (b), and C$^{18}$O (c). 
Our adopted method for calculating line optical depth is described in Section \ref{sec:3-4}.
\restartappendixnumbering
\section{The uncertainty of our derived values of ionisation rates}\label{Appendix-B}
\begin{figure*}[htbp]
\begin{center}
\includegraphics[scale=0.35]{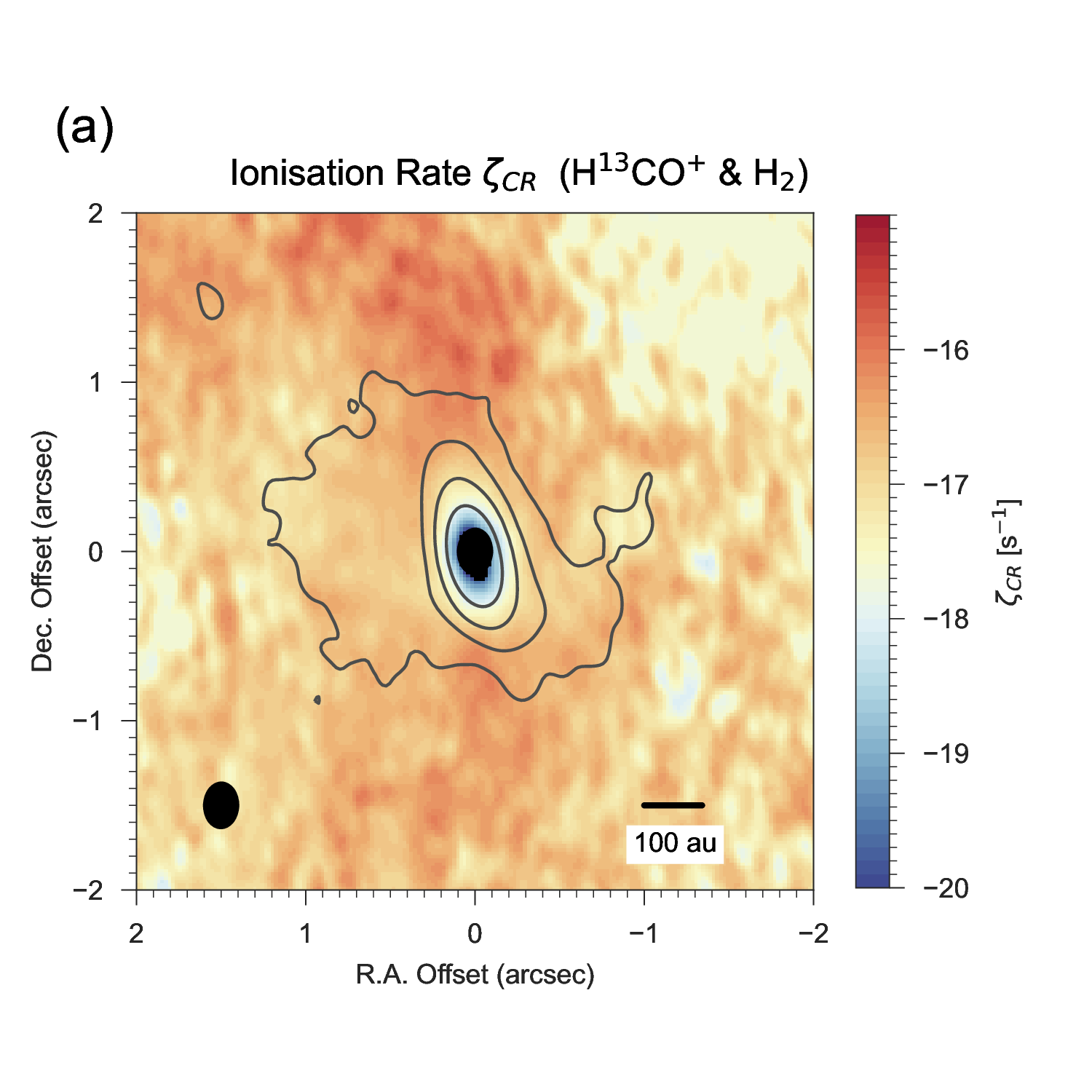}
\includegraphics[scale=0.35]{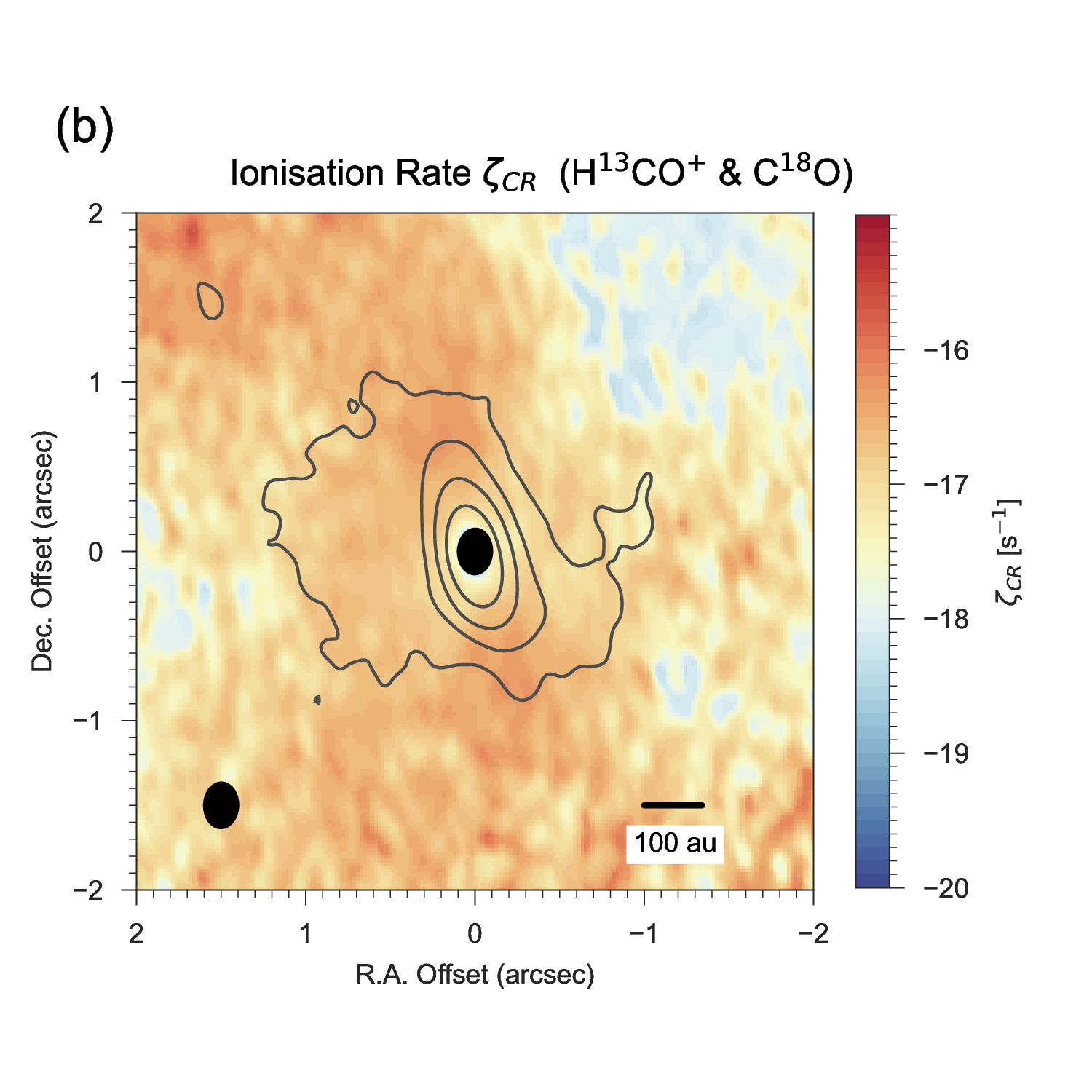}
\end{center}
\hspace{1cm}
\caption{
Same as bottom left and right panels of Figure \ref{Figure8}, except assuming $T_{\mathrm{ex}}=20$ K.
\vspace{0.2cm}
}\label{FigureB1}
\end{figure*}

\begin{figure*}[htbp]
\begin{center}
\includegraphics[scale=0.35]{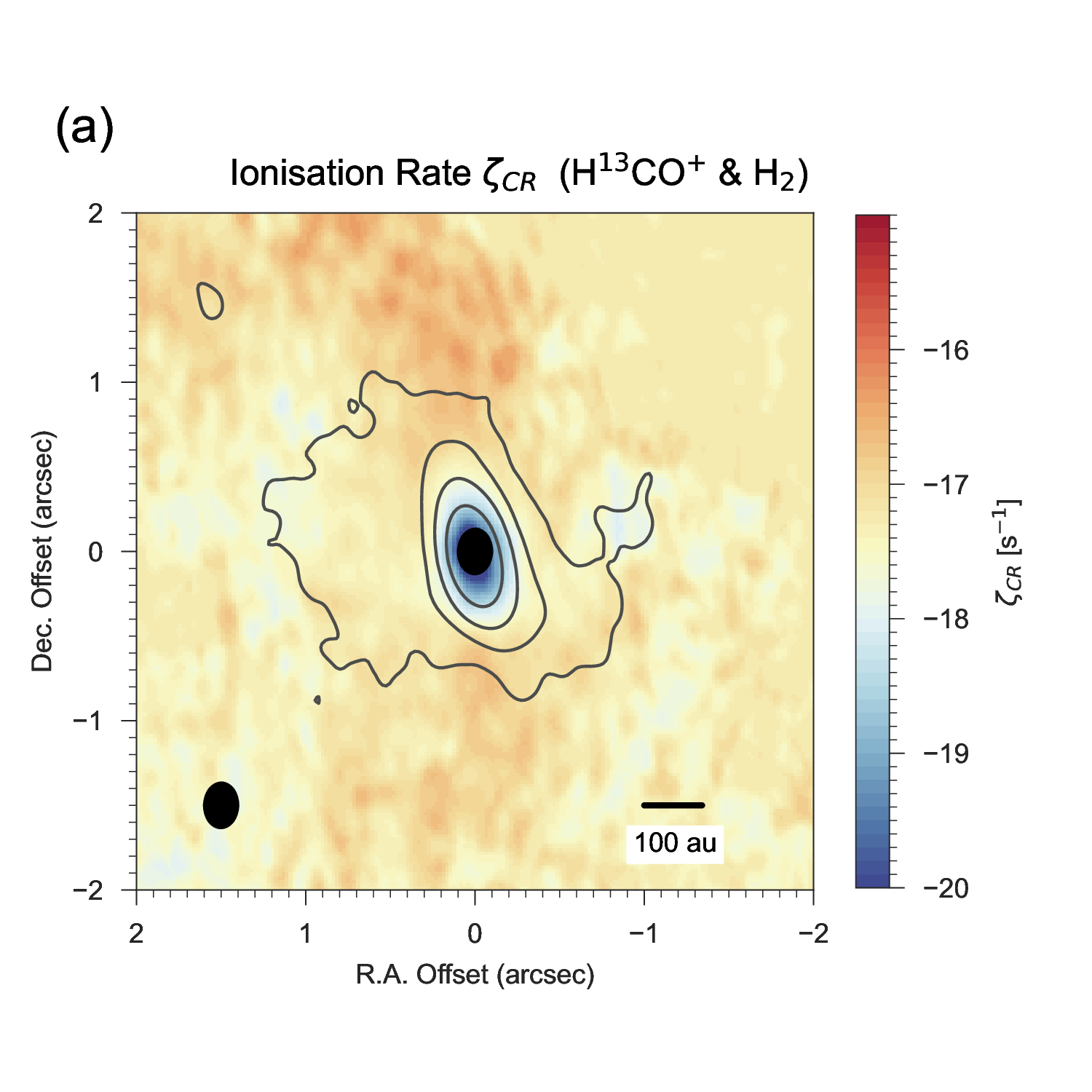}
\includegraphics[scale=0.35]{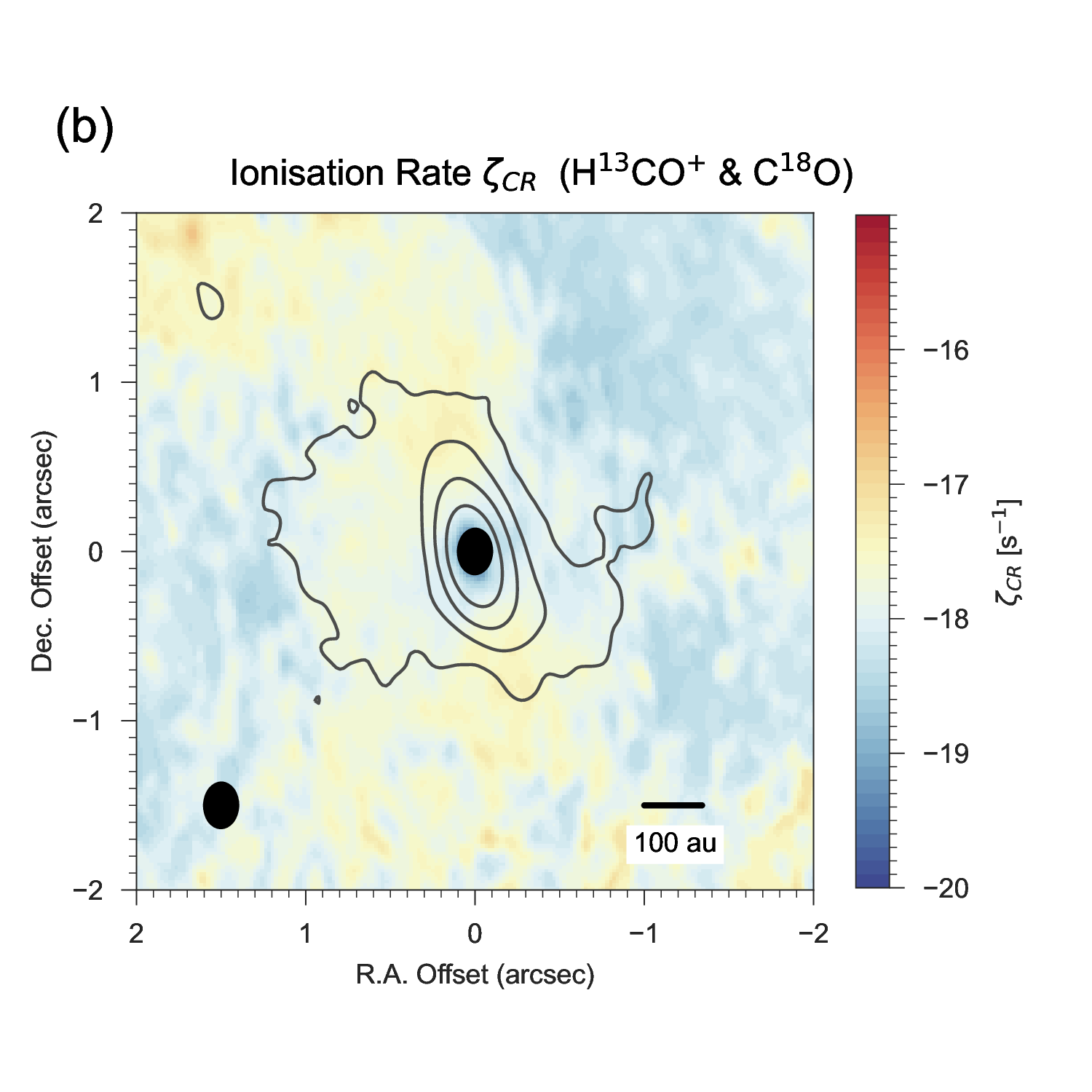}
\end{center}
\caption{
Same as bottom left and right panels of Figure \ref{Figure8}, except assuming $T_{\mathrm{ex}}=100$ K.
\vspace{0.2cm}
}\label{FigureB2}
\end{figure*}

\begin{figure*}[htbp]
\begin{center}
%
\includegraphics[scale=0.35]{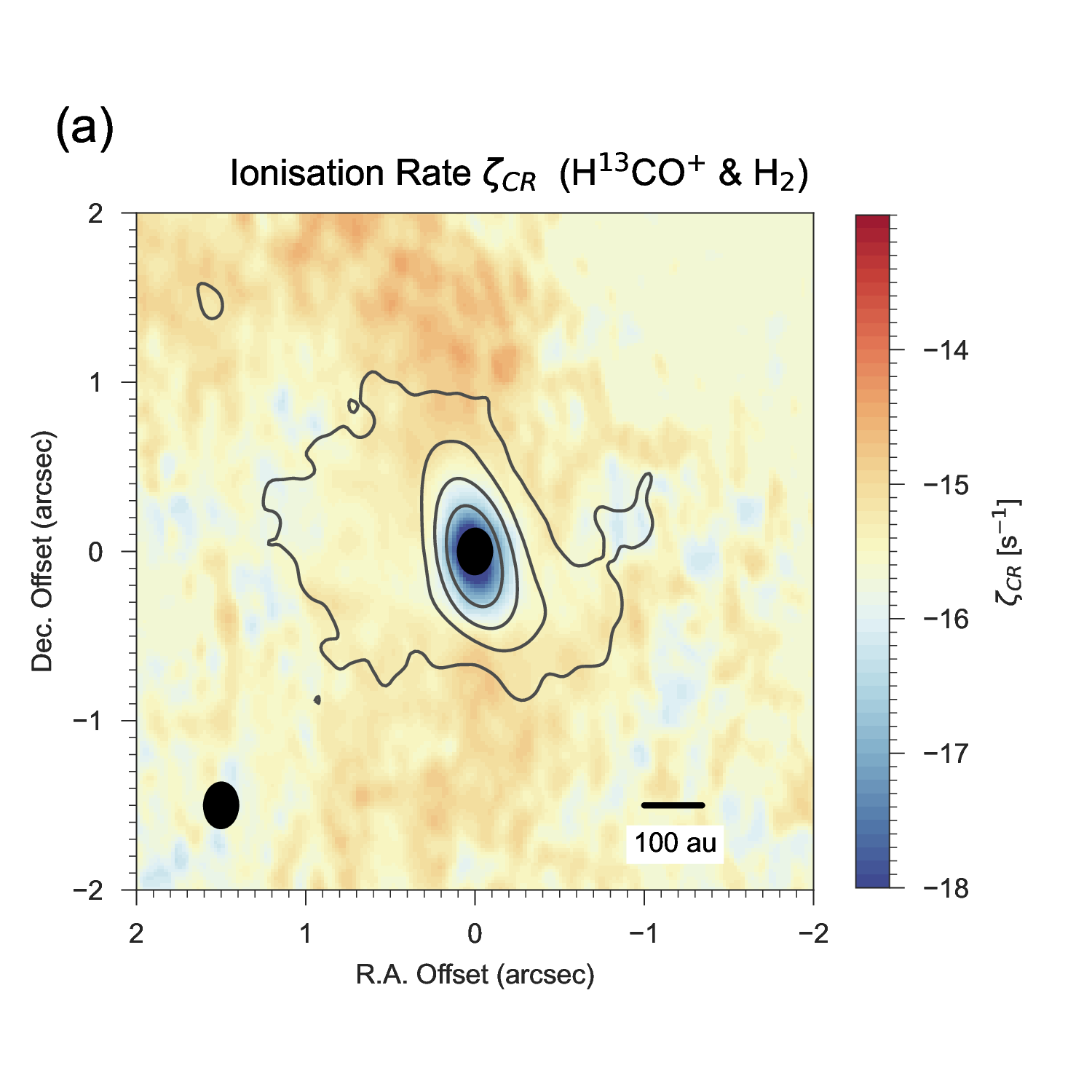}
\includegraphics[scale=0.35]{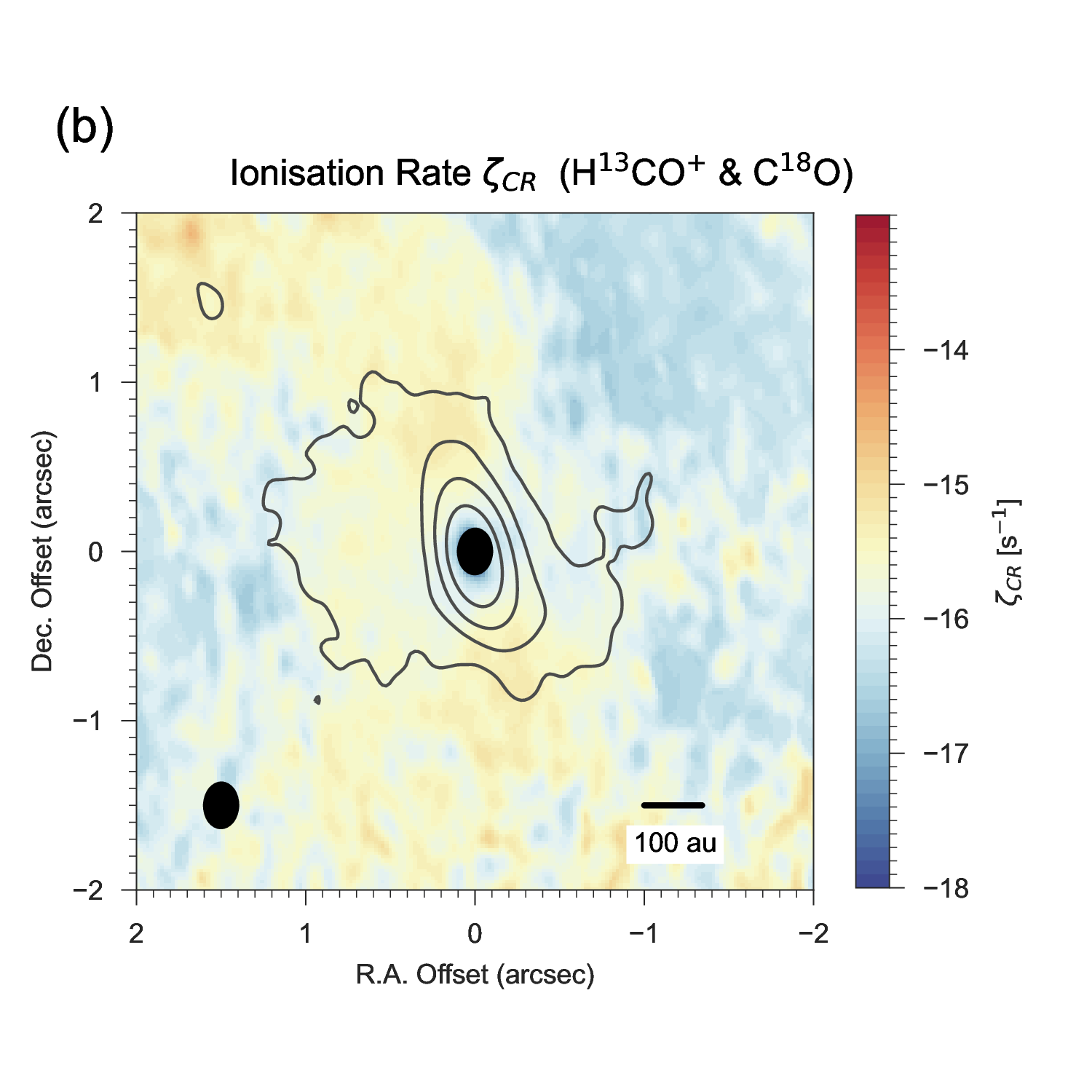}
\end{center}
\caption{
Same as bottom left and right panels of Figure \ref{Figure8}, except assuming $L=1000$ au.
We note that the ranges of the color scale for $\zeta_{\mathrm{CR}}$ are 100 times larger than that in Figures \ref{Figure8}.
\vspace{0.2cm}
}\label{FigureB3}
\end{figure*}

Figures \ref{FigureB1}-\ref{FigureB3} show the uncertainty of our derived values of ionisation rates $\zeta_{\mathrm{CR}}$ (see Section \ref{sec:4-2}), by changing $T_{\mathrm{ex}}$ (Figures \ref{FigureB1} and \ref{FigureB2}) and the depth of the structure $L$ (Figure \ref{FigureB3}).

\end{document}